\documentclass[oneside,11pt]{article}

\usepackage{times,url,mathrsfs}
\usepackage{amsmath,wrapfig,color,bigfoot}
\usepackage{amsfonts}
\usepackage{graphicx}
\usepackage{subfigure}
\newtheorem{theorem}{Theorem}
\newtheorem{lemma}{Lemma}

\begin{document}

\title{\LARGE Sparse Recovery with Very Sparse Compressed Counting\vspace{0.5in}}
\author{ \textbf{Ping Li} \vspace{0.05in}\\
         Department of Statistics \& Biostatistics\\
         Department of Computer Science\\
       Rutgers University\\
          Piscataway, NJ 08854, USA\\
       \texttt{pingli@stat.rutgers.edu}\\\\\\
       \and
         \textbf{Cun-Hui Zhang} \vspace{0.05in}\\
         Department of Statistics \& Biostatistics\\
         Rutgers University\\
         Piscataway, NJ 08854, USA\\
         \texttt{czhang@stat.rutgers.edu}
       \and
         \textbf{Tong Zhang}\vspace{0.05in}\\
         Department of Statistics \& Biostatistics\\
         Rutgers University\\
         Piscataway, NJ 08854, USA\\
        \texttt{tongz@rci.rutgers.edu}
}

\date{}
\maketitle

\begin{abstract}
\noindent Compressed\footnote{Part of the content of this paper was submitted to a conference in May 2013.} sensing (sparse signal recovery) often encounters nonnegative data (e.g., images). Recently \cite{Report:CCCS} developed the methodology of using (dense) {\em Compressed Counting} for recovering nonnegative $K$-sparse signals. In this paper, we adopt {\em very sparse Compressed Counting} for nonnegative signal recovery. Our design matrix is sampled from a maximally-skewed $\alpha$-stable distribution ($0<\alpha<1$), and we sparsify the design matrix so that on average $(1-\gamma)$-fraction of the  entries become zero. The idea is related to {\em very sparse stable random projections}~\cite{Proc:Li_Hastie_Church_KDD06,Proc:Li_KDD07}, the prior work for estimating summary statistics of the data.\\

\noindent In our theoretical analysis,  we show  that, when $\alpha\rightarrow0$, it suffices to use  $M= \frac{K}{1-e^{-\gamma K}}\log N/\delta$ measurements, so that with probability $1-\delta$, all coordinates can be recovered within $\epsilon$ additive precision, in one scan of the coordinates. If $\gamma = 1$ (i.e., dense design), then  $M = K\log N/\delta$. If $\gamma = 1/K$ or $2/K$ (i.e., very sparse design), then $M = 1.58K\log N/\delta$ or $M = 1.16K\log N/\delta$. This means the design matrix can be indeed very sparse at only a minor inflation of the sample complexity. \\

\noindent Interestingly, as $\alpha\rightarrow1$, the required number of measurements is essentially  $M = eK\log N/\delta$ provided $\gamma = 1/K$. It turns out that this complexity  $eK\log N/\delta$  (at $\gamma = 1/K$) is a general worst-case bound.

\end{abstract}

\newpage\clearpage

\section{Introduction}

In a recent paper~\cite{Report:CCCS}, we developed a new framework for {compressed sensing} (sparse signal recovery)~\cite{Article:Donoho_CS_JIT06,Article:Candes_Robust_JIT06}, by focusing on  nonnegative sparse signals, i.e.,  $\mathbf{x}\in\mathbb{R}^{N}$ and $x_i\geq 0, \forall\ i$. Note that real-world signals are often {nonnegative}.  The technique was based on {\em Compressed Counting (CC)}~\cite{Proc:Li_SODA09,Proc:Li_UAI09,Proc:Li_Zhang_COLT11}. In that framework, entries of the (dense) design matrix are sampled i.i.d. from an $\alpha$-stable maximally-skewed distribution. In this paper, we integrate the idea of {\em very sparse stable random projections}~\cite{Proc:Li_Hastie_Church_KDD06,Proc:Li_KDD07} into the procedure, to develop {\em very sparse compressed counting for compressed sensing}.\\

In this paper, our  procedure for compressed sensing first collects $M$ non-adaptive linear measurements
\begin{align}
y_j = \sum_{i=1}^N x_i \left[s_{ij}r_{ij}\right],\hspace{0.5in} j = 1, 2, ..., M
\end{align}
Here, $s_{ij}$ is the $(i,j)$-th entry of the design matrix with $s_{ij}\sim S(\alpha,1,1)$ i.i.d, where $S(\alpha,1,1)$ denotes an $\alpha$-stable maximally-skewed (i.e., skewness  = 1) distribution with unit scale.  Instead of using a dense design matrix, we randomly sparsify $(1-\gamma)$-fraction of the entries of the design matrix to be zero, i.e.,
\begin{align}
r_{ij} = \left\{\begin{array}{ll}
1 & \text{ with prob. } \gamma \\
0 & \text{ with prob. } 1-\gamma
\end{array}\right.\ \ \ i.i.d.
\end{align}
And any $s_{ij}$ and  $r_{ij}$ are also independent. \\

In the decoding phase, our proposed estimator of the $i$-th coordinate $x_i$ is  simply
\begin{align}
\hat{x}_{i,min,\gamma} = \min_{j\in T_i} \frac{y_j}{s_{ij}r_{ij}}
\end{align}
where $T_i$ is the set of nonzero entries in the $i$-th row of the design matrix, i.e.,
\begin{align}
T_i = \{j,\ 1\leq j\leq M, \ r_{ij}= 1\}
\end{align}
Note that the size of the set $|T_i|\sim Binomial(M,\gamma)$.\\

To analyze the sample complexity (i.e., the required number of measurements),  we need to study the following error probability
\begin{align}
&\mathbf{Pr}\left(\hat{x}_{i,min,\gamma}> x_i + \epsilon\right)
\end{align}
from which we can derive the sample complexity by using the following inequality
\begin{align}
N\mathbf{Pr}\left(\hat{x}_{i,min,\gamma}> x_i + \epsilon\right)\leq \delta
\end{align}
so that any $x_i$ can be estimated within $(x_i,\ \ x_i+\epsilon)$ with a probability (at least) $1-\delta$. \\

\noindent\textbf{Main Result 1}: As $\alpha\rightarrow0+$, the required number of measurements is
\begin{align}
M =\frac{1}{-\log \left[1-\frac{1}{K+1}\left(1-(1-\gamma)^{K+1}\right)\right]} \log N/\delta
\end{align}
which can essentially be written as
\begin{align}\label{eqn_M0+_appr}
M = \frac{K}{1-e^{-\gamma K }} \log N/\delta
\end{align}
If $\gamma = 1/K$, then the required $M$ is about $1.58K\log N/\delta$. If $\gamma = 2/K$, then  $M$ is about $1.16K\log N/\delta$. In other words, we can  use a very sparse design matrix and the required number of measurements will only be inflated slightly, if we choose to use a  small $\alpha$.\\

Indeed, using $\alpha\rightarrow0+$ achieves the smallest complexity. However, there will be a numerical issue if $\alpha$ is too small.  To see this, consider the approximate mechanism for generating $S(\alpha,1,1)$ by using $1/U^{1/\alpha}$, where $U\sim unif(0,1)$. If $\alpha=0.05$, then we have to compute $(1/U)^{20}$, which may potentially create numerical problems. In our Matlab simulations,  we do not notice obvious numerical issues  with $\alpha=0.05$ (or even smaller). However, if a device (e.g., camera or other hand-held device)  has a limited precision and/or memory, then we expect that we must use a larger $\alpha$, away from 0. \\

\noindent\textbf{Main Result 2}: \ \ If $x_i>\epsilon$ whenever $x_i>0$, then as $\alpha\rightarrow1-$, the required number of measurements is
\begin{align}
M =  \frac{1}{-\log\left(1-\frac{1}{K+1}\left(1-\frac{1}{K+1}\right)^K\right)}\log N/\delta,\hspace{0.5in} \text{with } \ \gamma=\frac{1}{K+1}
\end{align}
This complexity bound can essentially be written as
\begin{align}
M = eK\log N/\delta,\hspace{0.5in} \text{with } \ \gamma=\frac{1}{K}
\end{align}

Interestingly, this result $eK\log N/\delta$ (with $\gamma=1/K$) is the general worse-case  bound.

\section{A Simulation Study}

\begin{figure}[h!]
\begin{center}
\mbox{
\includegraphics[width=3in]{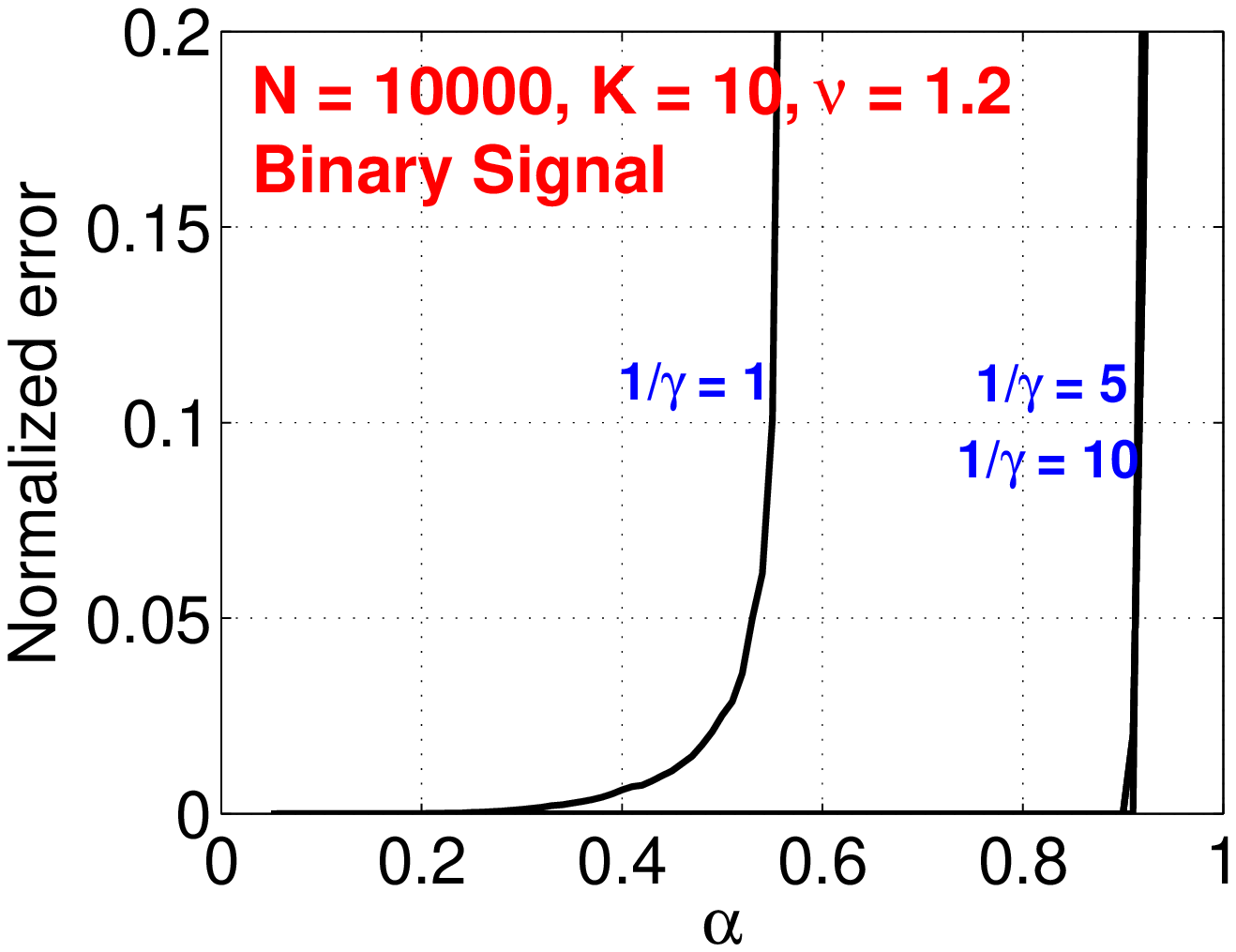}
\includegraphics[width=3in]{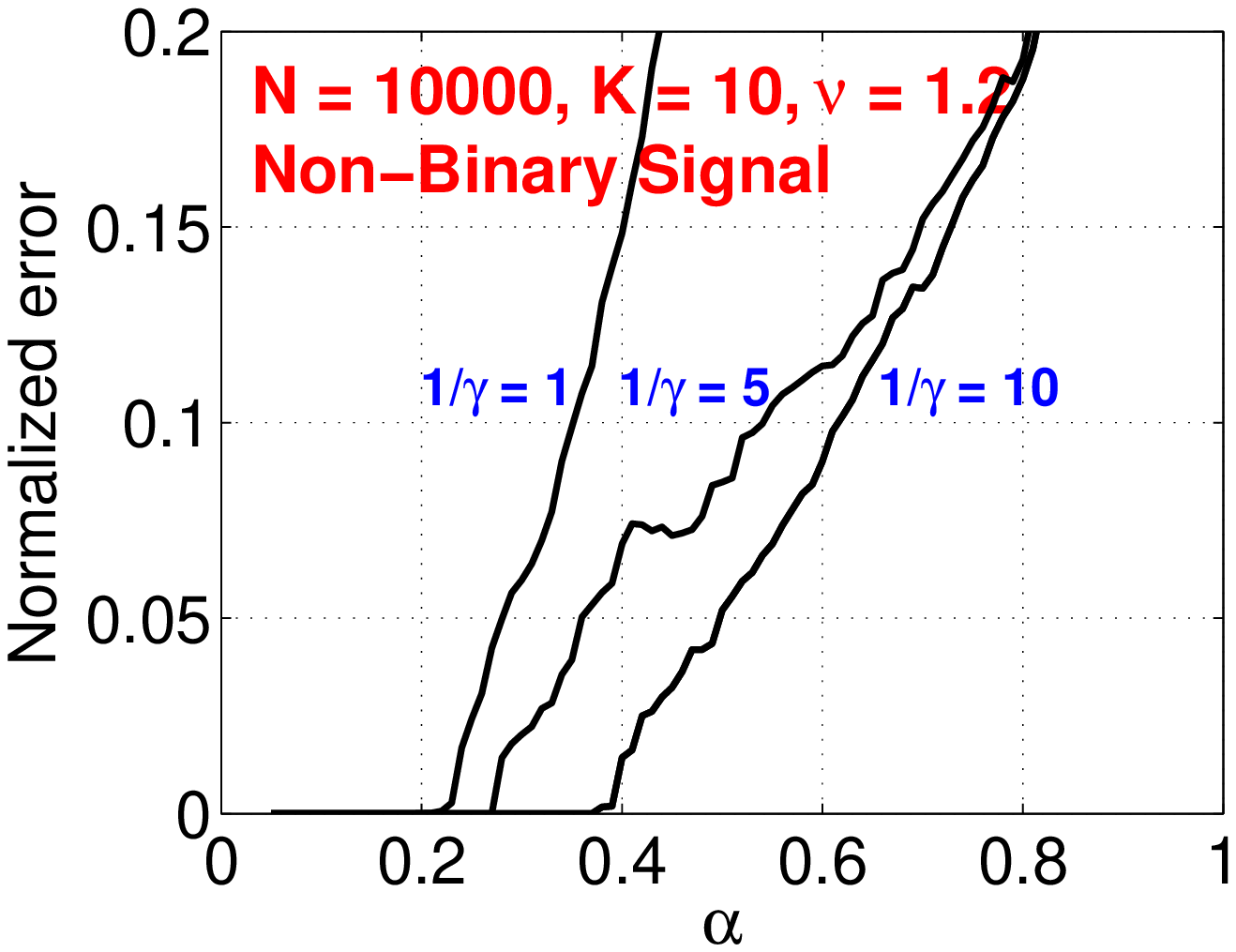}
}

\mbox{
\includegraphics[width=3in]{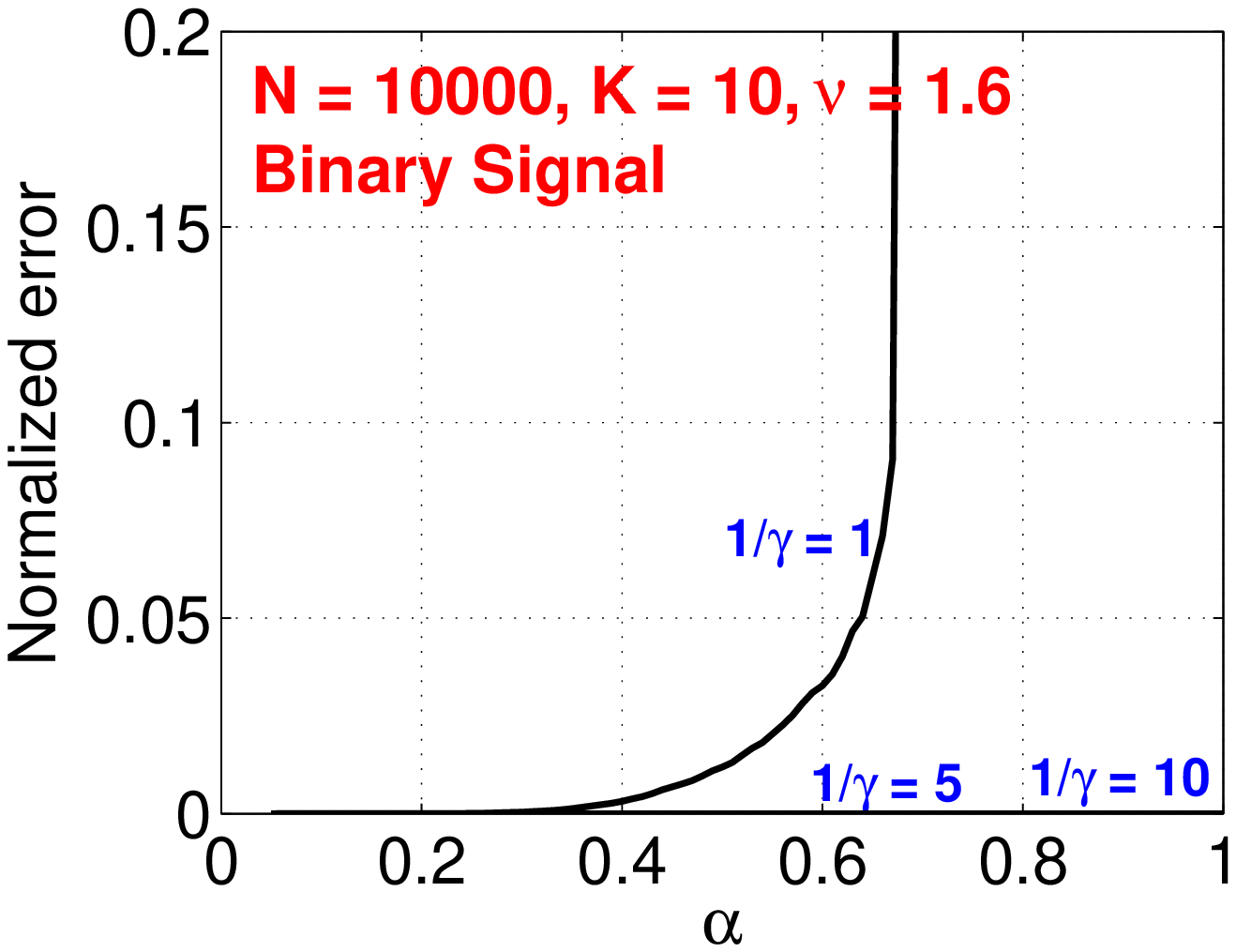}
\includegraphics[width=3in]{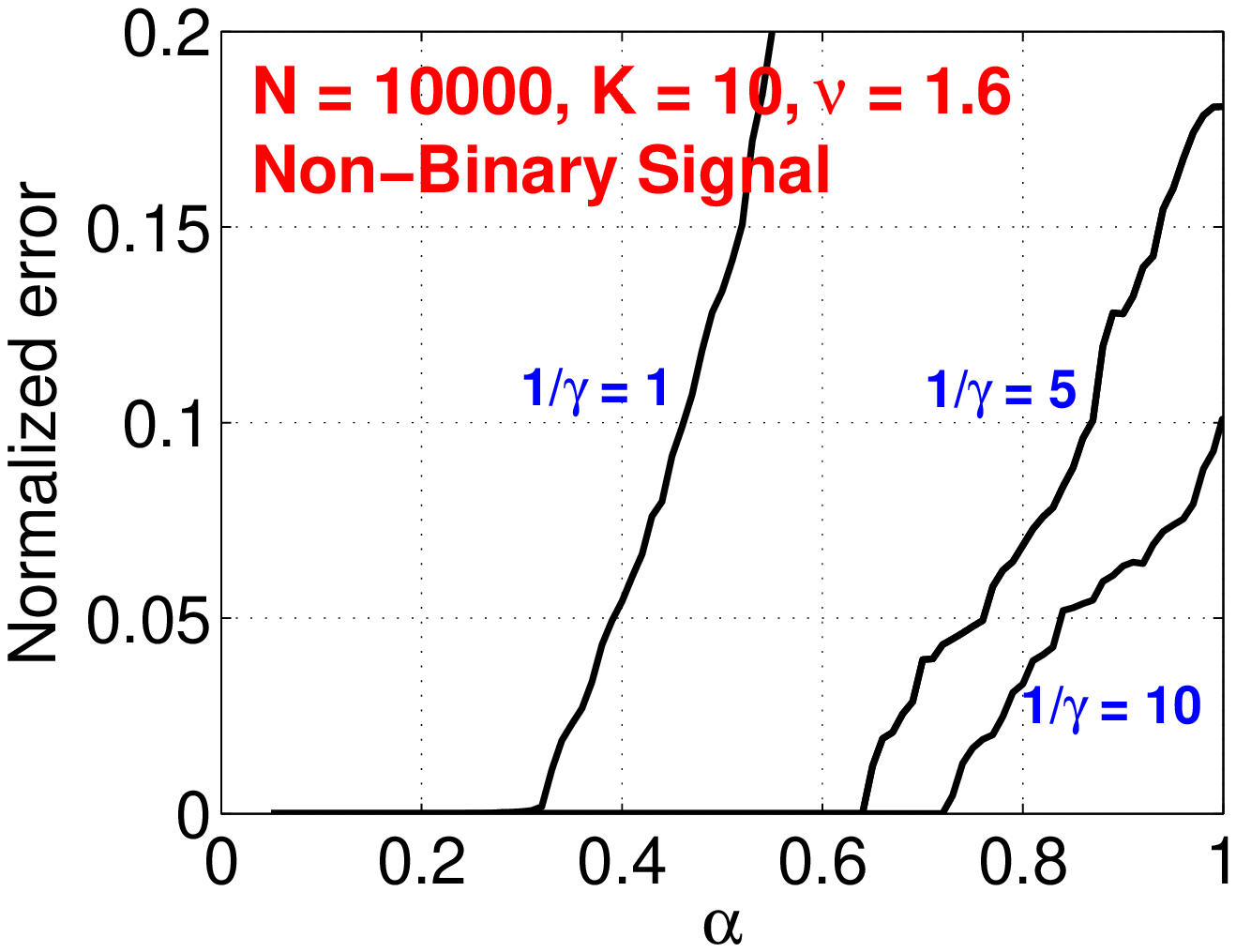}
}

\mbox{
\includegraphics[width=3in]{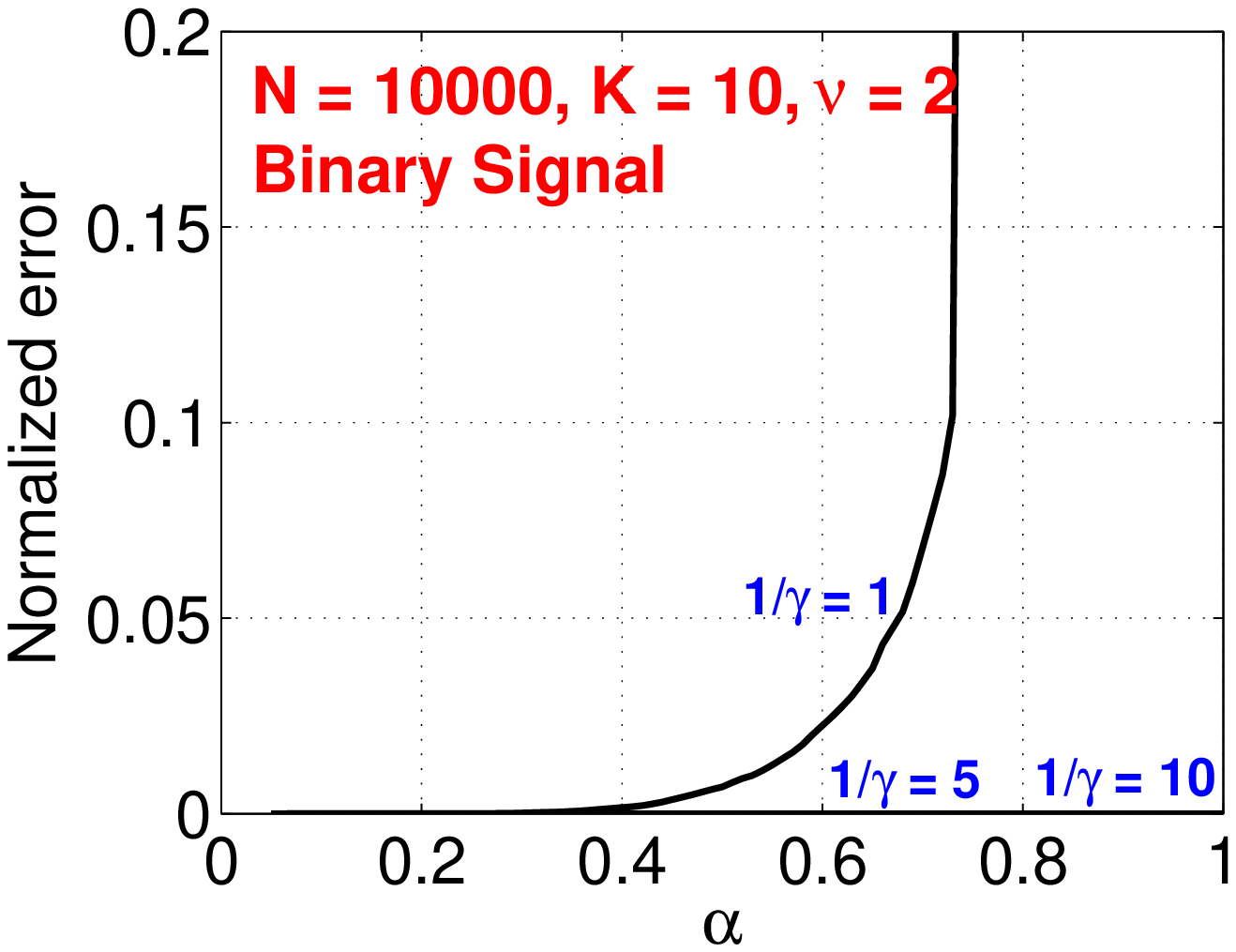}
\includegraphics[width=3in]{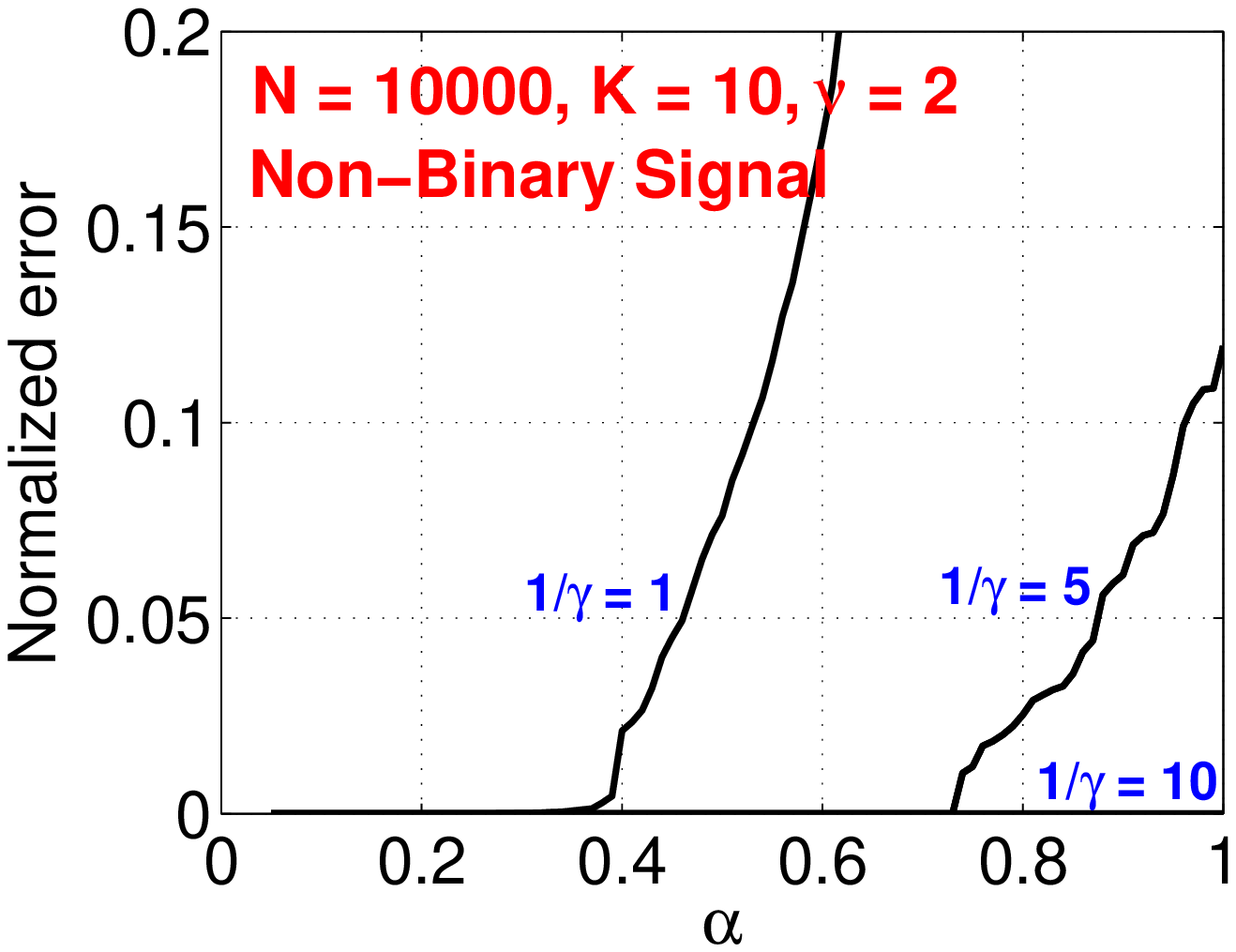}
}

\end{center}
\vspace{-0.2in}
\caption{Normalized estimation errors (\ref{eqn_error}) with $N=10000$ and $K=10$. }\label{fig_N10000K10}
\end{figure}

\begin{figure}[h!]
\begin{center}
\mbox{
\includegraphics[width=3in]{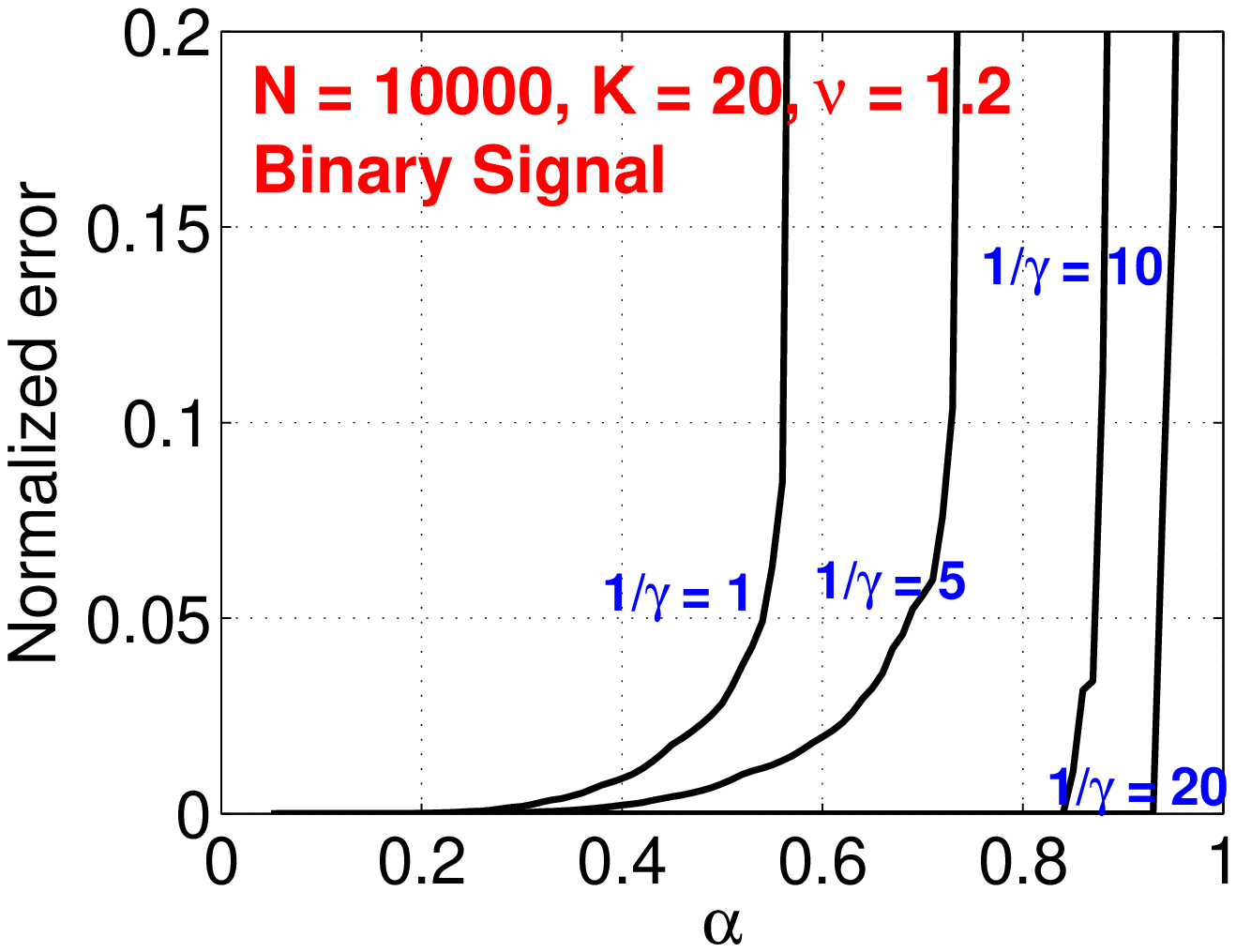}
\includegraphics[width=3in]{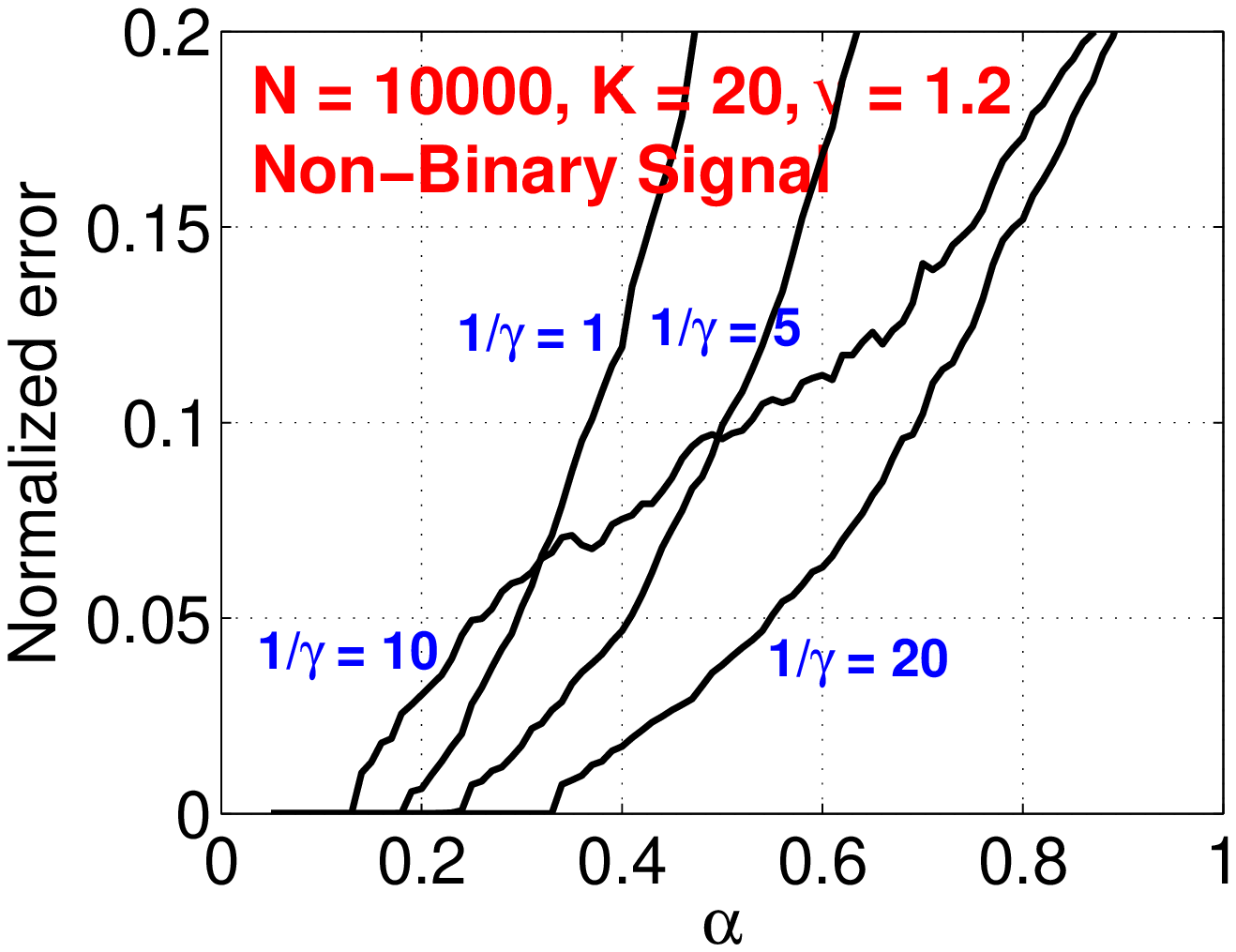}
}

\mbox{
\includegraphics[width=3in]{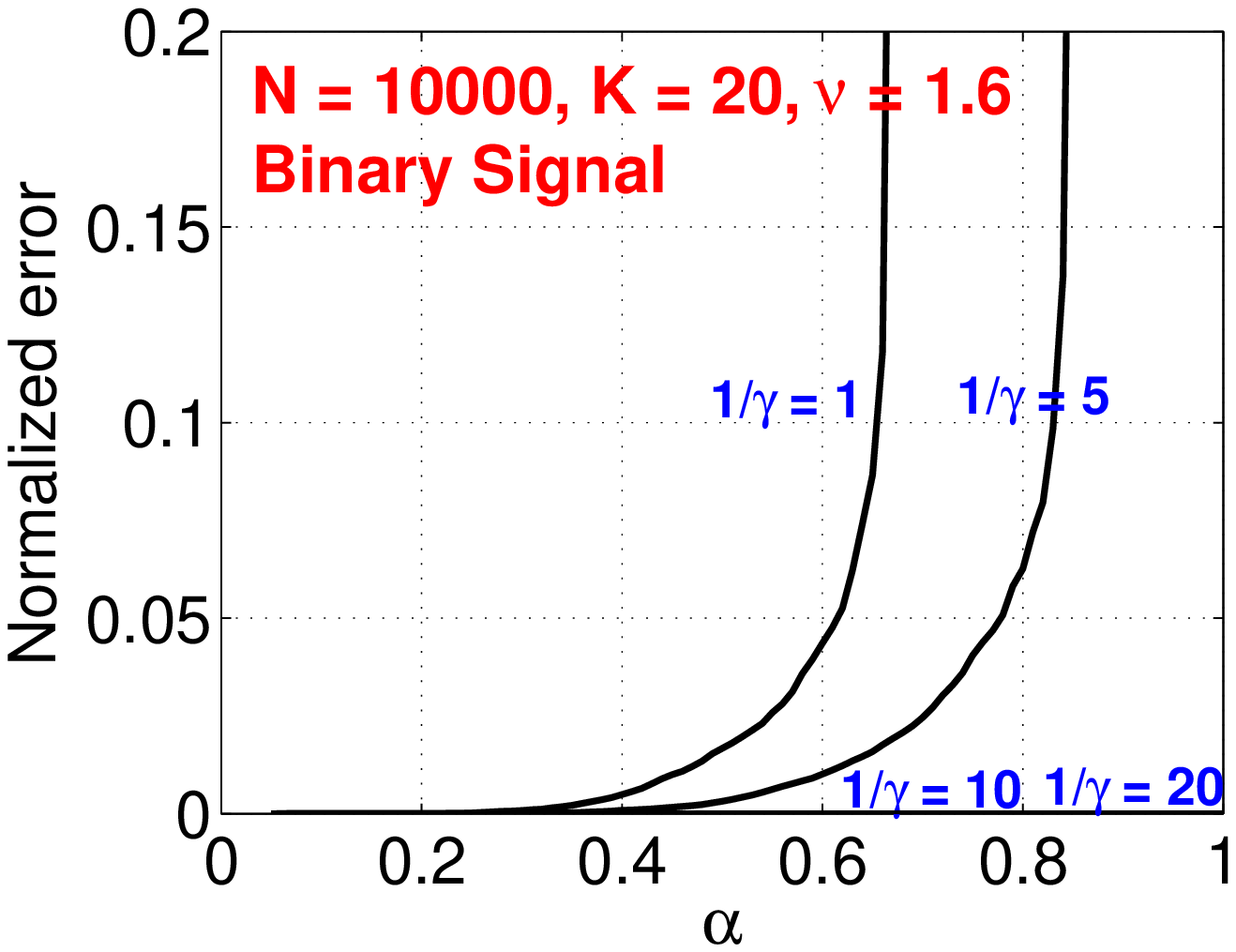}
\includegraphics[width=3in]{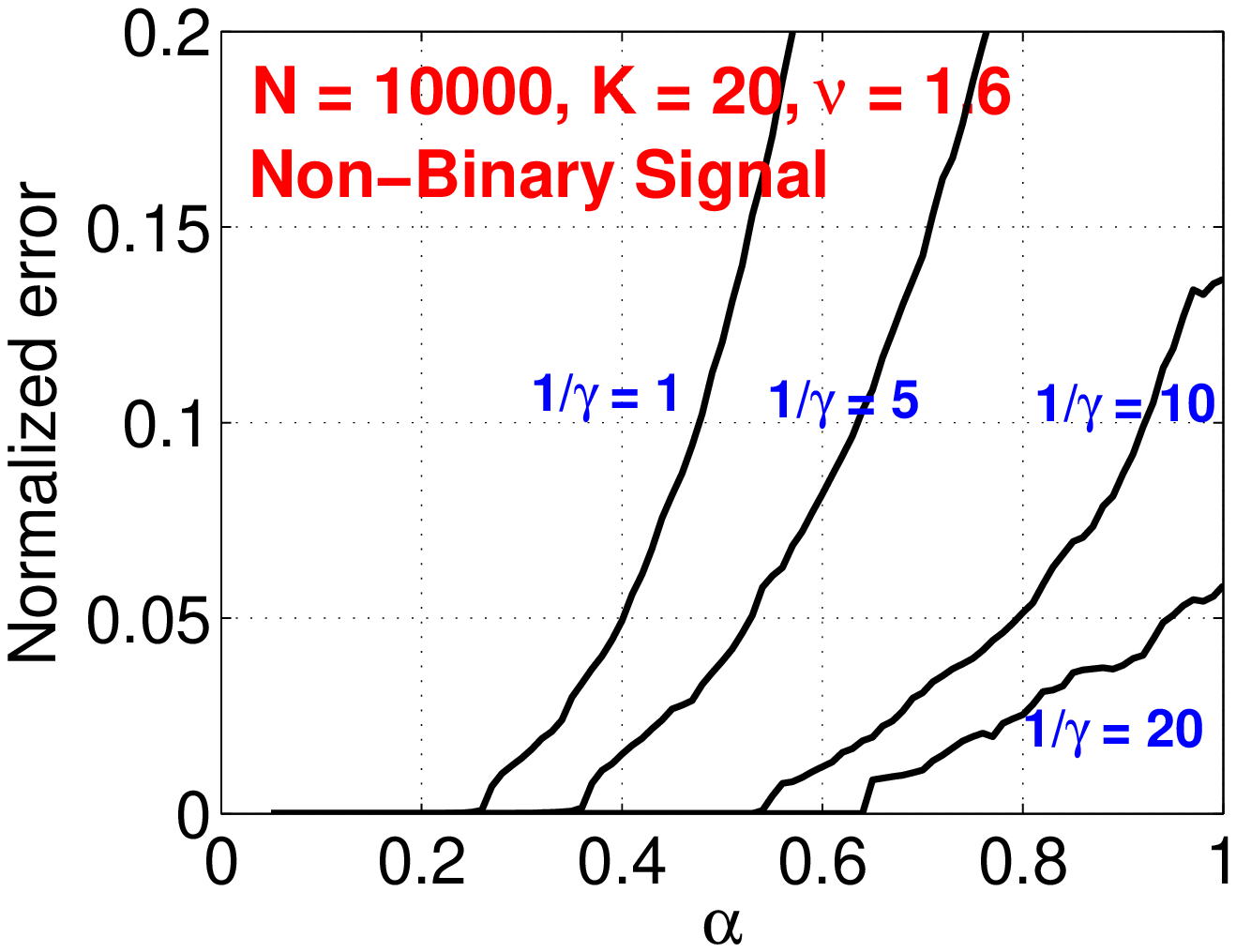}
}

\mbox{
\includegraphics[width=3in]{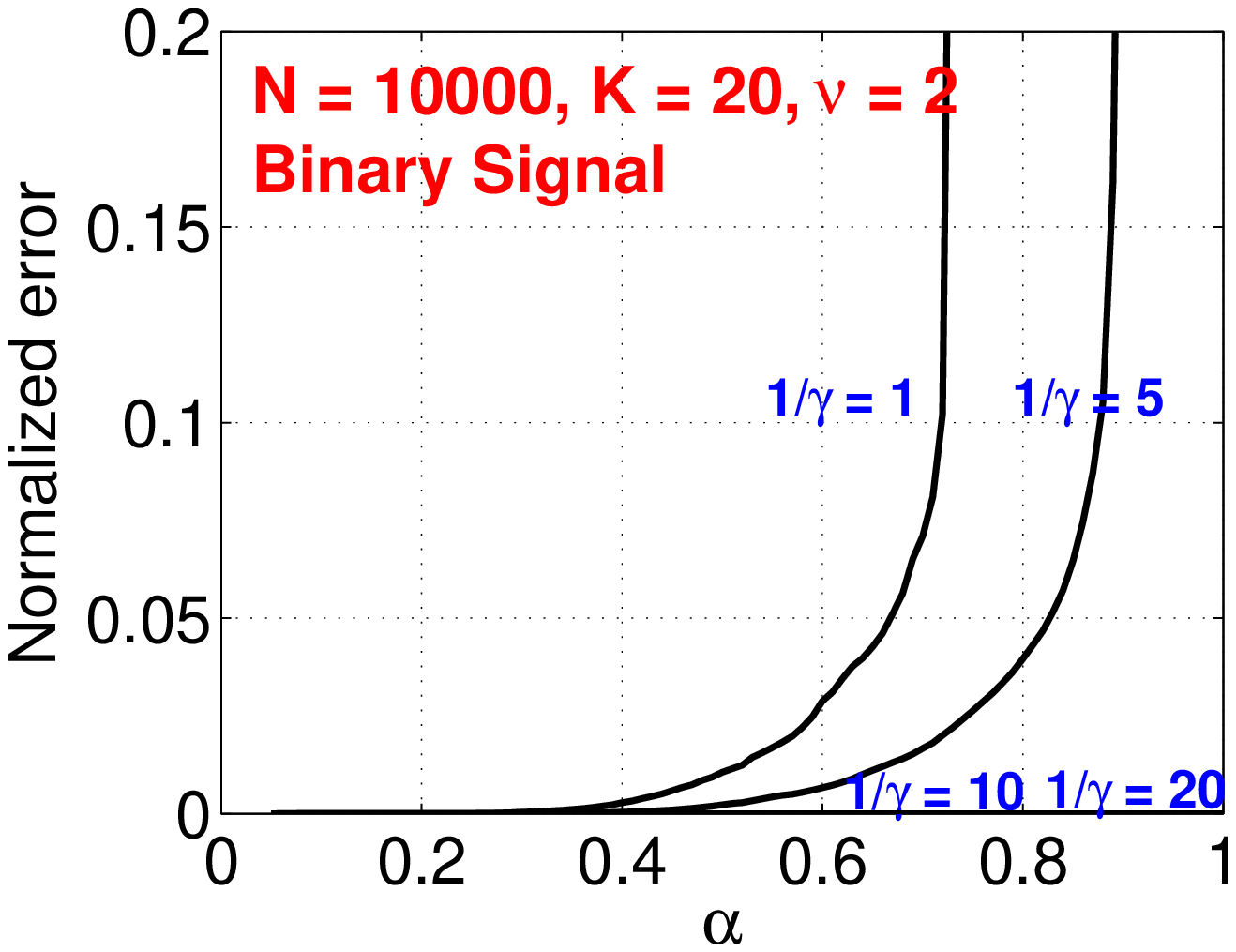}
\includegraphics[width=3in]{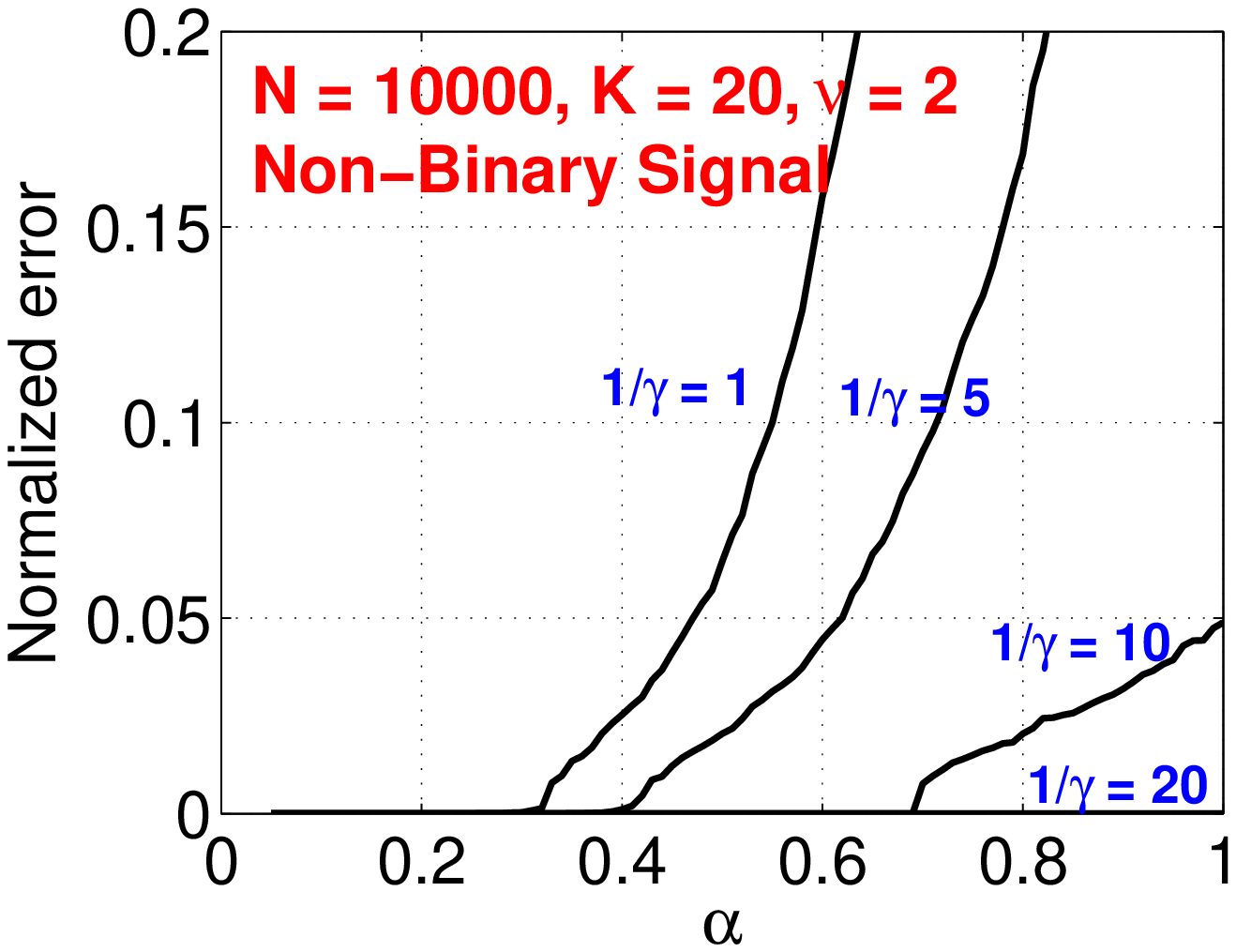}
}

\end{center}
\vspace{-0.2in}
\caption{Normalized estimation errors (\ref{eqn_error}) with $N=10000$ and $K=20$.}\label{fig_N10000K20}
\end{figure}

\begin{figure}[h!]
\begin{center}
\mbox{
\includegraphics[width=3in]{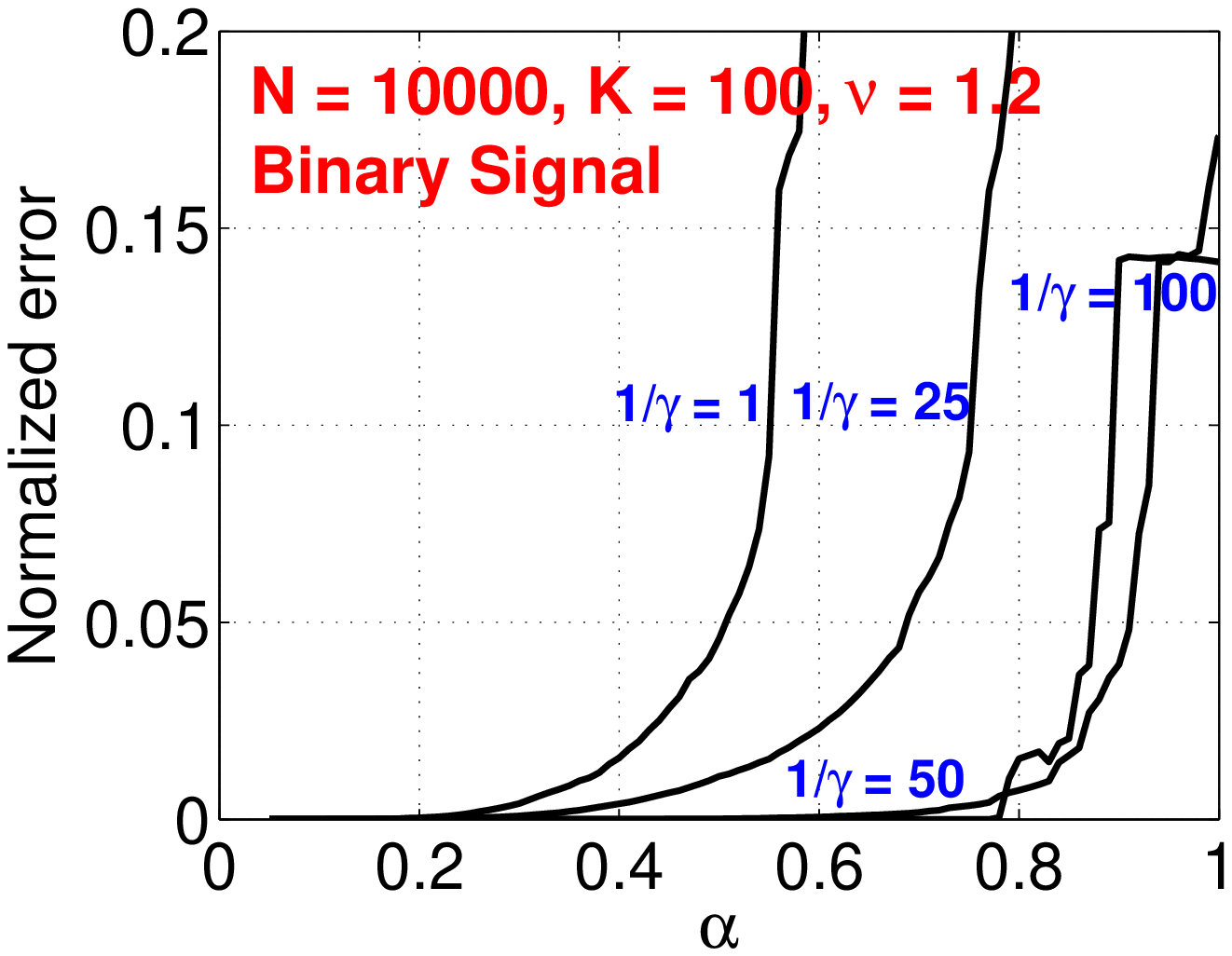}
\includegraphics[width=3in]{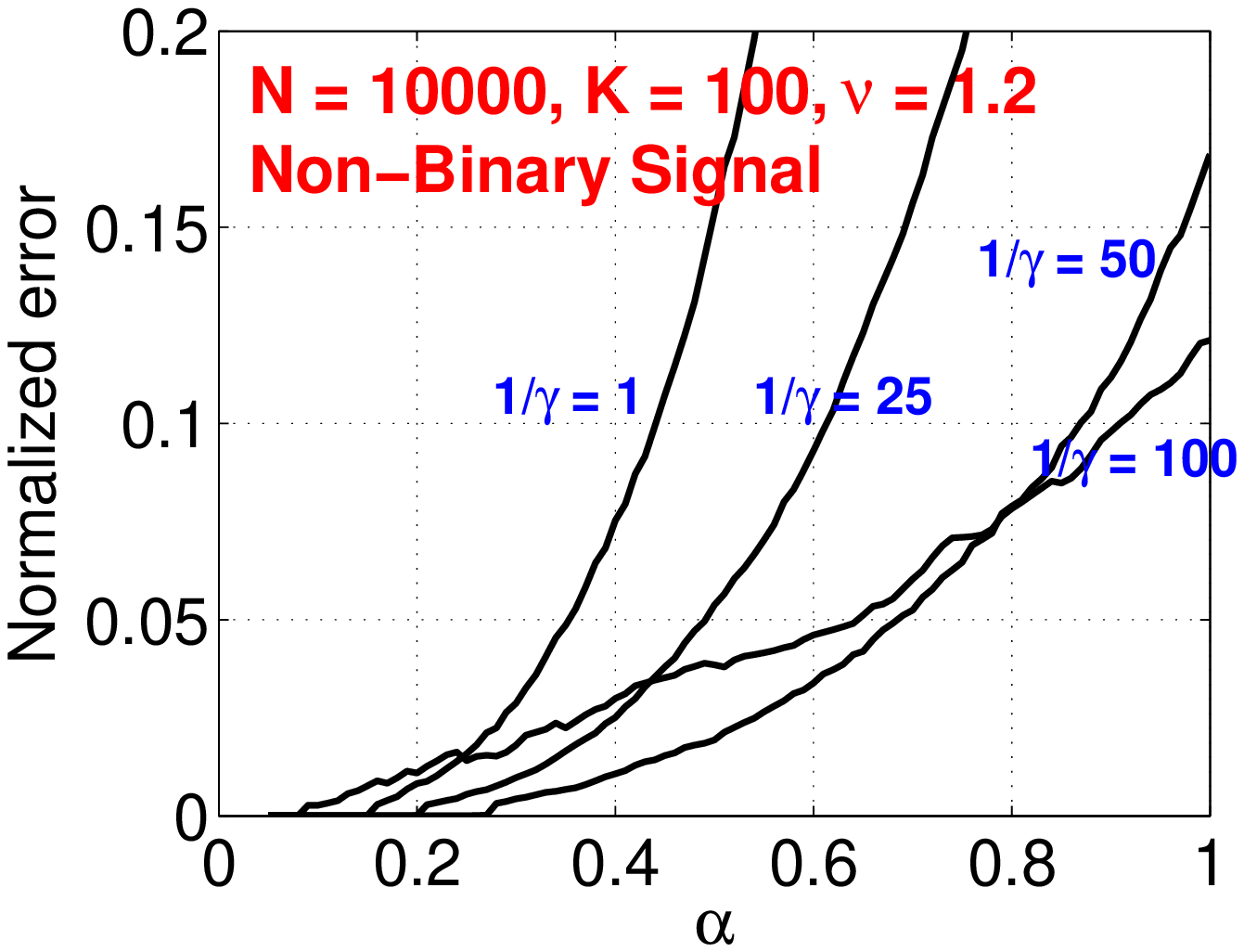}
}

\mbox{
\includegraphics[width=3in]{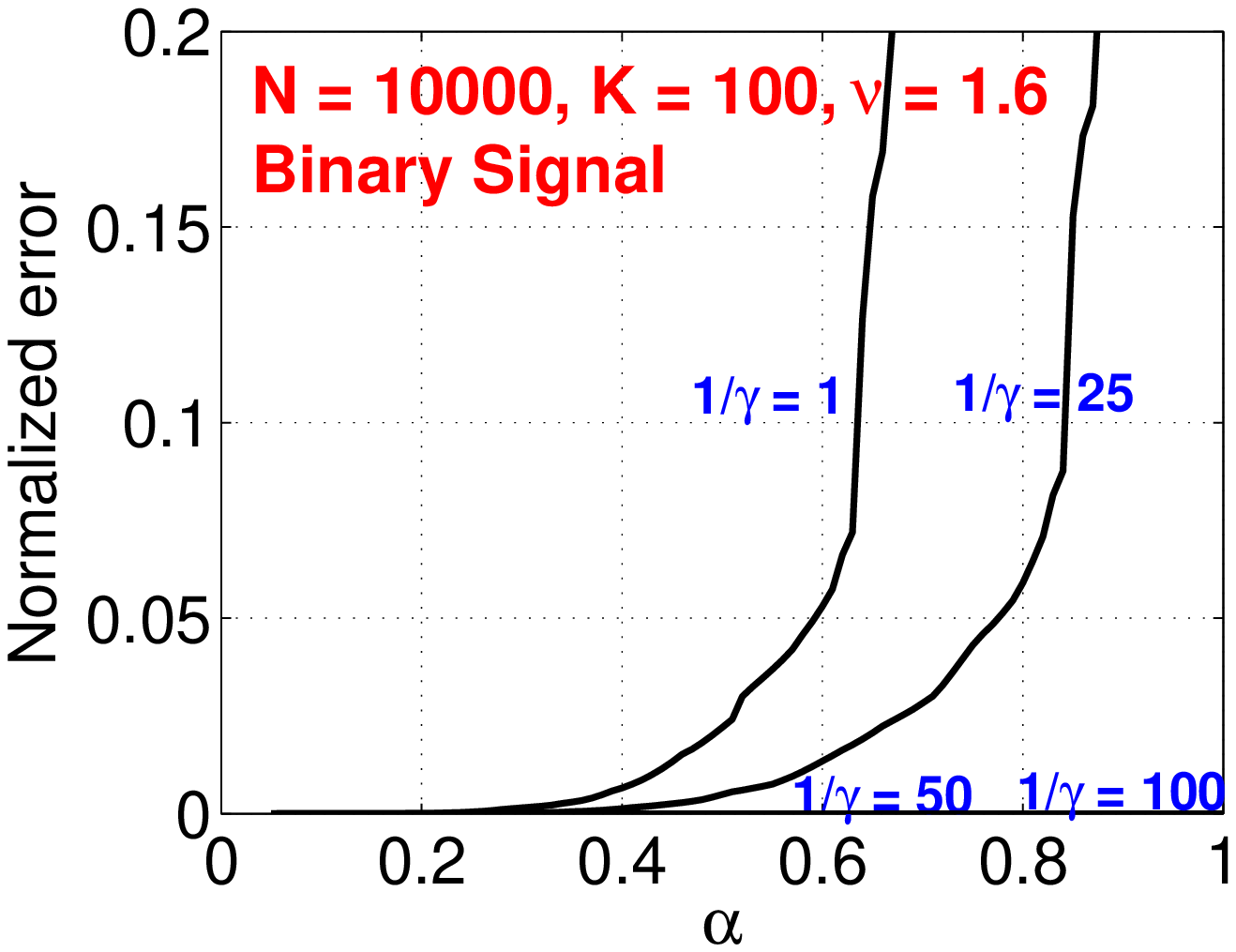}
\includegraphics[width=3in]{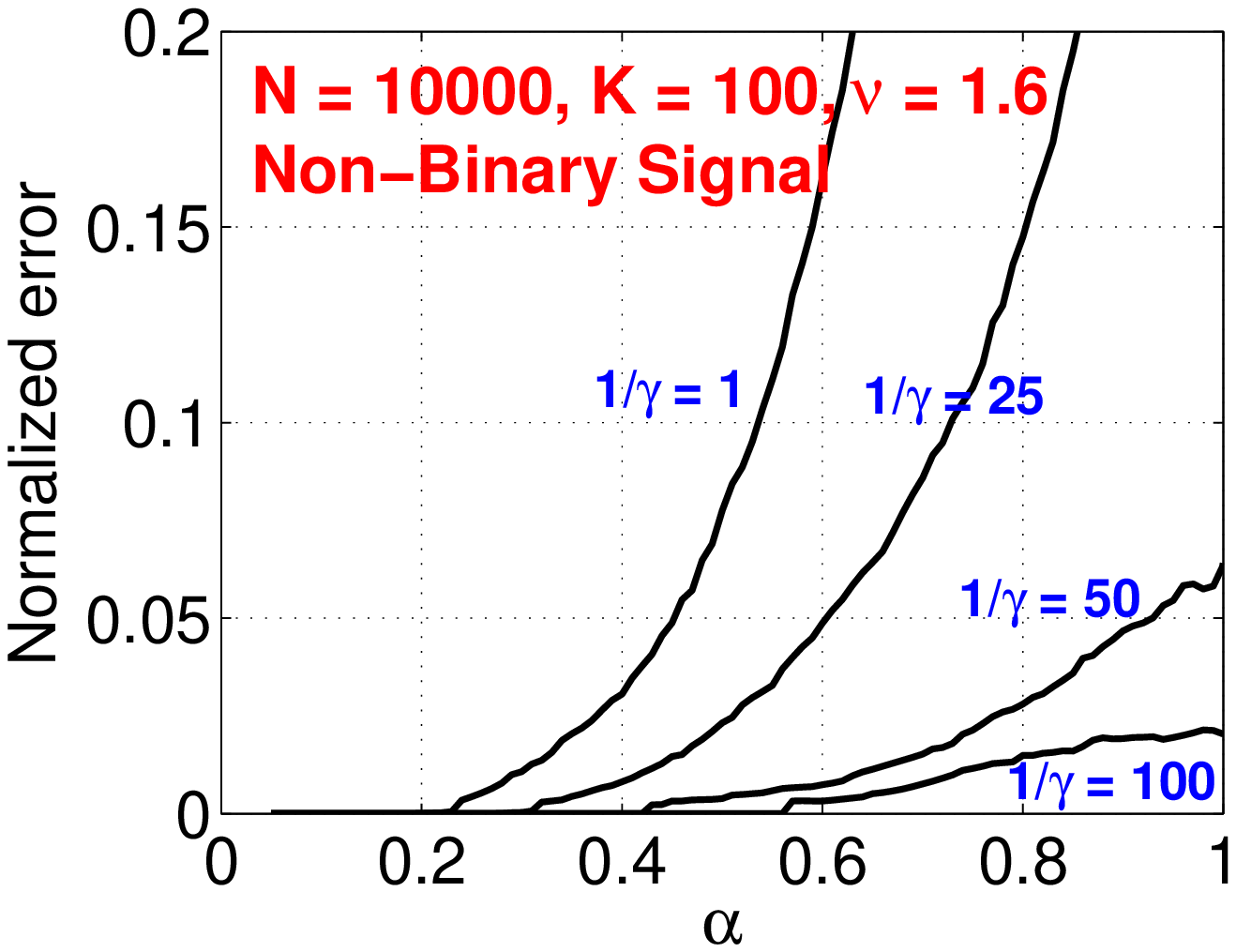}
}

\mbox{
\includegraphics[width=3in]{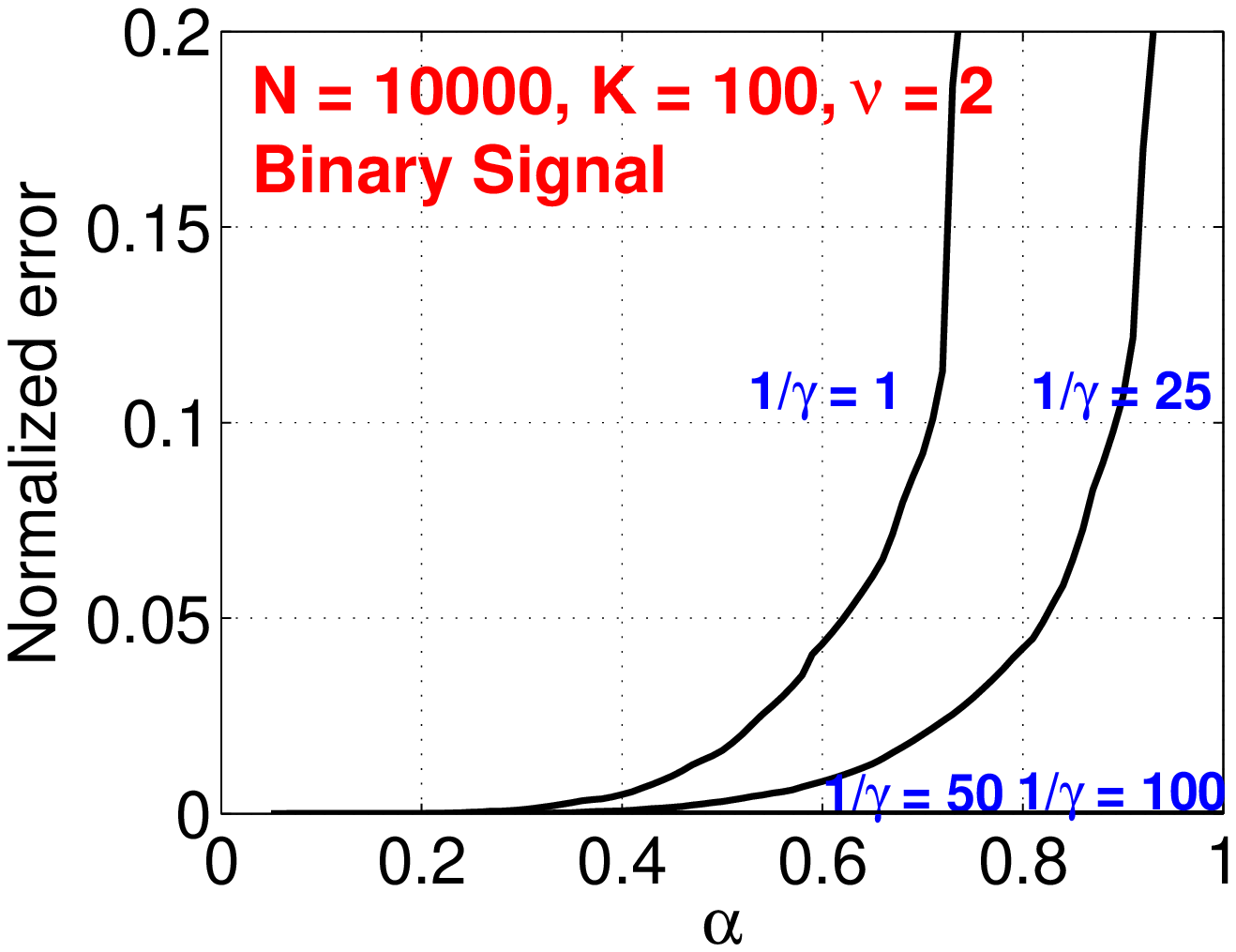}
\includegraphics[width=3in]{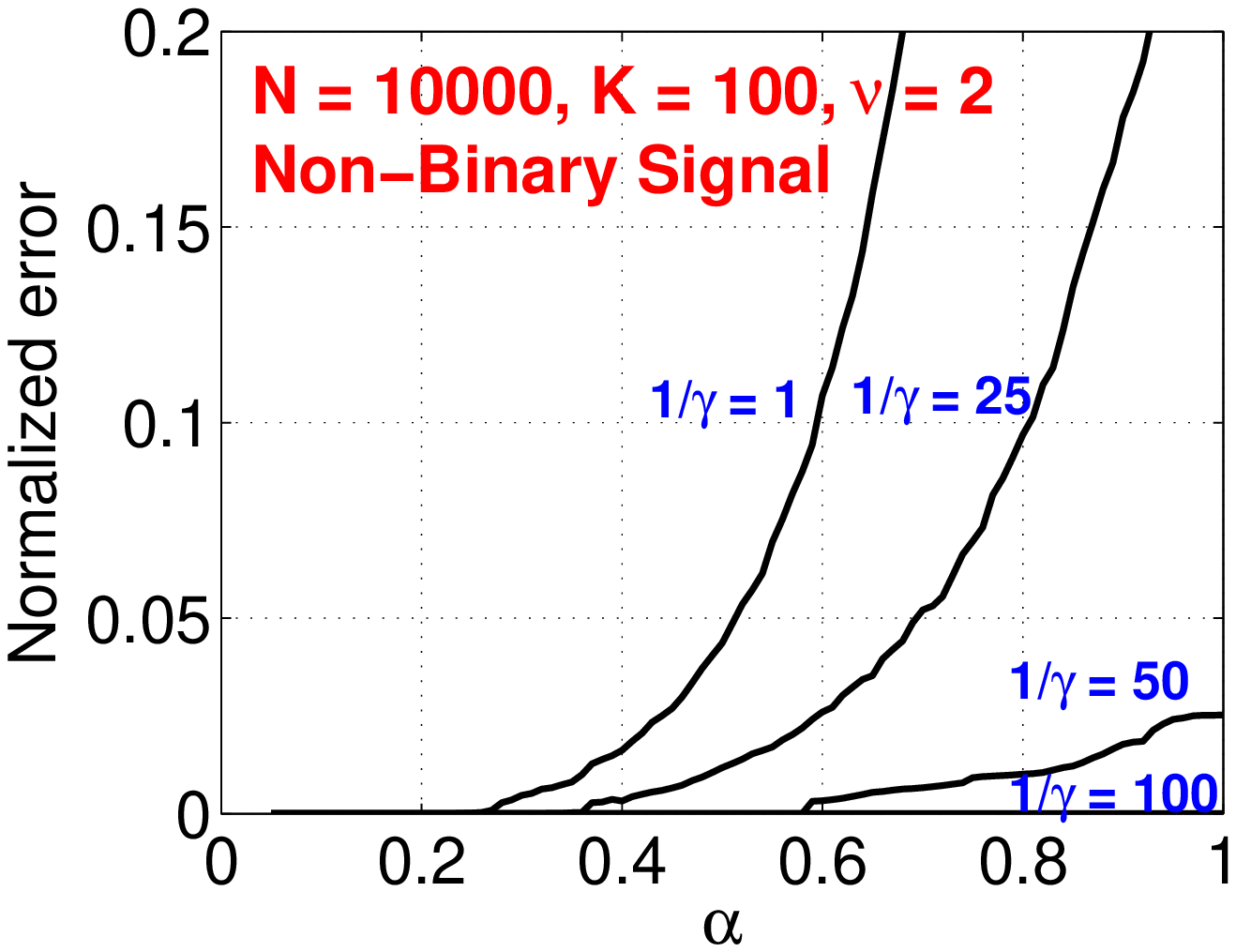}
}

\end{center}
\vspace{-0.2in}
\caption{Normalized estimation errors (\ref{eqn_error}) with $N=10000$ and $K=100$.}\label{fig_N10000K100}
\end{figure}

\begin{figure}[h!]
\begin{center}
\mbox{
\includegraphics[width=3in]{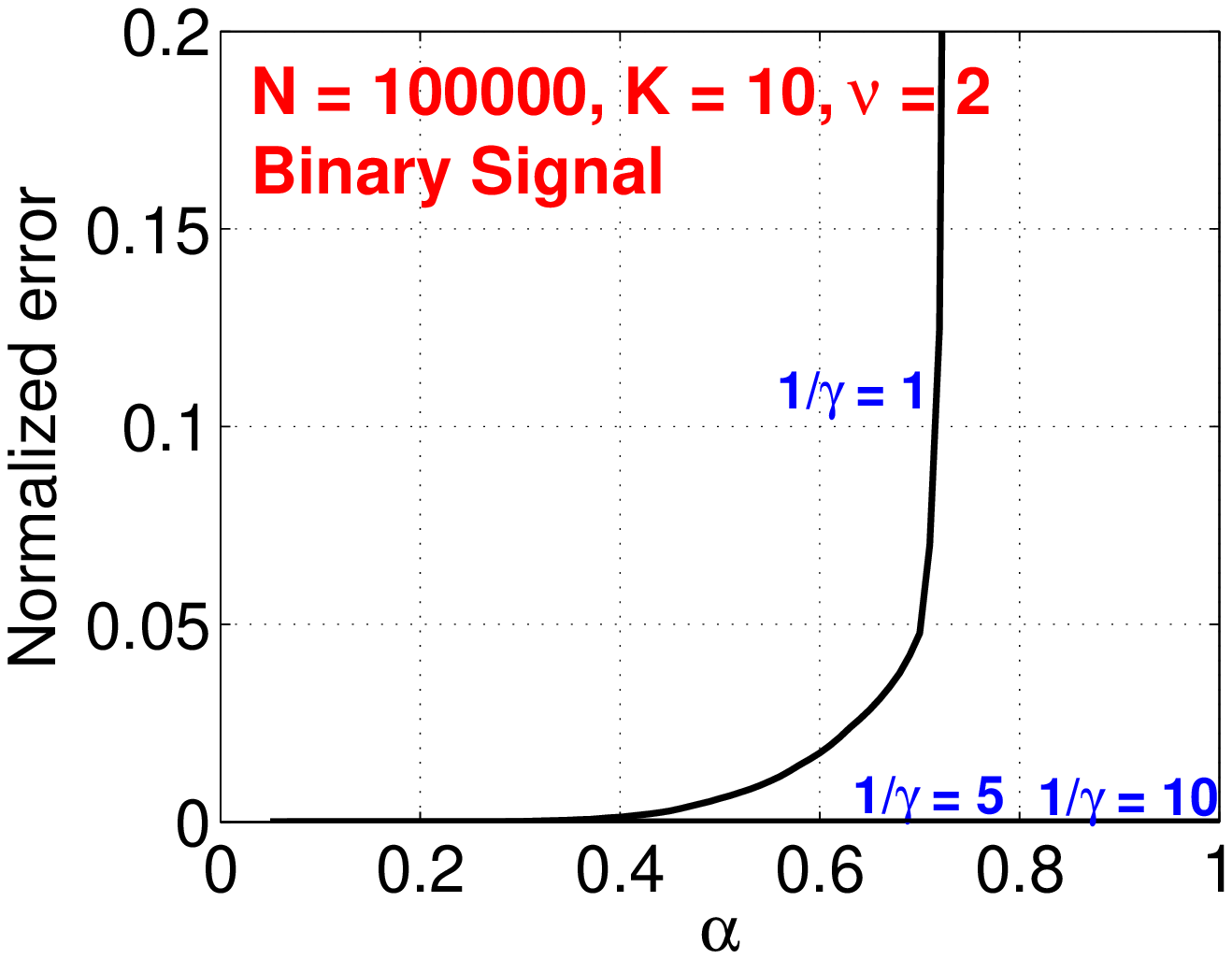}
\includegraphics[width=3in]{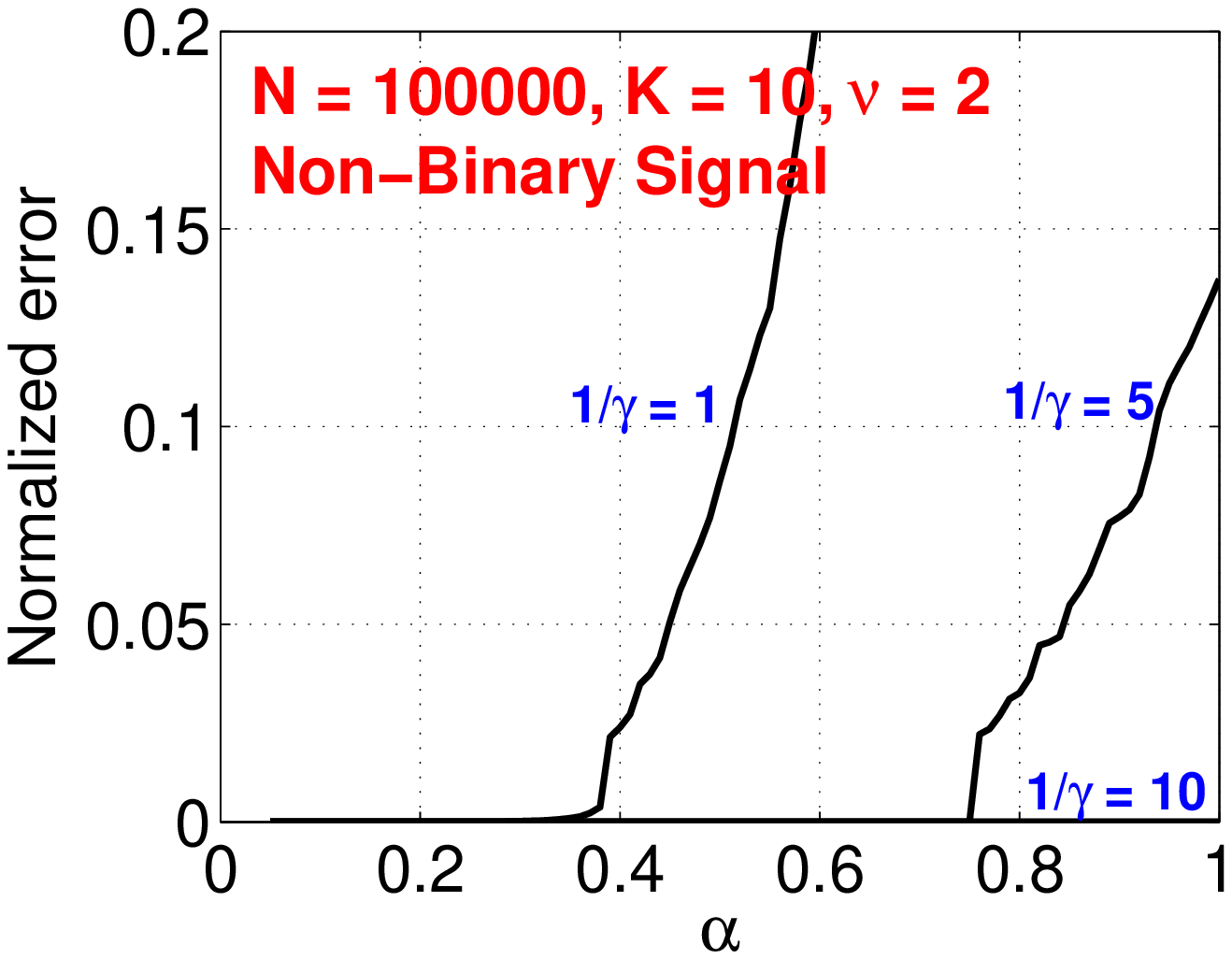}
}

\mbox{
\includegraphics[width=3in]{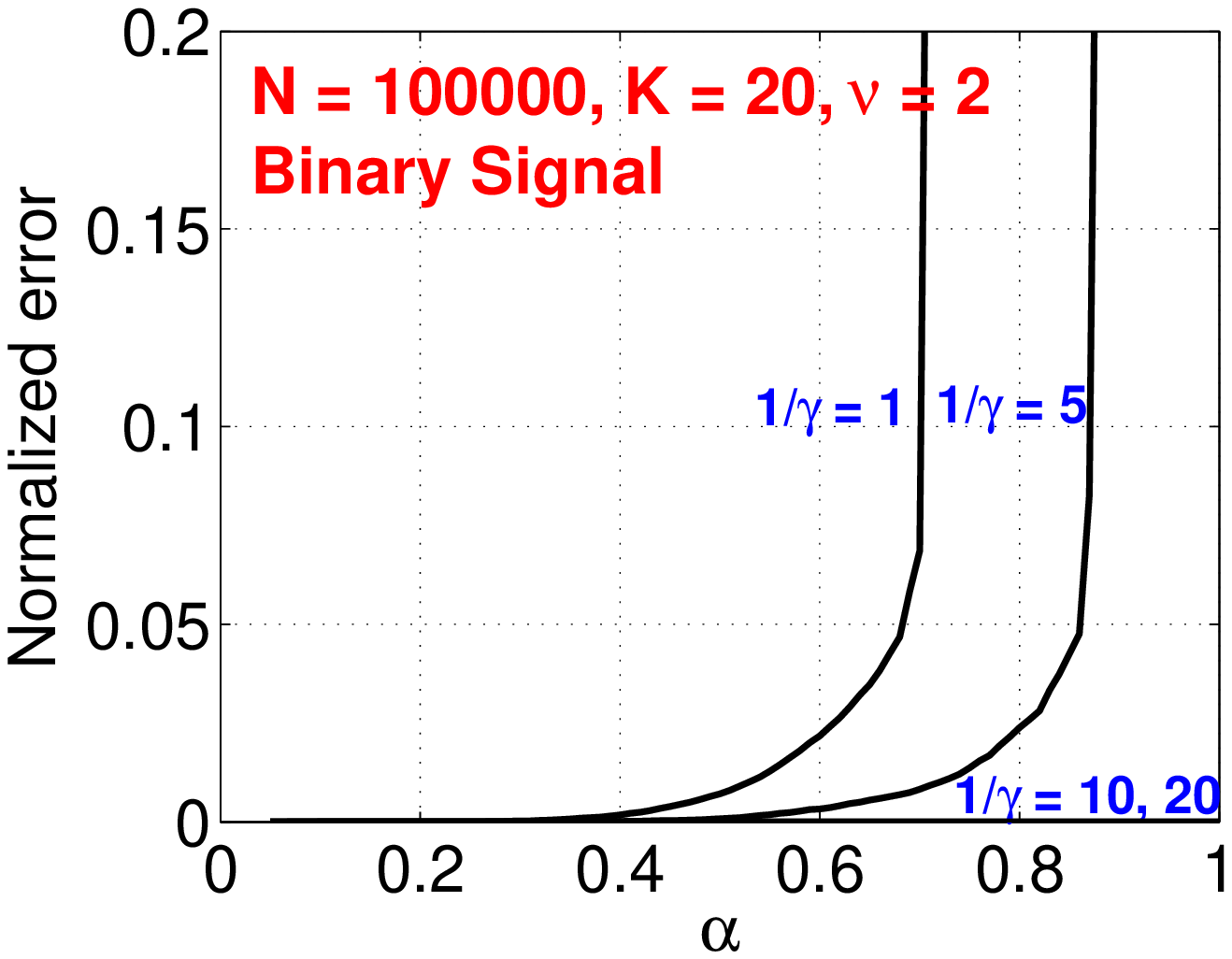}
\includegraphics[width=3in]{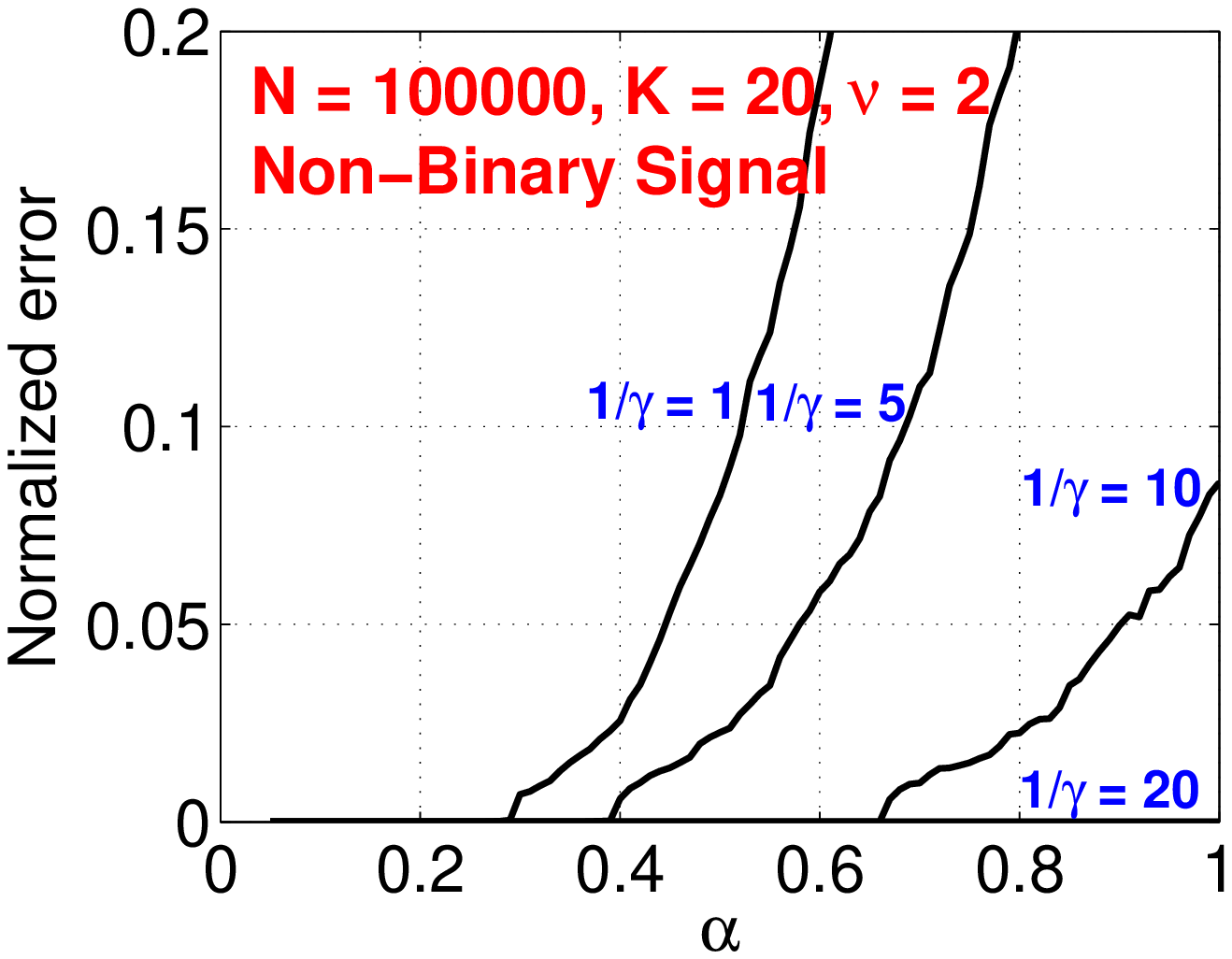}
}

\mbox{
\includegraphics[width=3in]{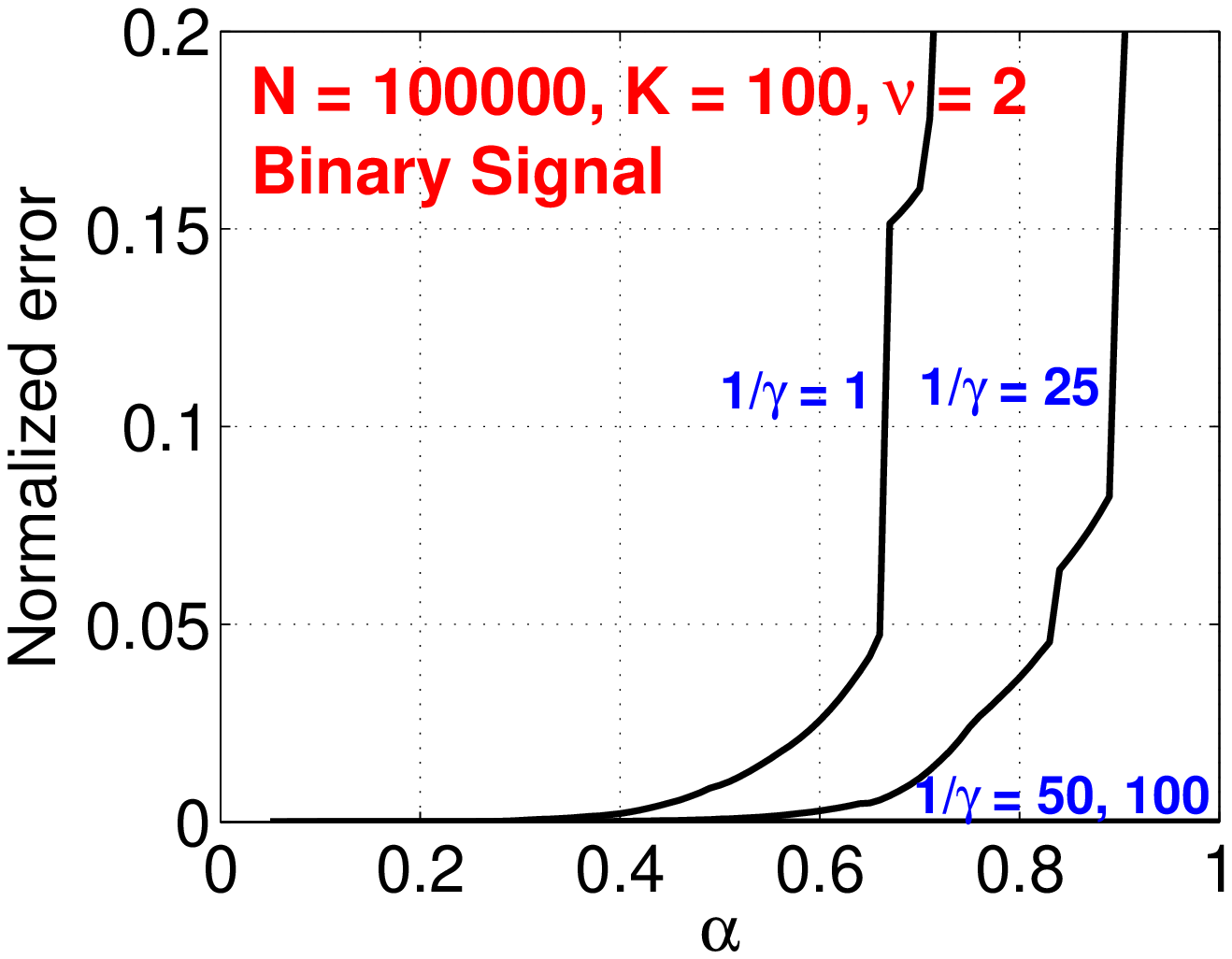}
\includegraphics[width=3in]{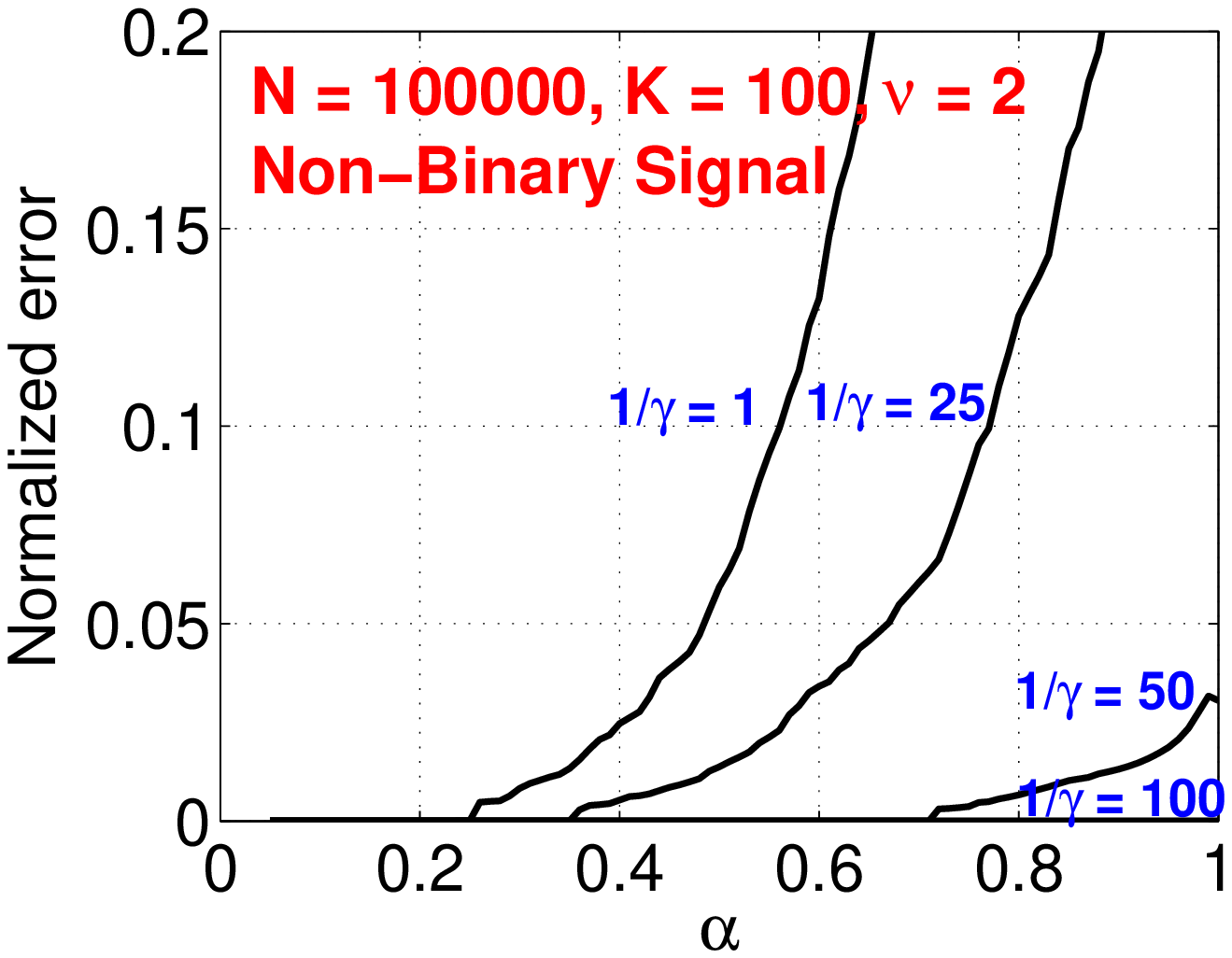}
}

\end{center}
\vspace{-0.2in}
\caption{Normalized estimation errors (\ref{eqn_error}) with $N=100000$ and $\nu=2$.}\label{fig_N100000v2}
\end{figure}

We\footnote{This  report does not include  comparisons with the SMP algorithm~\cite{Proc:SMP08,Article:Gilbert_IEEE10}, as we can not run the code  from \url{http://groups.csail.mit.edu/toc/sparse/wiki/index.php?title=Sparse_Recovery_Experiments}, at the moment. We will provide the comparisons after we are able to execute the code. We thank the communications with the author of~\cite{Proc:SMP08,Article:Gilbert_IEEE10}.}
 consider two types of signals. To generate ``binary signal'', we  randomly select $K$ (out of $N$) coordinates to be 1.   For ``non-binary signal'', we assign the values of  $K$ randomly selected nonzero coordinates according to $|N(0,5^2)|$.  The number of measurements is determined by
\begin{align}
M = \nu K \log N/\delta
\end{align}
where $N\in\{10000, 100000\}$, $\delta = 0.01$ and $\nu \in \{1.2, 1.6, 2\}$. We report the normalized recovery errors:
\begin{align}\label{eqn_error}
\text{Normalized Error} = \sqrt{\frac{\sum_{i=1}^N (x_i - \text{estimated } x_i)^2}{\sum_{i=1}^N x_i^2}}
\end{align}
We experiment with all possible values of $1/\gamma\in\{1, 2, 3, ..., K\}$, although we only plot a few selected $\gamma$ values in Figures~\ref{fig_N10000K10} to~\ref{fig_N100000v2}. For each combination $(\gamma, N, \nu)$, we conduct 100 simulations and report the median errors. The results confirm our theoretical analysis. When $\nu$ is small (i.e., less measurements), we need to choose a small $\alpha$ in order to achieve perfect recovery. When $\nu$ is large (i.e., more measurements), we can use a larger $\alpha$. Also, the simulations  confirm that, in general, we can choose a very sparse design.

\clearpage\newpage

\section{Analysis}

Recall, we collect our measurements as
\begin{align}
y_j = \sum_{i=1}^N x_i s_{ij} r_{ij},\ \ \  j = 1, 2, ..., M
\end{align}
where $s_{ij}\sim S(\alpha,1,1)$ i.i.d.  and
\begin{align}
r_{ij} = \left\{\begin{array}{ll}
1 & \text{ with prob. } \gamma \\
0 & \text{ with prob. } 1-\gamma
\end{array}\right.\ \ \ i.i.d.
\end{align}
And any $s_{ij}$ and  $r_{ij}$ are also independent. Our proposed estimator is  simply
\begin{align}
\hat{x}_{i,min,\gamma} = \min_{j\in T_i} \frac{y_j}{s_{ij}r_{ij}}
\end{align}
where $T_i$ is the set of nonzero entries in the $i$-th row of $S$, i.e.,
\begin{align}
T_i = \{j,\ 1\leq j\leq M, \ r_{ij}= 1\}
\end{align}

Conditional on $r_{ij}=1$,
\begin{align}
\left.\frac{y_j}{s_{ij}r_{ij}}\right|r_{ij}=1  = \frac{\sum_{t=1}^Nx_ts_{tj}r_{tj}}{s_{ij}}=x_i + \frac{\sum_{t\neq i}^Nx_ts_{tj}r_{tj}}{s_{ij}} = x_i + \left(\eta_{ij}\right)^{1/\alpha} \frac{S_2}{S_1}
\end{align}
where $S_1, S_2\sim S(\alpha,1,1)$, i.i.d., and
\begin{align}
\eta_{ij} = \sum_{t\neq i}^N \left(x_t r_{tj}\right)^\alpha= \sum_{t\neq i}^N x_t^\alpha r_{tj}
\end{align}
Note that
\begin{align}
E(\eta_{ij}) = \gamma\sum_{t\neq i}^N x_{tj}^\alpha\leq \gamma\sum_{t=1}^N x_{tj}^\alpha,\hspace{0.3in} \lim_{\alpha\rightarrow0+} E(\eta_{ij}) \leq \gamma K
\end{align}
When the signals are binary, i.e., $x_i\in\{0,\ 1\}$, we have
\begin{align}
\eta_{ij} \sim \left\{
\begin{array}{ll}
Binomial(K,\ \gamma)  &\text{ if } \ \   x_i = 0 \\
Binomial(K-1,\ \gamma) &\text{ if } \ \ x_i=1
\end{array}
\right.
\end{align}

The key in our theoretical analysis is  the distribution of the ratio of two independent stable random variables.  Here, we consider $S_1, S_2 \sim S(\alpha,1,1)$, i.i.d., and  define
\begin{align}
&F_\alpha(t)
= \mathbf{Pr}\left(\left({S_2}/{S_1}\right)^{\alpha/(1-\alpha)}\leq t\right),\hspace{0.2in} t\geq 0
\end{align}

There is a standard procedure to  sample from $S(\alpha,1,1)$~\cite{Article:Chambers_JASA76}. We  first generate an exponential random variable with mean 1, $w \sim \exp(1)$,  and a uniform random variable $u \sim unif \left(0, \pi\right)$, and then compute
\begin{align}\label{eqn_CMS}
\frac{\sin\left(\alpha u\right)}{\left[\sin u \cos\left(\alpha\pi/2 \right)
\right]^{\frac{1}{\alpha}}} \left[\frac{\sin\left( u - \alpha u\right)}{w}
\right]^{\frac{1-\alpha}{\alpha}} \sim S(\alpha,1,1)
\end{align}

\begin{lemma}~\cite{Report:CCCS}\label{lem_F}
For any $t\geq 0$, $S_1, S_2 \sim S(\alpha,1,1)$, i.i.d.,
\begin{align}
&F_\alpha(t)
= \mathbf{Pr}\left(\left({S_2}/{S_1}\right)^{\alpha/(1-\alpha)}\leq t\right)= \frac{1}{\pi^2}\int_0^{\pi}\int_0^{\pi} \frac{1}{1+Q_\alpha/t}du_1 du_2
\end{align}
where
\begin{align}
Q_\alpha = \left[\frac{\sin\left(\alpha u_2\right)}{\sin\left(\alpha u_1\right)}\right]^{\alpha/(1-\alpha)}\left[\frac{\sin u_1}{\sin u_2
}\right]^{\frac{1}{1-\alpha}} \frac{\sin\left( u_2 - \alpha u_2\right)}{\sin\left( u_1 - \alpha u_1\right)}
\end{align}
In particular,
\begin{align}
&\lim_{\alpha\rightarrow 0+} F_\alpha(t) = \frac{1}{1+1/t},\hspace{0.5in}
F_{0.5}(t) = \frac{2}{\pi}\tan^{-1}\sqrt{t}\hspace{0.5in}\hfill\Box
\end{align}

\end{lemma}

\subsection{Error Probability}

The following Lemma derives the general formula (\ref{eqn_Err_gamma}) for the error probability in terms of an expectation, which in general does not have a close-form solution. Nevertheless, when $\alpha=0+$ and $\alpha=0.5$, we can derive two convenient upper bounds, (\ref{eqn_Err0+_Up}) and (\ref{eqn_Err0.5_Up}), respectively, which however are not  tight.

\begin{lemma}\label{lem_Err_gamma}
\begin{align}\label{eqn_Err_gamma}
\mathbf{Pr}\left(\hat{x}_{i,min,\gamma}> x_i + \epsilon\right) =& \left[1- \gamma E\left\{F_\alpha\left(\left(\frac{\epsilon^\alpha}{\eta_{ij}}\right)^{1/(1-\alpha)}\right)\right\}\right]^M
\end{align}
When $\alpha\rightarrow0+$, we have
\begin{align}
\mathbf{Pr}\left(\hat{x}_{i,min,\gamma}> x_i + \epsilon\right) \leq& \left[1- \frac{1}{1/\gamma+K-1+1_{x_i=0}}\right]^M\\\label{eqn_Err0+_Up}
\leq&\left[1- \frac{1}{1/\gamma+K}\right]^M
\end{align}
When $\alpha=0.5$, we have
\begin{align}
\mathbf{Pr}\left(\hat{x}_{i,min,\gamma}> x_i + \epsilon\right) \leq& \left[1-\gamma\frac{2}{\pi}\tan^{-1}\left(\frac{\sqrt{\epsilon}}{\gamma\sum_{t\neq i}^N x_t^{1/2}}\right)  \right]^M\\\label{eqn_Err0.5_Up}
\leq&\left[1-\gamma\frac{2}{\pi}\tan^{-1}\left(\frac{\sqrt{\epsilon}}{\gamma\sum_{t=1}^N x_t^{1/2}}\right)  \right]^M
\end{align}

\noindent\textbf{Proof:}\hspace{0.2in} See Appendix~\ref{app_lem_Err_gamma}.$\hfill\Box$\\

\end{lemma}

It turns out, when $\alpha=0+$, we can precisely evaluate the expectation (\ref{eqn_Err_gamma}) and derive an accurate complexity bound (\ref{eqn_Err0+}) in Lemma~\ref{lem_Err_gamma0+}.
\begin{lemma}\label{lem_Err_gamma0+}
As $\alpha\rightarrow0+$,  we have
\begin{align}\label{eqn_Err0+}
\mathbf{Pr}\left(\hat{x}_{i,min,\gamma}> x_i + \epsilon\right)=&\left[1-\frac{1}{K+1_{x_i=0}}\left(1-(1-\gamma)^{K+1_{x_i=0}}\right)\right]^M\\
\leq&\left[1-\frac{1}{K+1}\left(1-(1-\gamma)^{K+1}\right)\right]^M\\
\leq&\left[1-\frac{1}{1/\gamma + K}\right]^M
\end{align}

\noindent\textbf{Proof:}\hspace{0.2in} See Appendix~\ref{app_lem_Err_gamma0+}.$\hfill\Box$

\end{lemma}

\subsection{Sample Complexity when $\alpha\rightarrow0+$}
Based on the precise error probability  (\ref{eqn_Err0+}) in Lemma~\ref{lem_Err_gamma0+}, we can derive the sample complexity bound from
\begin{align}
&(N-K)\left[1-\frac{1}{K+1}\left(1-(1-\gamma)^{K+1}\right)\right]^M+
K\left[1-\frac{1}{K}\left(1-(1-\gamma)^{K}\right)\right]^M\leq \delta
\end{align}
Because $\left[1-\frac{1}{K}\left(1-(1-\gamma)^{K}\right)\right]^M \leq \left[1-\frac{1}{K+1}\left(1-(1-\gamma)^{K+1}\right)\right]^M$, it suffices to let
\begin{align}\notag
N\left[1-\frac{1}{K+1}\left(1-(1-\gamma)^{K+1}\right)\right]^M \leq \delta
\end{align}
This immediately leads to the sample complexity result for $\alpha\rightarrow0+$ in Theorem~\ref{thm_M0+}. \\

\begin{theorem}\label{thm_M0+}
As $\alpha\rightarrow0+$, the required number of measurements is
\begin{align}\label{eqn_M0+}
M =\frac{1}{-\log \left[1-\frac{1}{K+1}\left(1-(1-\gamma)^{K+1}\right)\right]} \log N/\delta
\end{align}$\hfill\Box$
\end{theorem}

\begin{figure}[h!]
\begin{center}
\includegraphics[width=3in]{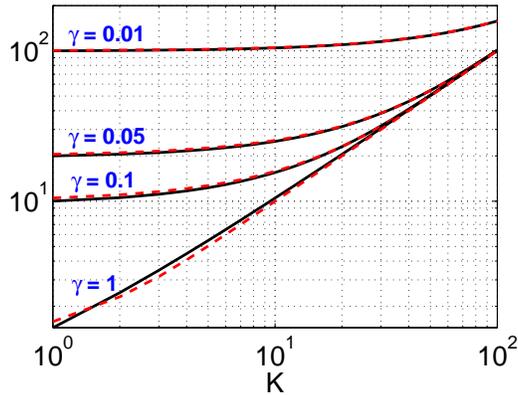}
\end{center}
\vspace{-0.2in}
\caption{Solid curves: $\frac{1}{-\log \left[1-\frac{1}{K+1}\left(1-(1-\gamma)^{K+1}\right)\right]}$. Dashed curves: $\frac{K}{1-e^{-\gamma K }}$. The difference between (\ref{eqn_M0+}) and (\ref{eqn_M0+_appr}) is very small even for small $K$. For large $K$, both terms approach $K$.}\label{fig_M0appr}
\end{figure}

\noindent\textbf{Remark:}\ \  The required number of measurements (\ref{eqn_M0+}) can essentially be written as
\begin{align}\label{eqn_M0+_appr}
M = \frac{K}{1-e^{-\gamma K }} \log N/\delta
\end{align}
The difference between (\ref{eqn_M0+}) and (\ref{eqn_M0+_appr}) is very small even when $K$ is small, as shown in Figure~\ref{fig_M0appr}. Let $\lambda = \gamma K$. If $\lambda = 1$ (i.e., $\gamma = 1/K$), then the required $M$ is about $1.58K\log N/\delta$. If $\lambda =2$ (i.e., $\gamma = 2/K$), then  $M$ is about $1.16K\log N/\delta$. In other words, we can use a very sparse design matrix and the required number of measurements is only inflated slightly.

\newpage\clearpage

\subsection{Worst-Case Sample Complexity}

\begin{theorem}\label{thm_worst_case}
If we choose $\gamma = \frac{1}{K+1}$, then it suffices to choose the number of measurements by
\begin{align}\label{eqn_Mworst}
M=\frac{1}{-\log \left[1-\frac{1}{K+1}\left(1-\frac{1}{K+1}\right)^K\right]}\log N/\delta
\end{align}
\noindent\textbf{Proof}:\ \ \ See Appendix~\ref{app_thm_worst_case}.$\hfill\Box$
\end{theorem}

\noindent\textbf{Remark}: \ \ The worst-case complexity (\ref{eqn_Mworst}) can essentially be written as
\begin{align}
M = eK\log N/\delta,\hspace{0.3in} \text{ if } \gamma = 1/K
\end{align}
where $e= 2.7183...$ The previous analysis of sample complexity for $\alpha\rightarrow0+$ says that if $\gamma = 1/K$, it suffices to let $M = 1.58K\log N/\delta$, and if $\gamma = 2/K$, it suffices to let $M = 1.15K\log N/\delta$. This means that the worst-case analysis is quite conservative and the choice $\gamma=1/K$ is not optimal for general $\alpha\in(0,1)$. \\

Interestingly, it turns out that the worst-case sample complexity is attained when $\alpha\rightarrow1-$.

\subsection{Sample Complexity when $\alpha=1-$}

\begin{theorem}

For a $K$-sparse signal whose nonzero coordinates are larger than $\epsilon$, i.e.,  $x_i>\epsilon$ if $x_i>0$. If we choose $\gamma = \frac{1}{K+1}$,  as $\alpha\rightarrow1-$,
  it suffices to choose the  number of measurements  by
\begin{align}
M =  \frac{1}{-\log\left(1-\frac{1}{K+1}\left(1-\frac{1}{K+1}\right)^K\right)}\log N/\delta
\end{align}
\noindent\textbf{Proof:}\hspace{0.2in} The proof can be directly inferred from the  proof of Theorem~\ref{thm_worst_case} at $\alpha=1-$.$\hfill\Box$\\
\end{theorem}

\noindent\textbf{Remark}:\hspace{0.2in} Note that, if the assumption $x_i>\epsilon$ whenever $x_i>0$ does not hold, then the required number of measurements will be  smaller.

%

\subsection{Sample Complexity Analysis for Binary Signals}

As this point, we know the precise sample complexities  for $\alpha=0+$ and $\alpha=1-$. And we also know the worst-case complexity. Nevertheless, it would be still interesting to study how the complexity varies as $\alpha$ changes between 0 and 1. While a precise analysis is  difficult, we can  perform an accurate analysis at least for binary signals, i.e., $x_i\in\{0,1\}$. For convenience, we first  re-write  the general error probability as
\begin{align}
\mathbf{Pr}\left(\hat{x}_{i,min,\gamma}> x_i + \epsilon\right) =\left[1- \frac{1}{K}\left(\gamma K\right) E\left\{F_\alpha\left(\left(\frac{\epsilon^\alpha}{\eta_{ij}}\right)^{1/(1-\alpha)}\right)\right\}\right]^M
\end{align}
For binary signals, we have $\eta_{ij}\sim Binomial(K-1+1_{x_i=0},\gamma)$. Thus, if $x_i=0$, then
\begin{align}\notag
H=H(\gamma,K;\epsilon,\alpha) \overset{\bigtriangleup}{=}& (\gamma K)E\left\{F_\alpha\left(\left(\frac{\epsilon^\alpha}{\eta_{ij}}\right)^{1/(1-\alpha)}\right)\right\}\\\label{eqn_H}
=& (\gamma K)\sum_{k=0}^{K} F_\alpha\left(\left(\frac{\epsilon^\alpha}{k}\right)^{1/(1-\alpha)}\right) \binom{K}{k}\gamma^k(1-\gamma)^{K-k}
\end{align}
The required number of measurements can be written as $\frac{1}{-\log\left(1-H/K\right)}\log N/\delta$, or essentially $\frac{K}{H}\log N/\delta$. We can compute $H(\gamma,K;\epsilon,\alpha)$ for  given $\gamma$, $K$, $\epsilon$, and $\alpha$, at least by simulations.

\section{Poisson Approximation for  Complexity Analysis with Binary Signals}

Again, the purpose is to study more precisely how the sample complexity varies with $\alpha\in(0,1)$, at least for binary signals. In this case, when $x_i=0$, we have $\eta_{ij}\sim Binomail(K,\gamma)$. Elementary statistics tells us that we can well approximate this binomial with a Poisson distribution with parameter $\lambda = \gamma K$ especially when $K$ is not small. Using the Poisson approximation, we can replace $H(\gamma,K;\epsilon,\alpha)$ in (\ref{eqn_H}) by $h(\lambda;\epsilon,\alpha)$ and re-write the error probability as
\begin{align}
\mathbf{Pr}\left(\hat{x}_{i,min,\gamma}> x_i + \epsilon\right) =& \left[1- \frac{1}{K}h(\lambda;\epsilon,\alpha)\right]^M
\end{align}
where
\begin{align}\notag
h(\lambda;\epsilon,\alpha) =& \lambda\sum_{k=0}^\infty F_\alpha\left(\left(\frac{\epsilon^\alpha}{k}\right)^{1/(1-\alpha)}\right)\frac{e^{-\lambda}\lambda^k}{k!}\\\label{eqn_h}
=&\lambda e^{-\lambda} +\lambda e^{-\lambda}\sum_{k=1}^\infty F_\alpha\left(\left(\frac{\epsilon^\alpha}{k}\right)^{1/(1-\alpha)}\right)\frac{\lambda^k}{k!}
\end{align}
which can be  computed numerically for any given $\lambda$ and $\epsilon$.\\

The required number of measurements can be computed from
\begin{align}
N\left[1- \frac{1}{K}h(\lambda;\epsilon,\alpha)\right]^M=\delta\Longleftrightarrow
M=\frac{\log N/\delta}{ -\log \left[1- \frac{1}{K}h(\lambda;\epsilon,\alpha)\right]}
\end{align}
for which it suffices to choose $M$ such that
\begin{align}
M= \frac{K}{h(\lambda;\epsilon,\alpha)} \log N/\delta
\end{align}

Therefore, we hope $h(\lambda;\epsilon,\alpha)$ should be as large as possible.

\subsection{Analysis for $\alpha=0.5$}

Before we demonstrate the results via Poisson approximation for general $0<\alpha<1$, we would like to illustrate the analysis particularly for $\alpha=0.5$, which is a case readers can more easily verify.


Recall when $\alpha=0.5$, the error probability can be written as
\begin{align}\notag
&\mathbf{Pr}\left(\hat{x}_{i,min,\gamma}> x_i + \epsilon\right)
=\left[1-\frac{1}{K}\left( \gamma K \right)E\left\{\frac{2}{\pi}\tan^{-1}\left(\left(\frac{\sqrt{\epsilon}}{\eta_{ij}}\right)\right)\right\}\right]^M = \left[1-\frac{1}{K}H(\gamma, K; \epsilon,0.5)\right]^M
\end{align}
where
\begin{align}
&H(\gamma, K; \epsilon,0.5) =(\gamma K)\frac{2}{\pi}\sum_{k=0}^K\tan^{-1}\left(\frac{\sqrt{\epsilon}}{k}\right)\binom{K}{k}\gamma^k(1-\gamma)^{K-k}
\end{align}
From Lemma~\ref{lem_Err_gamma}, in particular (\ref{eqn_Err0.5_Up}), we know there is a convenient lower bound of $H$:
\begin{align}
&H(\gamma, K; \epsilon,0.5) \geq H^{lower}(\gamma,K;\epsilon,0.5)= (\gamma K) \left\{\frac{2}{\pi}\tan^{-1}\left(\frac{\sqrt{\epsilon}}{\gamma K}\right)\right\}
= \lambda \frac{2}{\pi}\tan^{-1}\left(\frac{\sqrt{\epsilon}}{\lambda}\right)
\end{align}
We will compare the precise $H(\gamma, K; \epsilon,0.5)$ with its lower bound $H^{lower}(\gamma,K;\epsilon,0.5)$, along with the Poisson approximation:
\begin{align}
&H(\gamma, K; \epsilon,0.5)\approx h(\lambda; \epsilon,0.5) = \lambda e^{-\lambda }\frac{2}{\pi}\sum_{k=0}^\infty\tan^{-1}\left(\frac{\sqrt{\epsilon}}{k}\right)\frac{\lambda^k}{k!}
\end{align}

\begin{figure}[h!]
\begin{center}
\includegraphics[width = 3in]{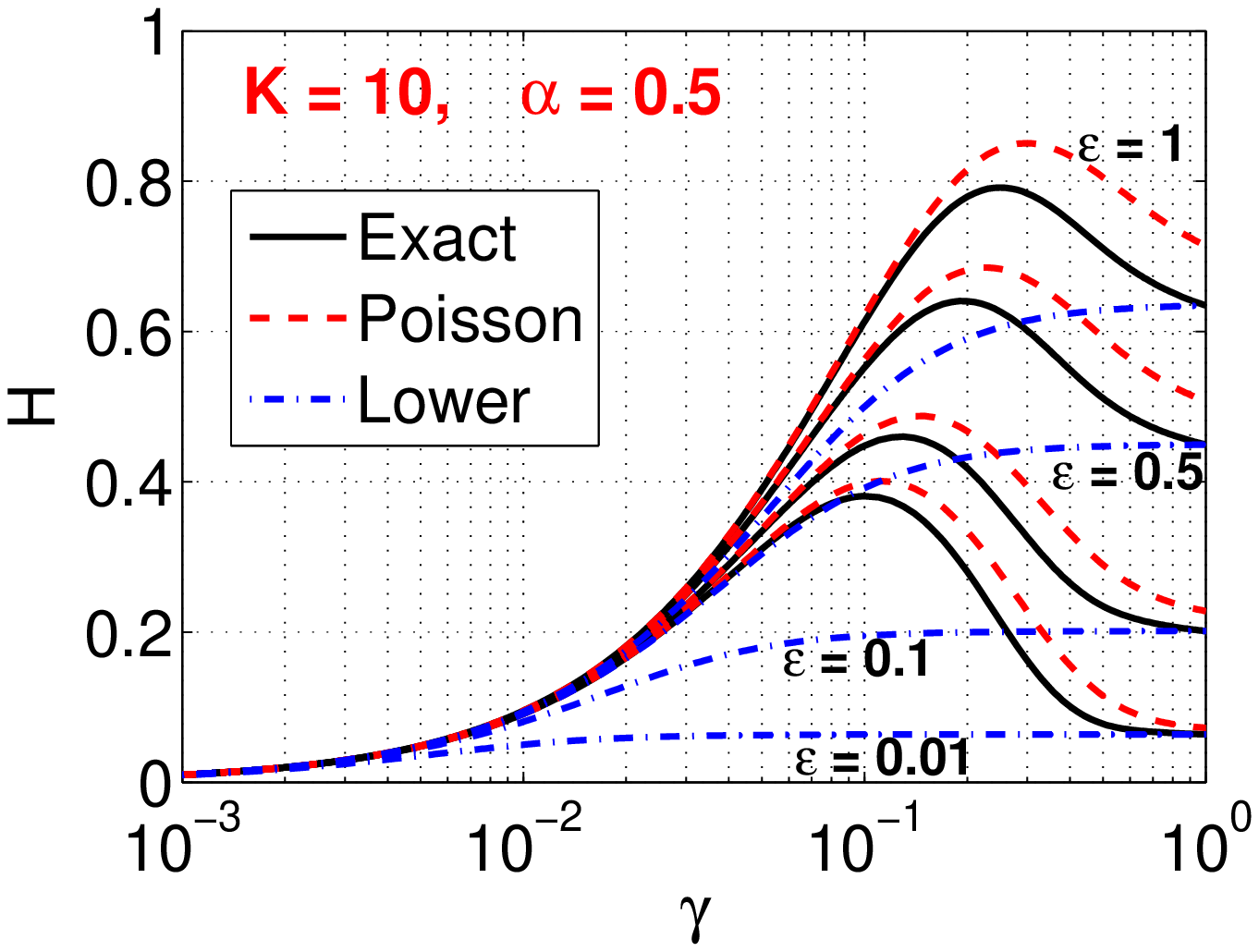}
\includegraphics[width = 3in]{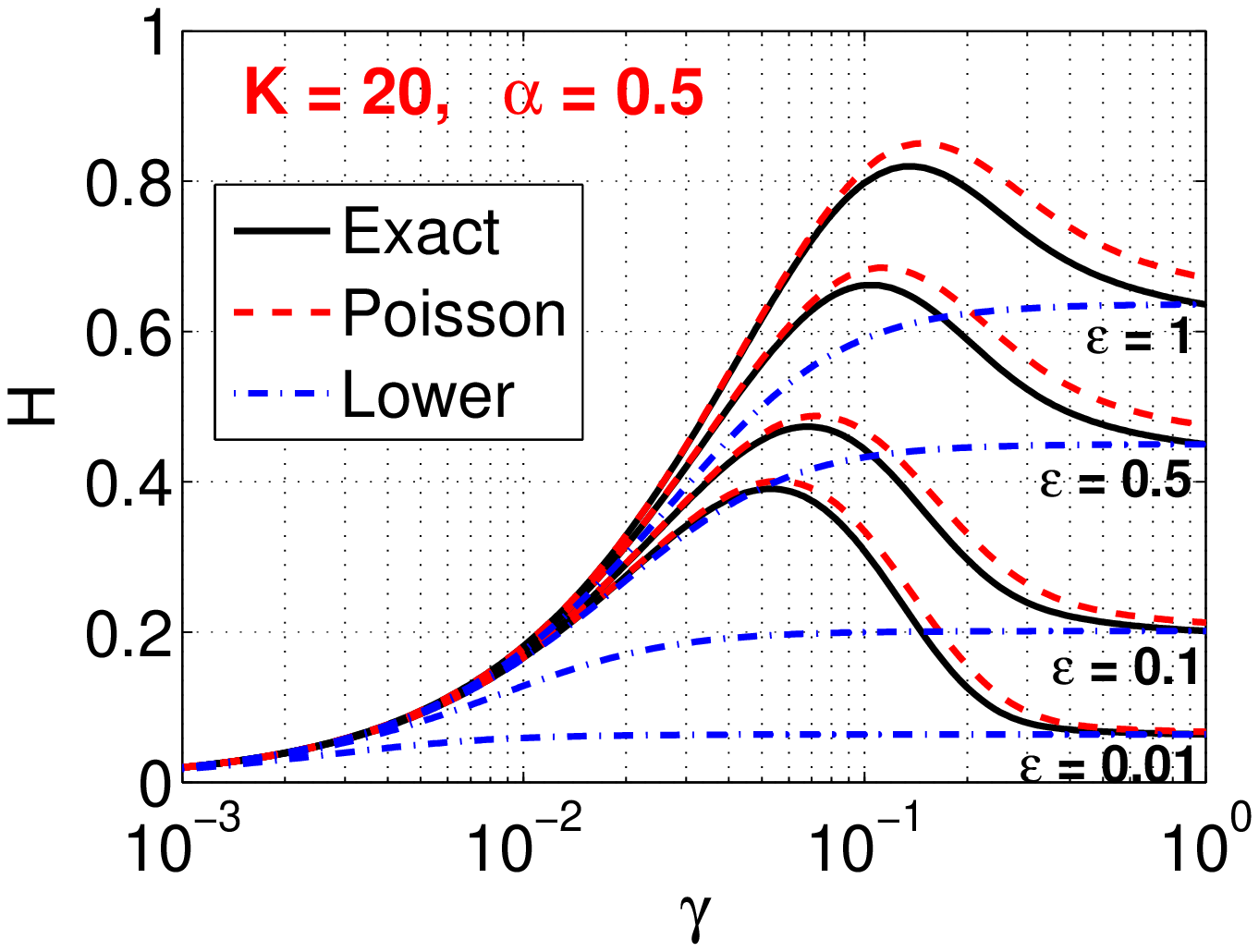}
\includegraphics[width = 3in]{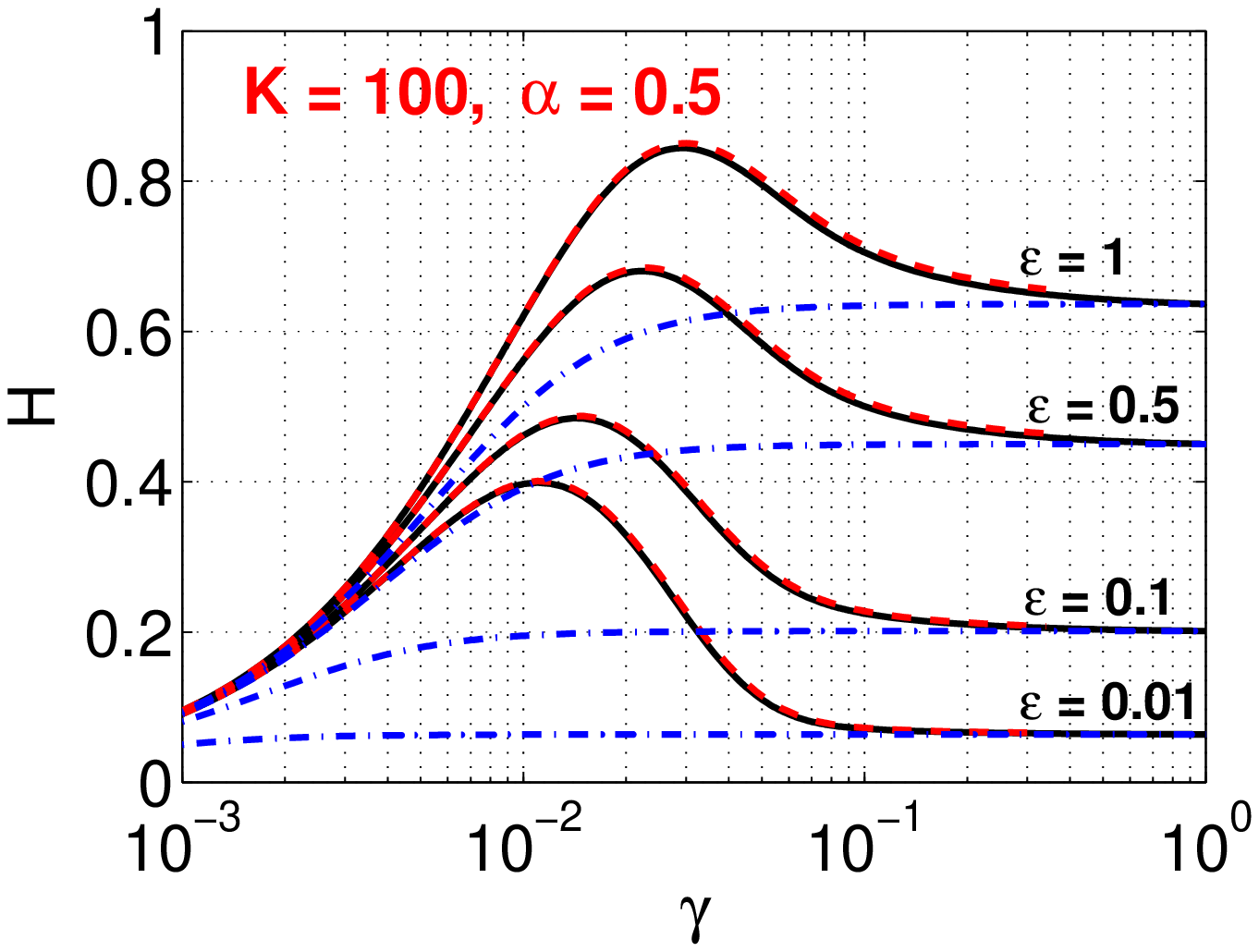}
\includegraphics[width = 3in]{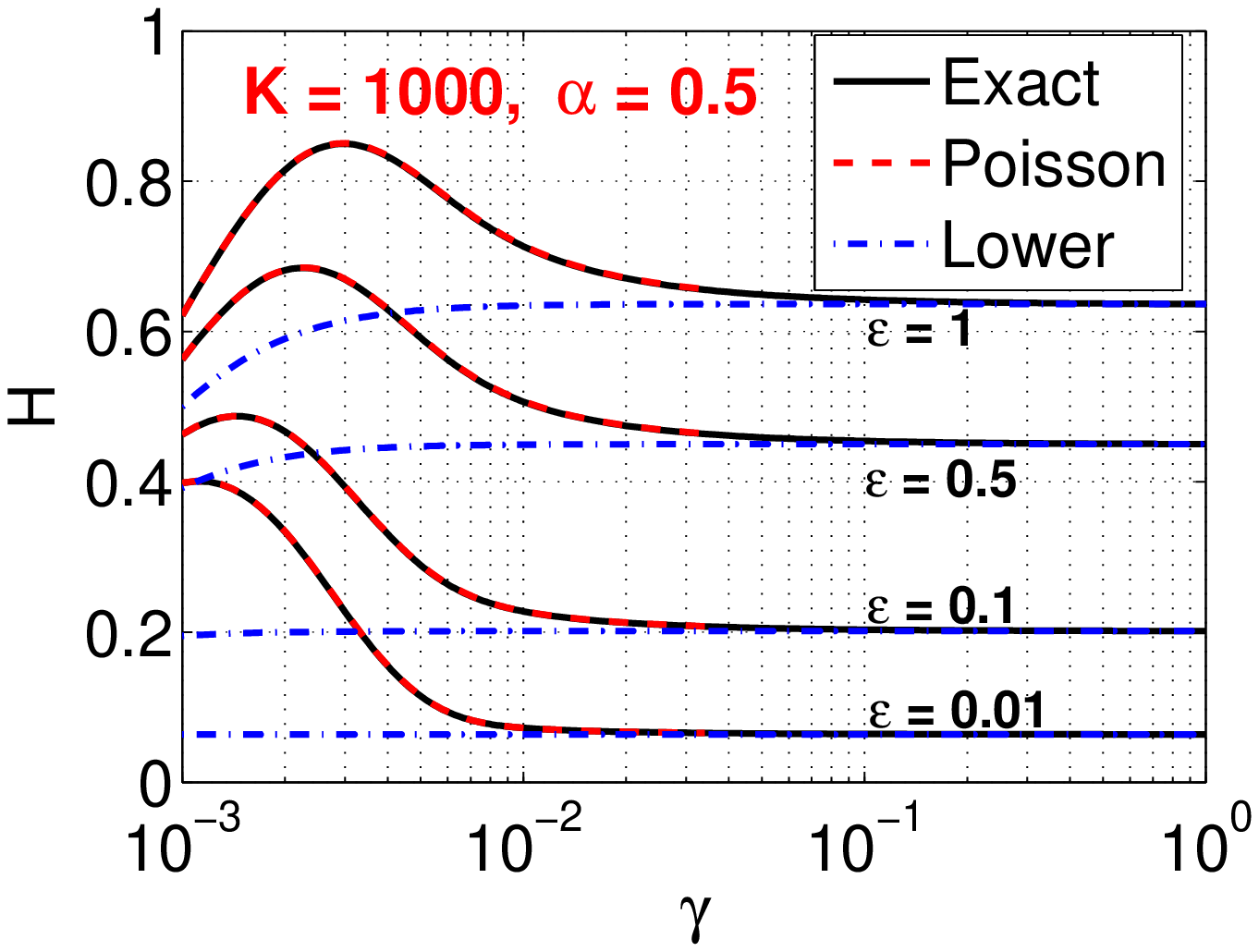}

\end{center}
\vspace{-0.2in}
\caption{$H(\gamma, K; \epsilon,0.5)$ at four different values of $\epsilon\in\{0.01, 0.1, 0.5, 1\}$. The exact $H$ and its Poisson approximation $h(\lambda;\epsilon,0.5)$ match very well unless $K$ is very small. The lower bound of $H$ is  conservative, especially when $\gamma$ is around the optimal value. For small $\epsilon$, the optimal $\gamma$ is around $1/K$.}\label{fig_H0.5}
\end{figure}

Figure~\ref{fig_H0.5} confirms that the Poisson approximation is very accurate unless $K$ is very small, while the lower bound is  conservative especially when $\gamma$ is around the optimal value. For small $\epsilon$, the optimal $\gamma$ is around $1/K$, which is consistent with the general worst-case complexity result.

\subsection{Poisson Approximation for General $0<\alpha<1$}

Once we are convinced that the Poisson approximation is reliable at least for $\alpha=0.5$,  we can use this tool to study for general $\alpha\in(0,1)$. Again, assume the Poisson approximation, we have
\begin{align}\notag
\mathbf{Pr}\left(\hat{x}_{i,min,\gamma}> x_i + \epsilon\right) =& \left[1- \frac{1}{K}h(\lambda;\epsilon,\alpha)\right]^M
\end{align}
where
\begin{align}\notag
h(\lambda;\epsilon,\alpha) =\lambda e^{-\lambda} +\lambda e^{-\lambda}\sum_{k=1}^\infty F_\alpha\left(\left(\frac{\epsilon^\alpha}{k}\right)^{1/(1-\alpha)}\right)\frac{\lambda^k}{k!}
\end{align}
The required number of measurements can be computed from $M= \frac{K}{h(\lambda;\epsilon,\alpha)} \log N/\delta$.

\begin{figure}[h!]
\begin{center}
\mbox{
\includegraphics[width = 2.3in]{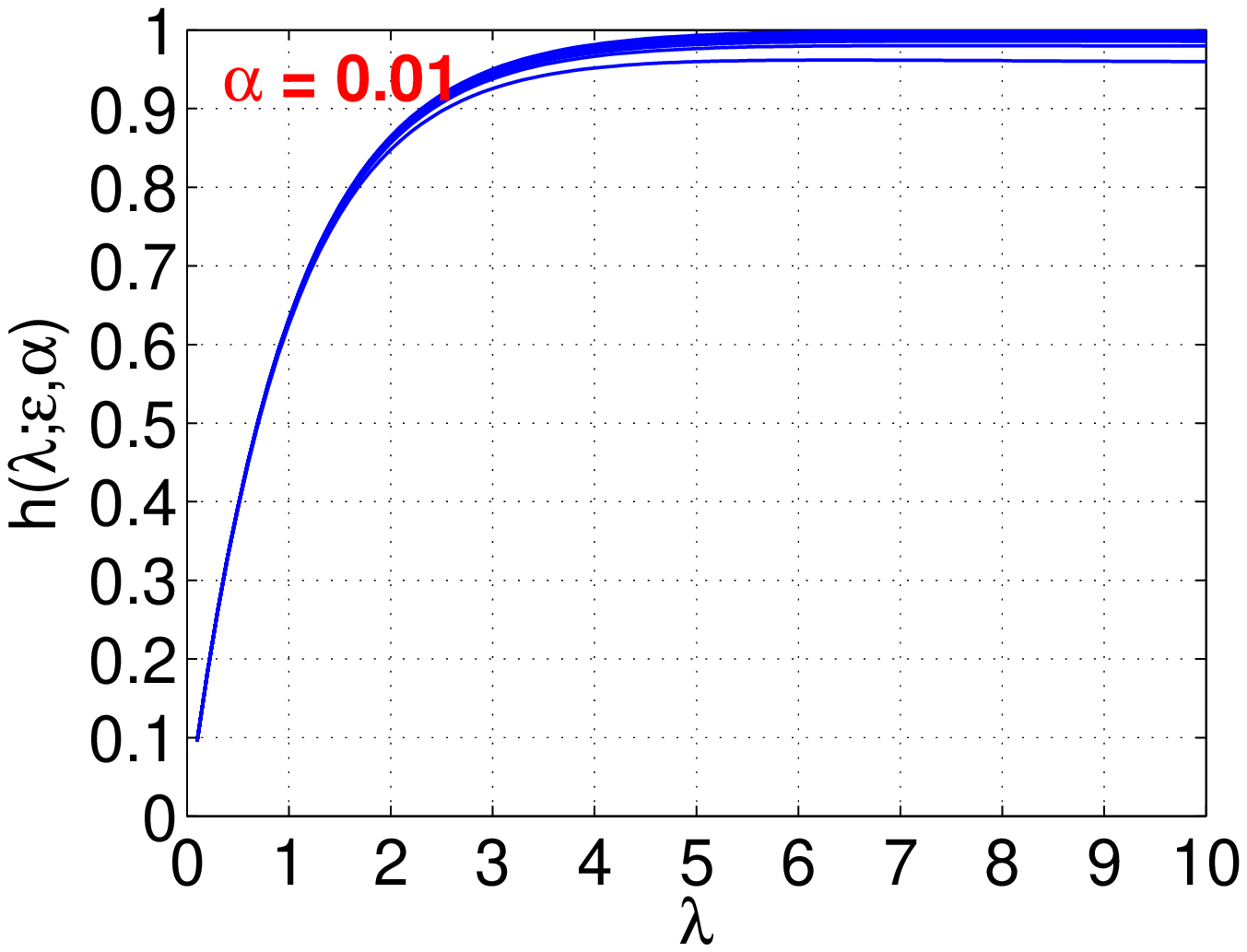}\hspace{-0.18in}
\includegraphics[width = 2.3in]{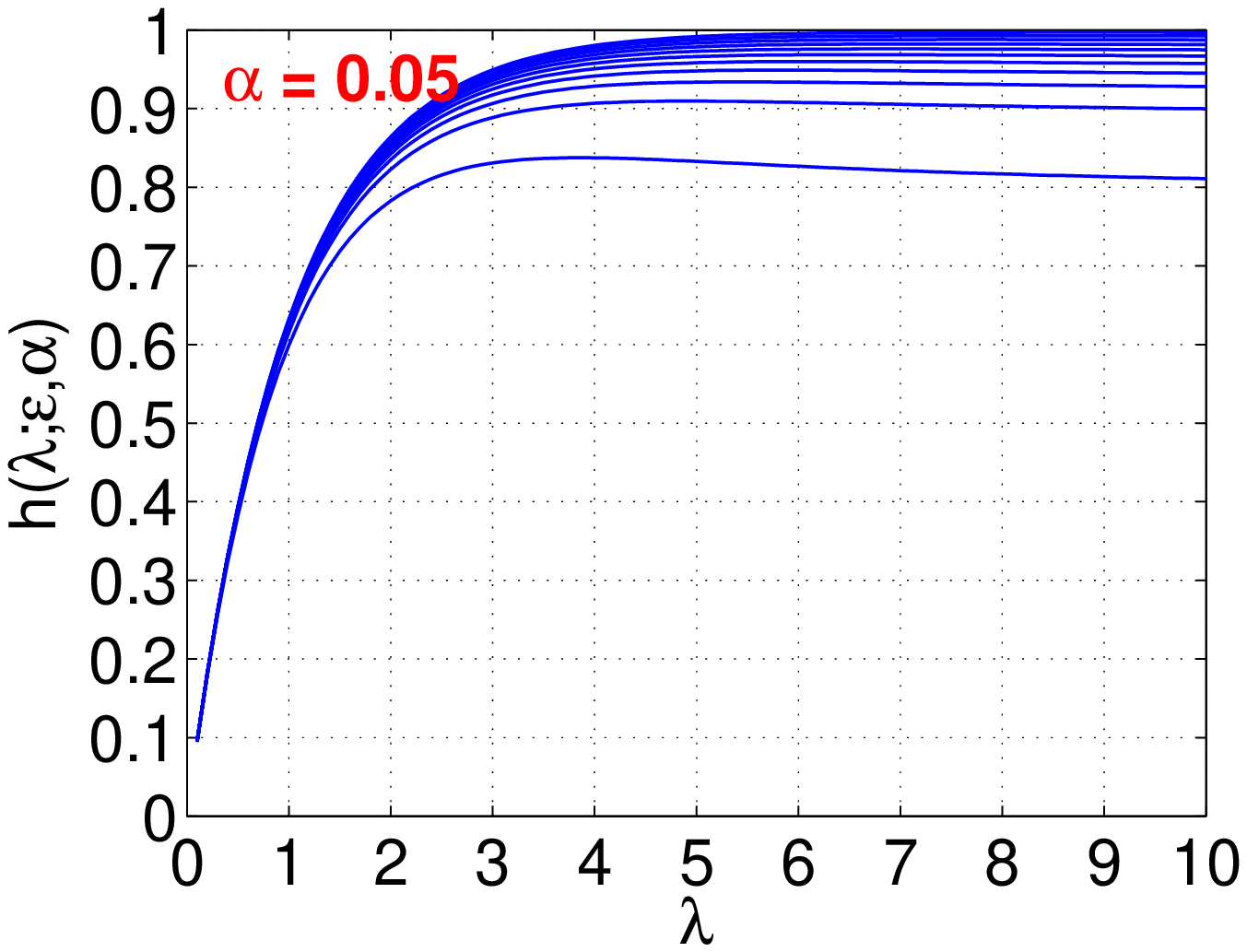}\hspace{-0.18in}
\includegraphics[width = 2.3in]{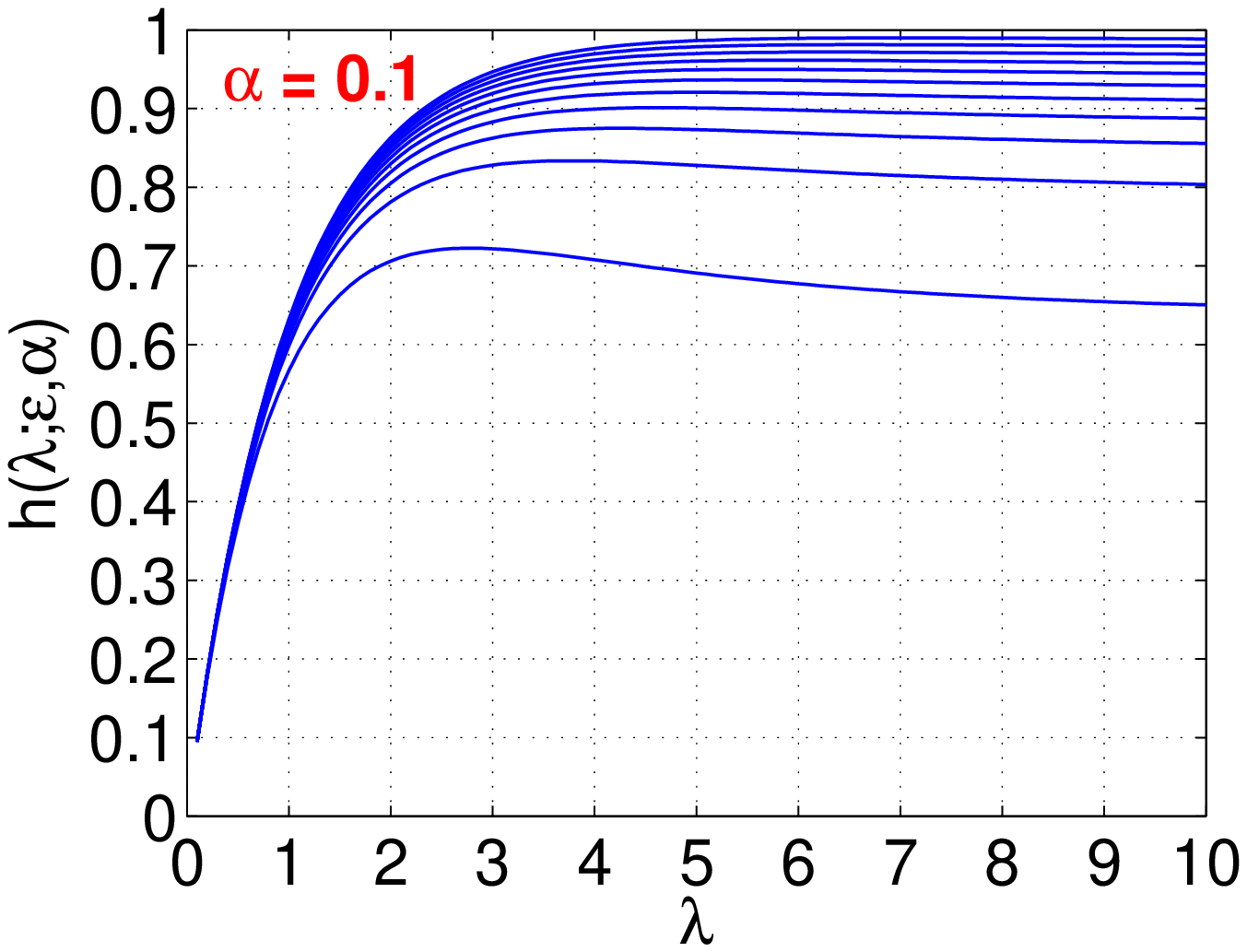}
}
\mbox{
\includegraphics[width = 2.3in]{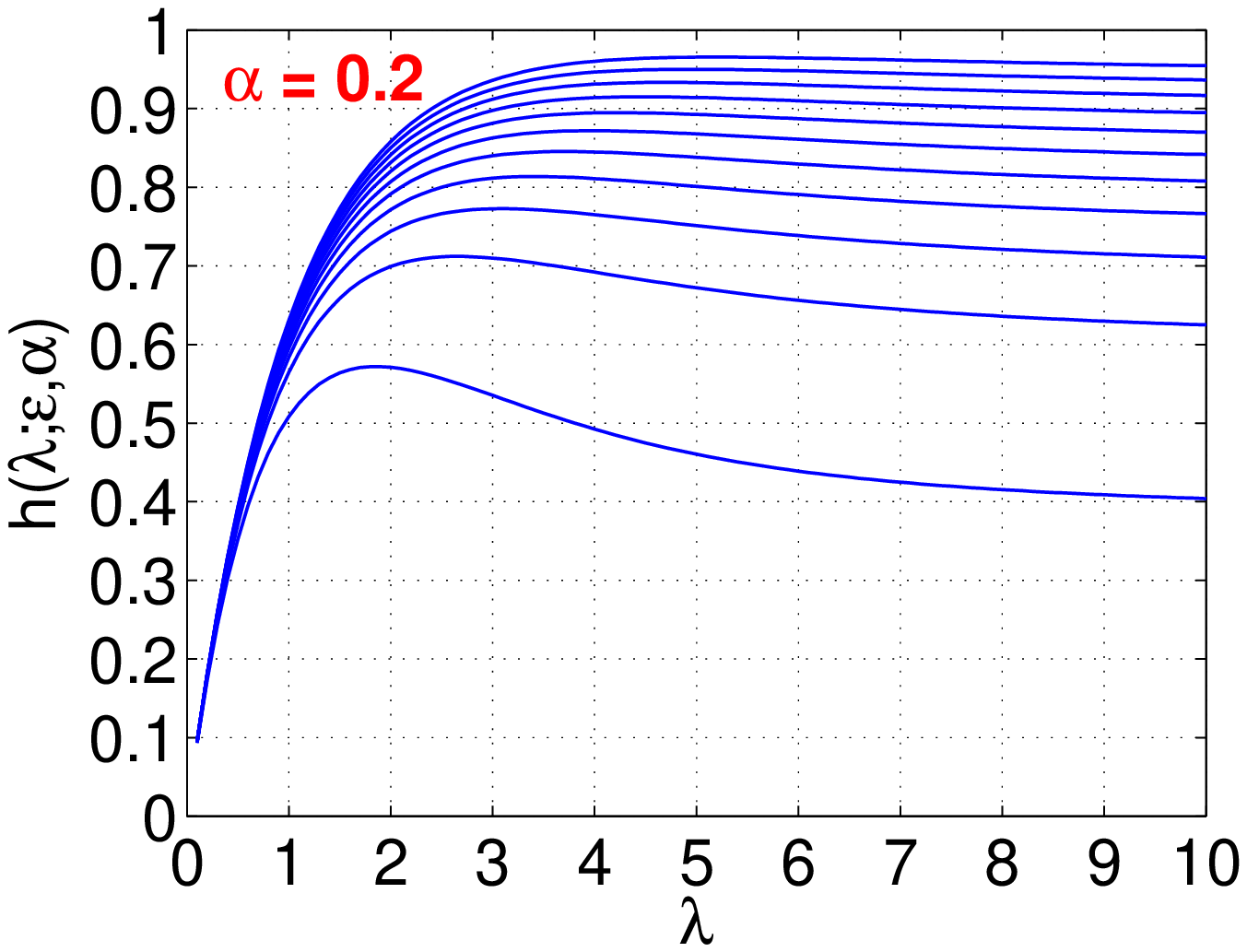}\hspace{-0.18in}
\includegraphics[width = 2.3in]{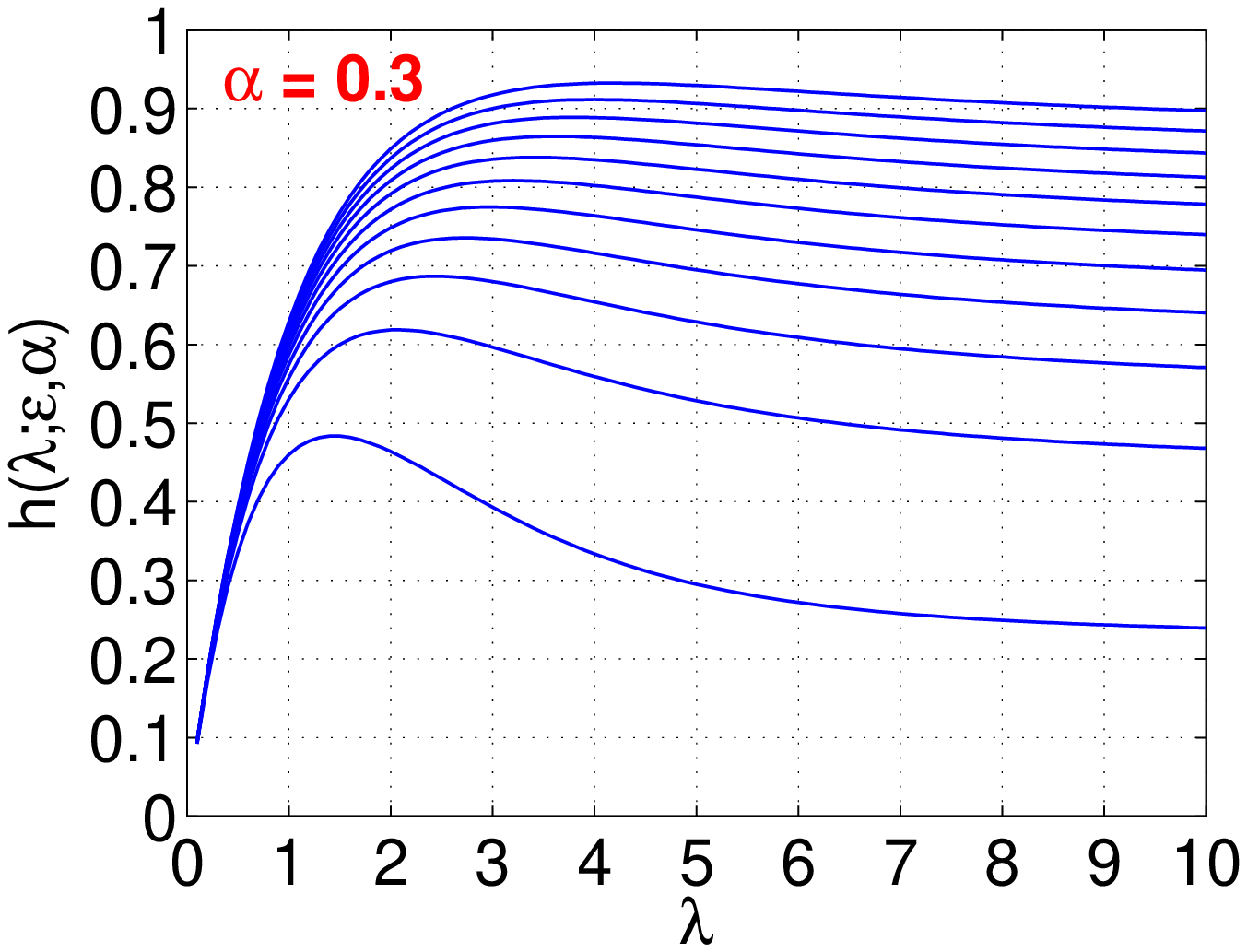}\hspace{-0.18in}
\includegraphics[width = 2.3in]{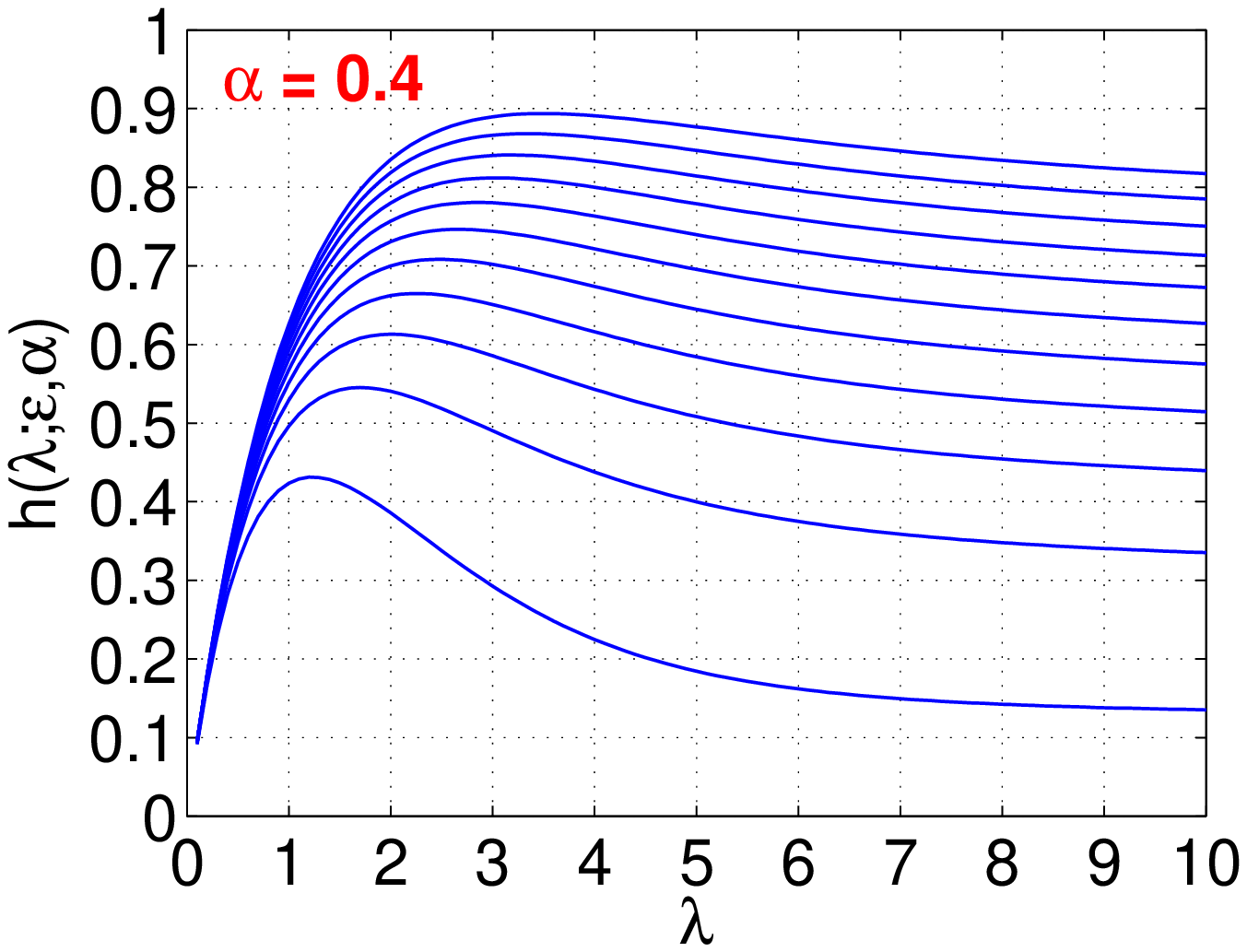}
}

\mbox{
\includegraphics[width = 2.3in]{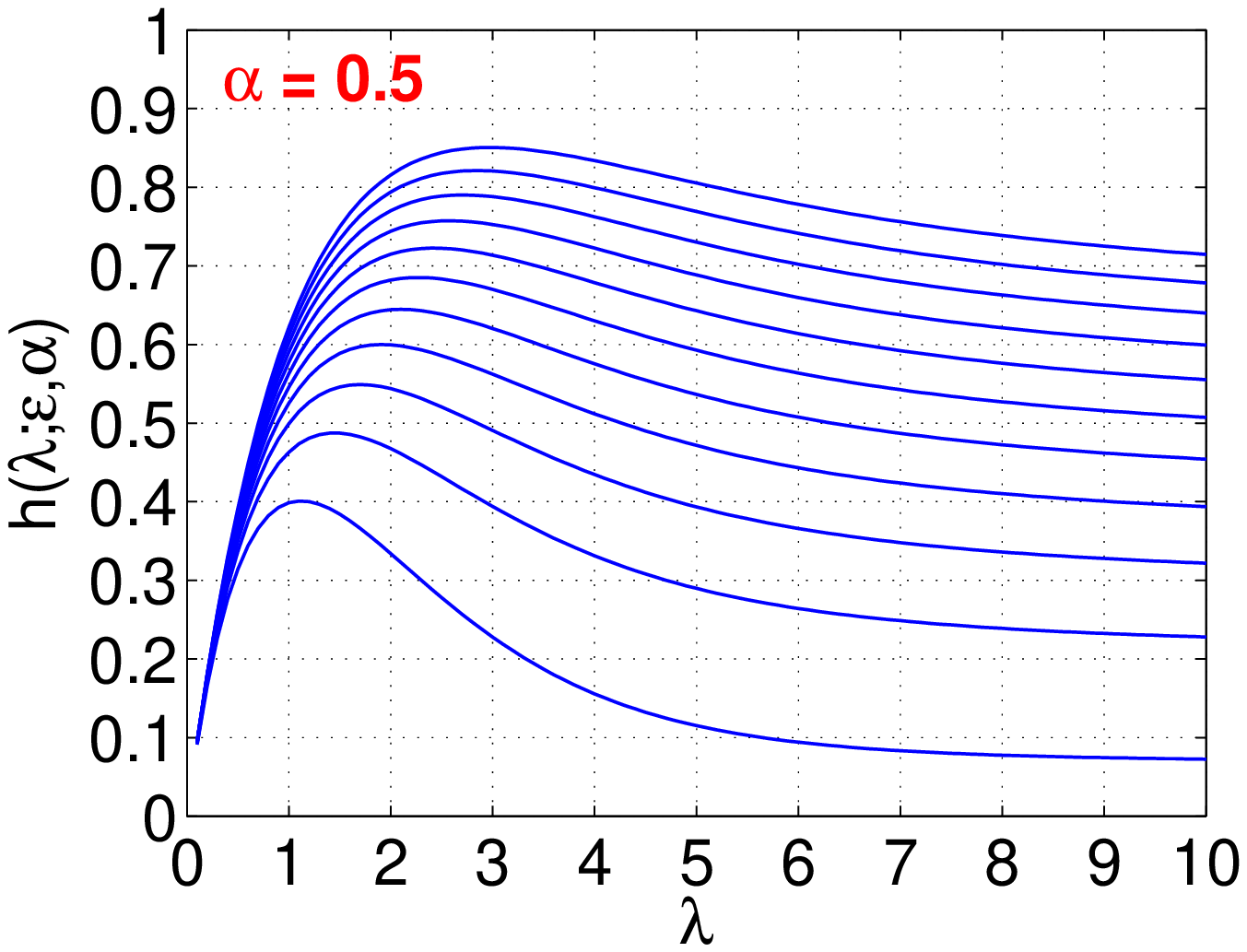}\hspace{-0.18in}
\includegraphics[width = 2.3in]{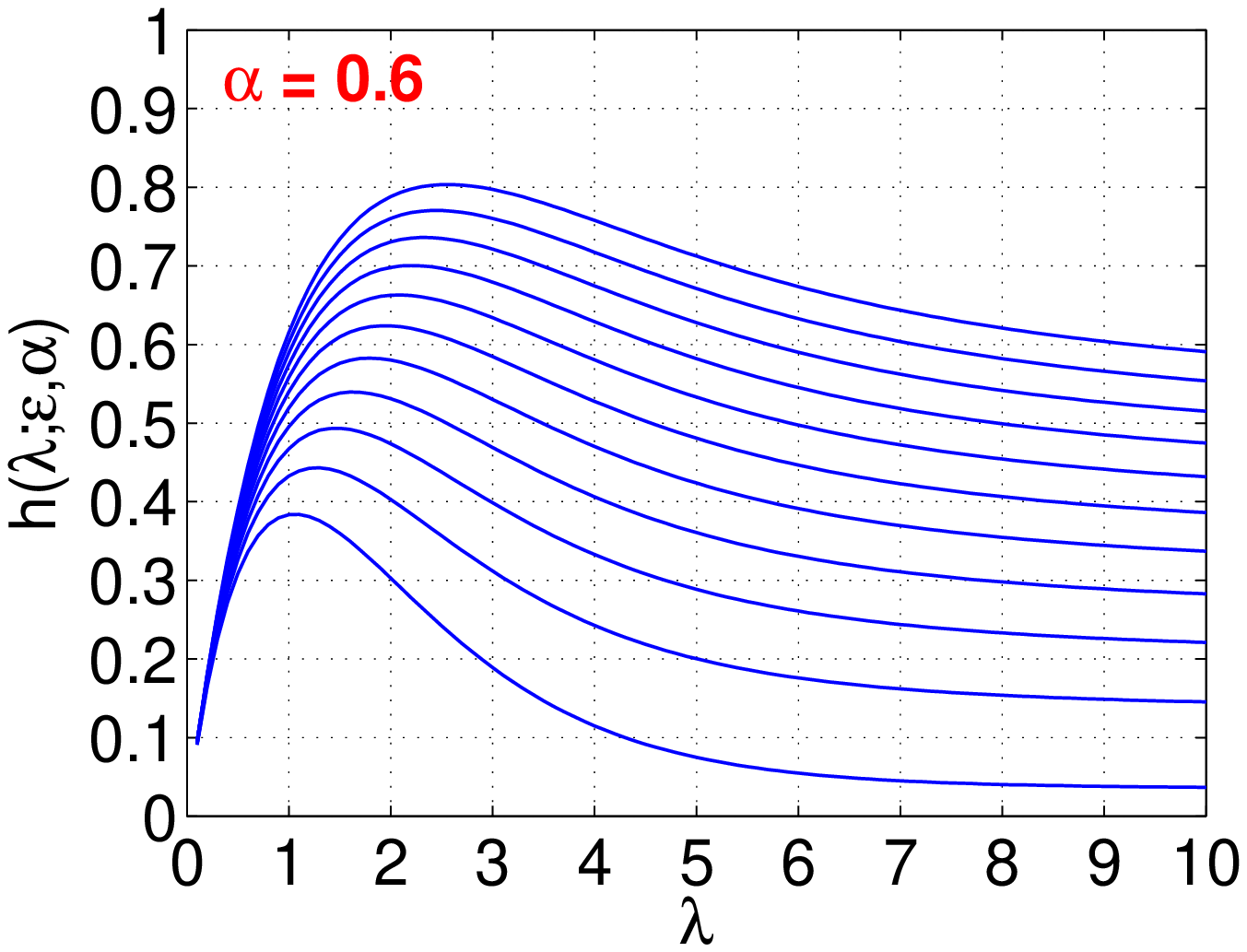}\hspace{-0.18in}
\includegraphics[width = 2.3in]{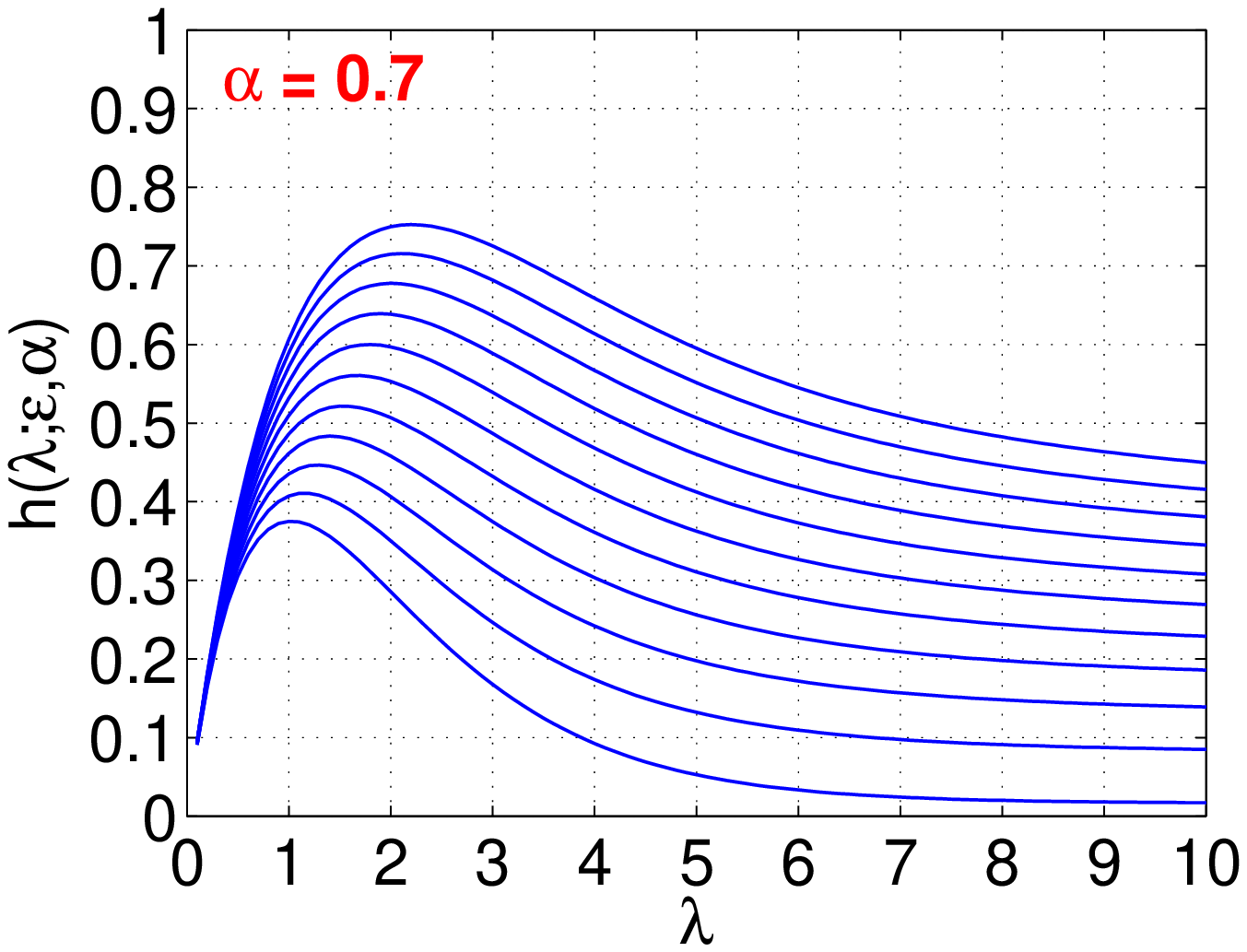}
}
\mbox{
\includegraphics[width = 2.3in]{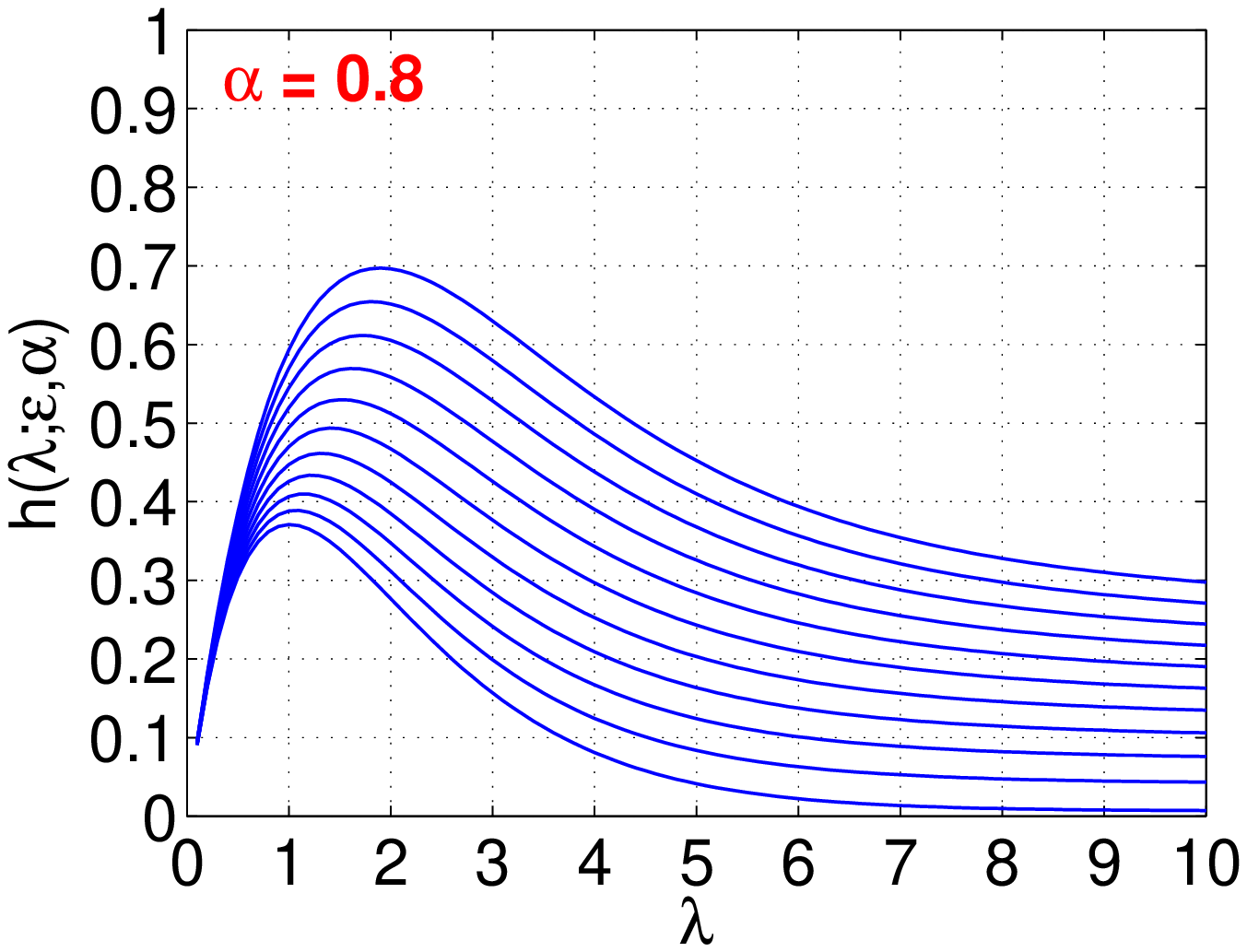}\hspace{-0.18in}
\includegraphics[width = 2.3in]{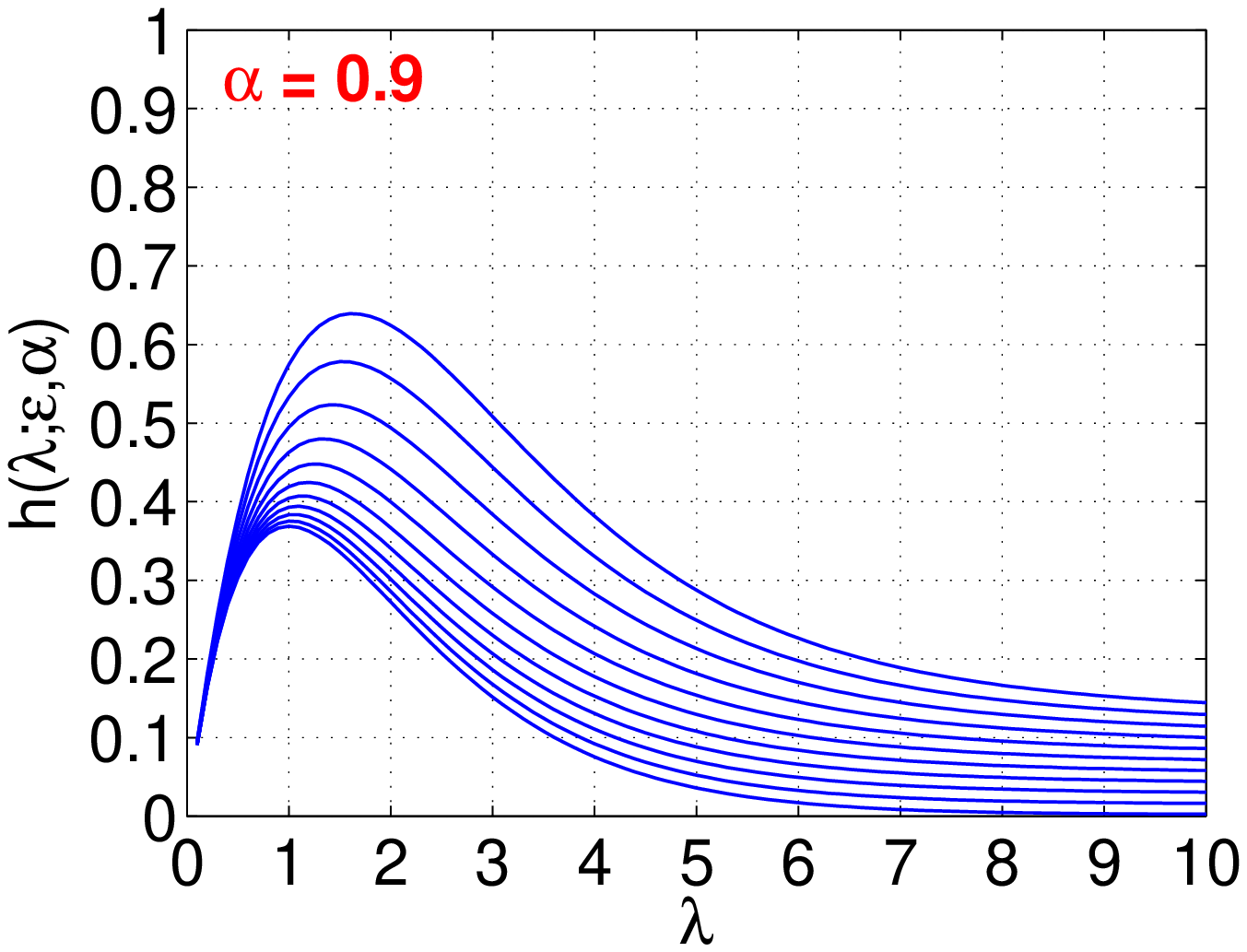}\hspace{-0.18in}
\includegraphics[width = 2.3in]{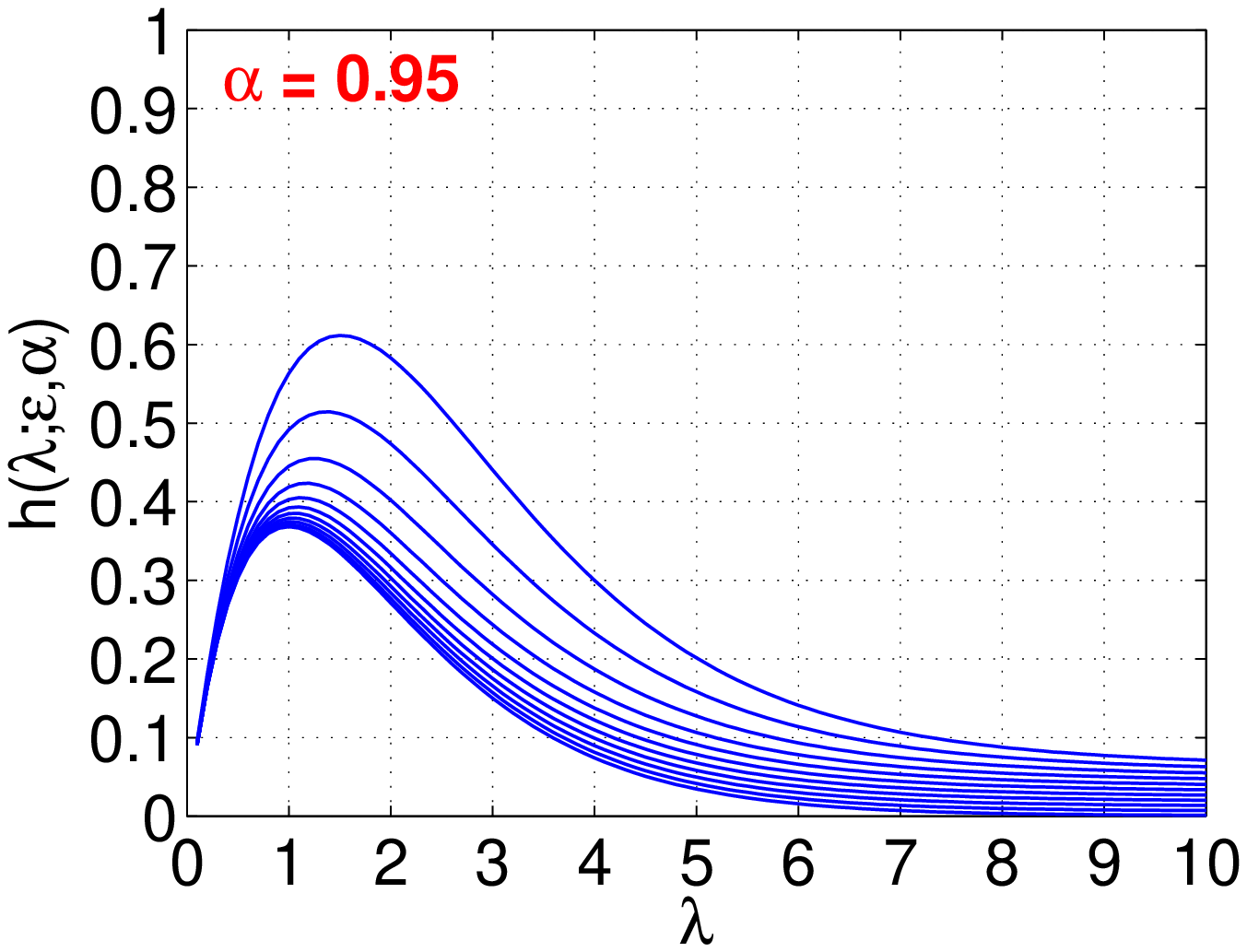}
}
\end{center}

\vspace{-0.25in}
\caption{$h(\lambda;\epsilon,\alpha)$ as defined in (\ref{eqn_h}) for selected $\alpha$ values ranging from $0.01$ to $0.95$. In each panel, each curve corresponds to an $\epsilon$ value, where $\epsilon\in\{0.01, 0.1, 0.2, 0.3, 0.4, 0.5, 0.6, 0.7, 0.8, 0.9, 1\}$ (from bottom to top). In each panel, the curve for $\epsilon = 0.01$ is the lowest and the curve for $\epsilon=1$ is the highest. }\label{fig_h}

\end{figure}

As shown in Figure~\ref{fig_h}, at fixed $\epsilon$ and $\lambda$, the optimal (highest) $h$ is larger when $\alpha$ is smaller. The optimal $h$ occurs at larger $\lambda$ when $\alpha$ is closer to zero and at smaller $\lambda$ when $\alpha$ is closer to 1.

\clearpage\newpage

\begin{figure}[h!]
\begin{center}

\mbox{
\includegraphics[width = 3in]{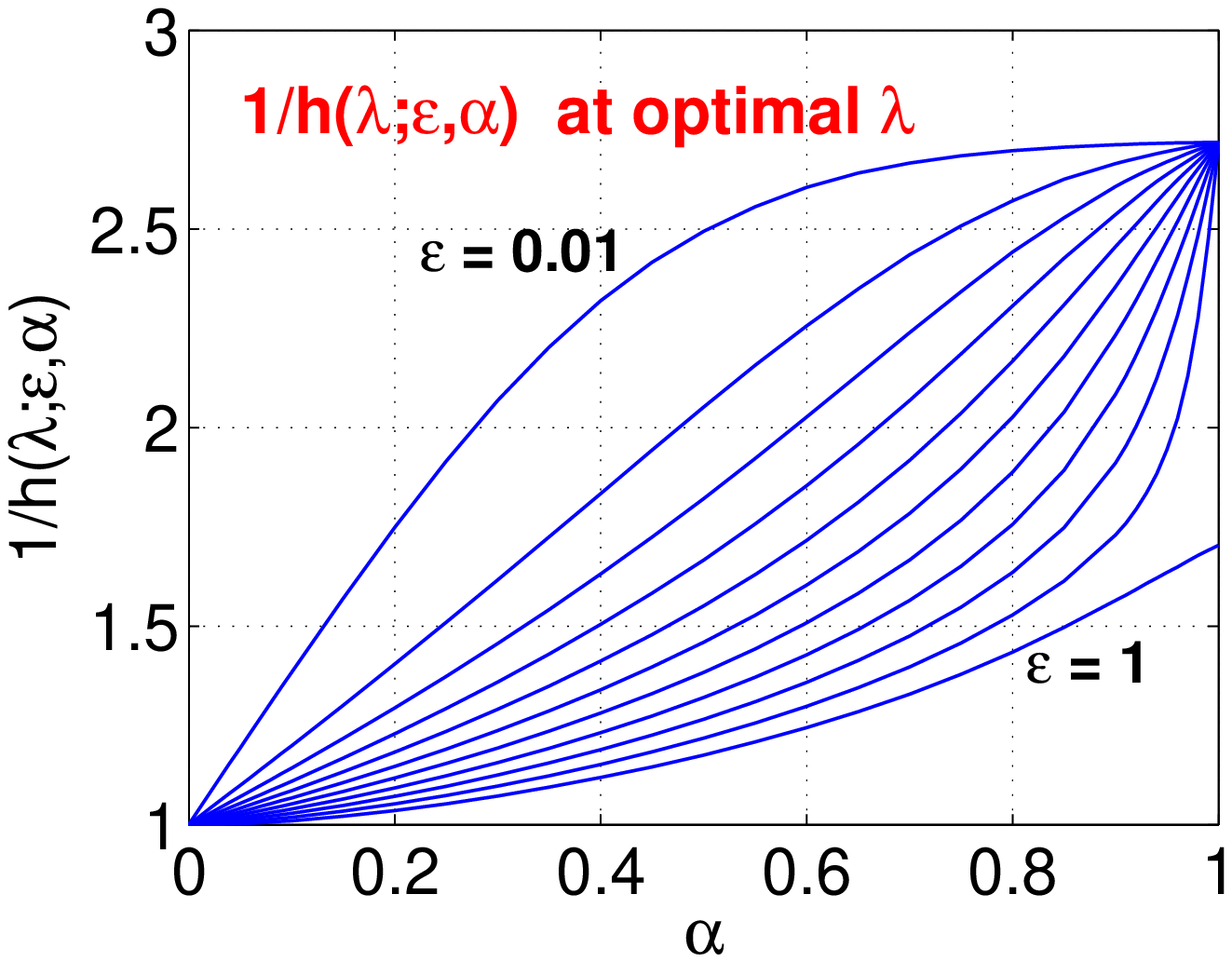}
\includegraphics[width = 3in]{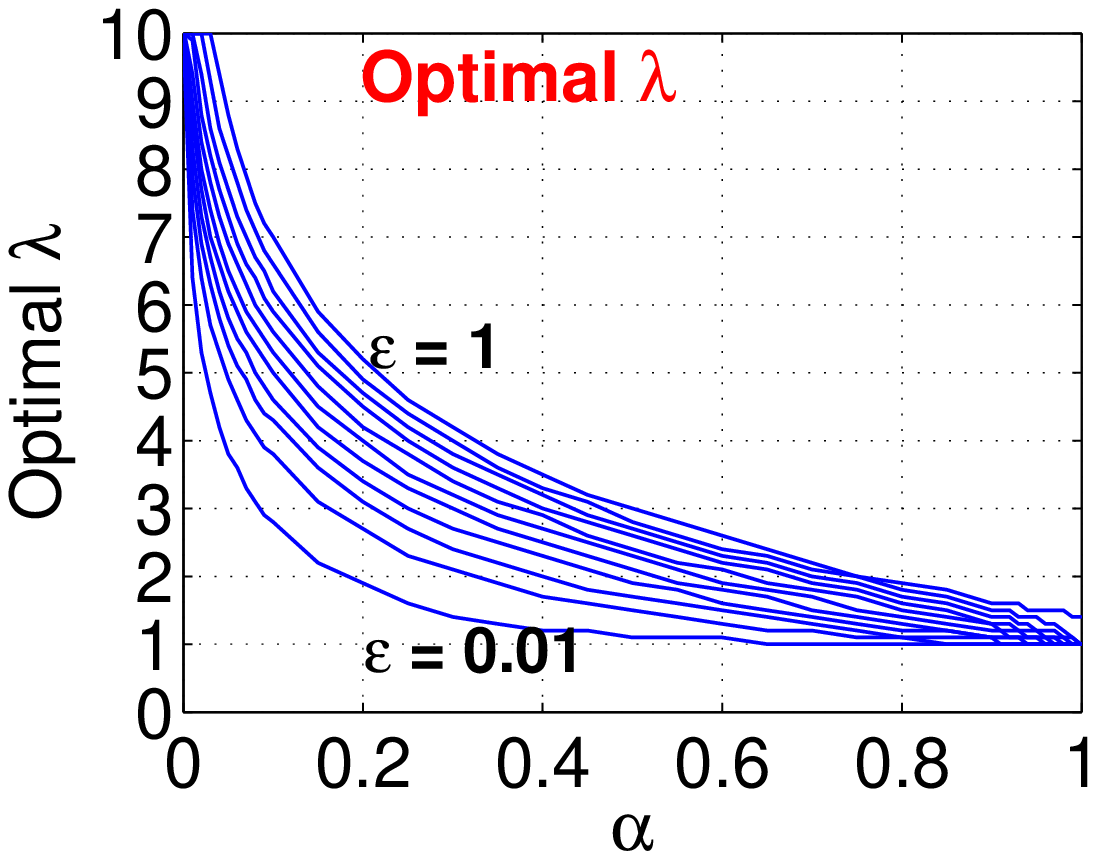}
}
\end{center}

\vspace{-0.25in}
\caption{Left Panel: $1/h(\lambda;\epsilon,\alpha)$ at the optimal $\lambda$ values. Right Panel: the optimal $\lambda$ values.}\label{fig_hlamopt}

\end{figure}

Figure~\ref{fig_hlamopt} plots the optimal (smallest) $1/h(\lambda;\epsilon,\alpha)$ values (left panel) and the optimal $\lambda$ values (right panel) which achieve the optimal $h$.

\vspace{0.2in}

\begin{figure}[h!]
\begin{center}

\mbox{
\includegraphics[width = 3in]{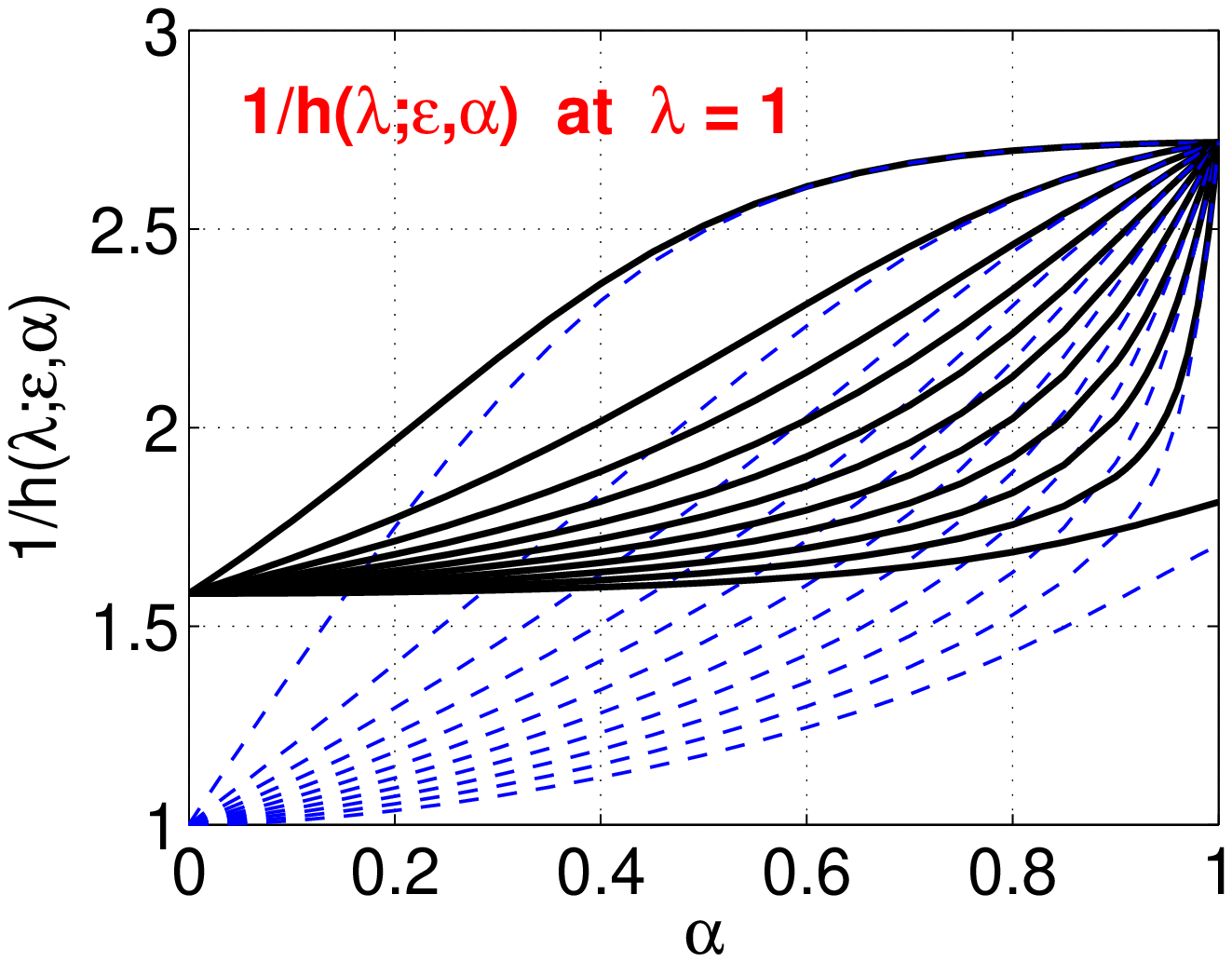}
\includegraphics[width = 3in]{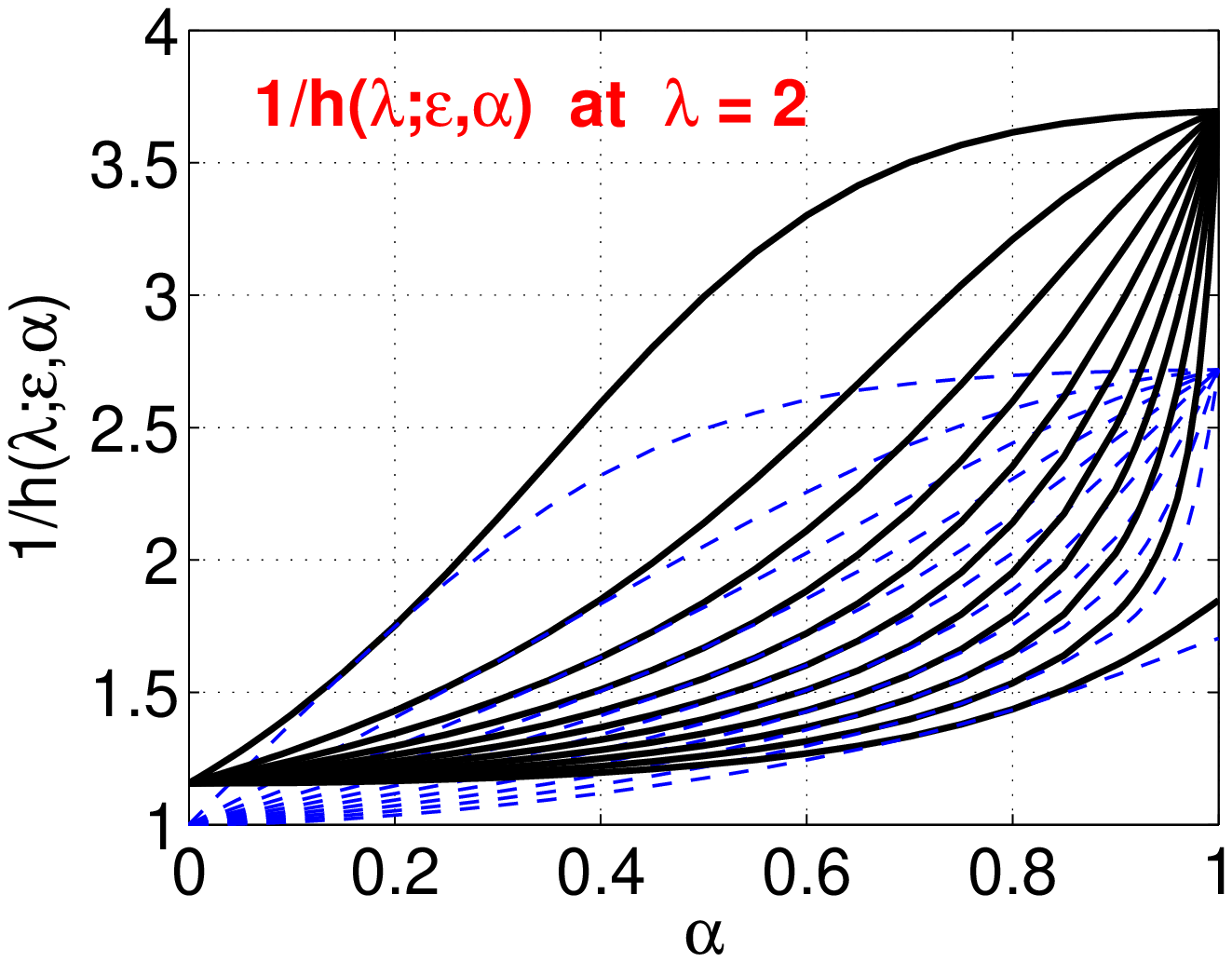}
}
\end{center}

\vspace{-0.25in}
\caption{$1/h(\lambda;\epsilon,\alpha)$ at the fixed  $\lambda=1$ (left panel) and $\lambda = 2$ (right panel). The dashed curves correspond to  $1/h(\lambda;\epsilon,\alpha)$ at the optimal $\lambda$ values.}\label{fig_hlam12}

\end{figure}

Figure~\ref{fig_hlam12} plots $1/h(\lambda;\epsilon,\alpha)$ for fixed $\lambda =1$ (left panel) and $\lambda=2$ (right panel), together with the optimal $1/h(\lambda;\epsilon,\alpha)$ values (dashed curves).

\subsection{Poisson Approximation for $\alpha\rightarrow1-$}

We now examine $h(\lambda;\epsilon,\alpha)$ closely at $\alpha=1-$, i.e., $\frac{1}{1-\alpha}\rightarrow\infty$.
\begin{align}\notag
h(\lambda;\epsilon,\alpha)
=\lambda e^{-\lambda} +\lambda e^{-\lambda}\sum_{k=1}^\infty F_\alpha\left(\left(\frac{\epsilon^\alpha}{k}\right)^{1/(1-\alpha)}\right)\frac{\lambda^k}{k!}
\end{align}
Interestingly, when $\epsilon=1$, only $k=0$ and $k=1$ will be useful, because otherwise $\left(\frac{\epsilon^\alpha}{k}\right)^{1/(1-\alpha)}\rightarrow\infty$ as $\Delta=1-\alpha\rightarrow0$. When $\epsilon<1$, then only $k=0$ is useful. Thus, we can write
\begin{align}
&h(\lambda;\epsilon<1,\alpha=1-) =\lambda e^{-\lambda}\\
&h(\lambda;\epsilon=1,\alpha=1-) =\lambda e^{-\lambda} + \lambda^2 e^{-\lambda}F_{1-}\left(1\right) = \lambda e^{-\lambda} + \lambda^2 e^{-\lambda}/2
\end{align}
Notes that $F_{1-}(1) = 1/2$ due to symmetry.\\

\begin{figure}[h!]
\begin{center}
\mbox{
\includegraphics[width = 2.3in]{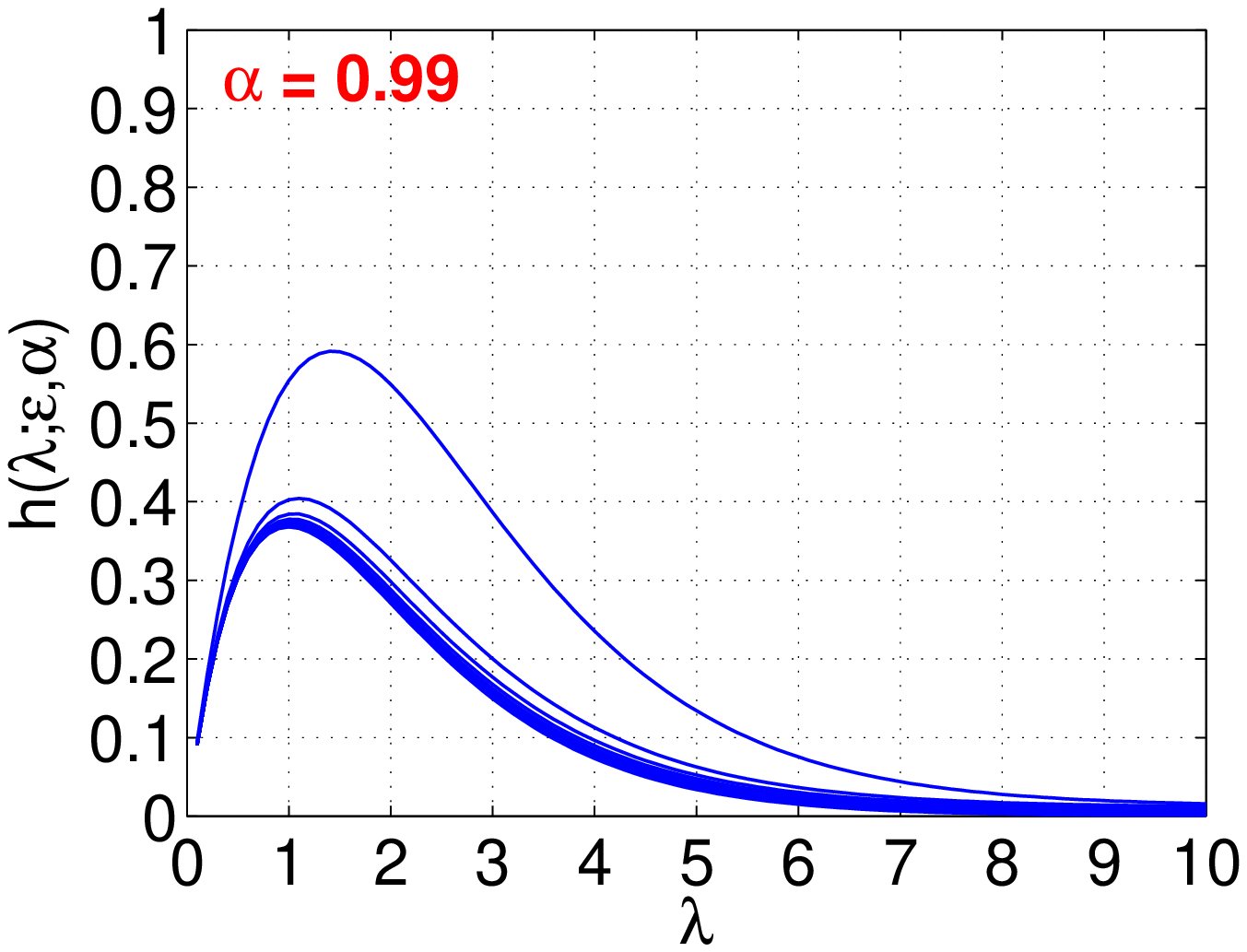}\hspace{-0.18in}
\includegraphics[width = 2.3in]{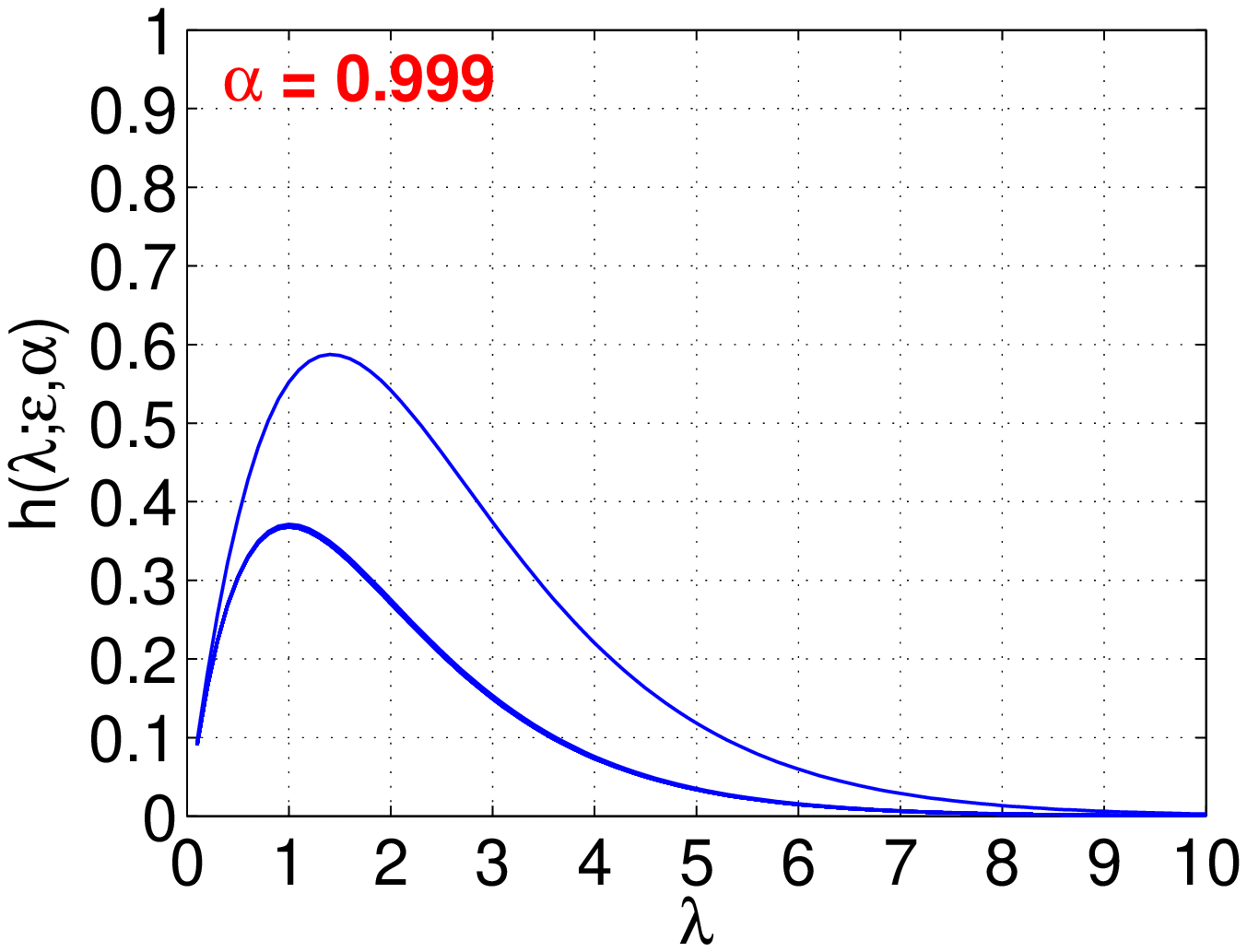}\hspace{-0.18in}
\includegraphics[width = 2.3in]{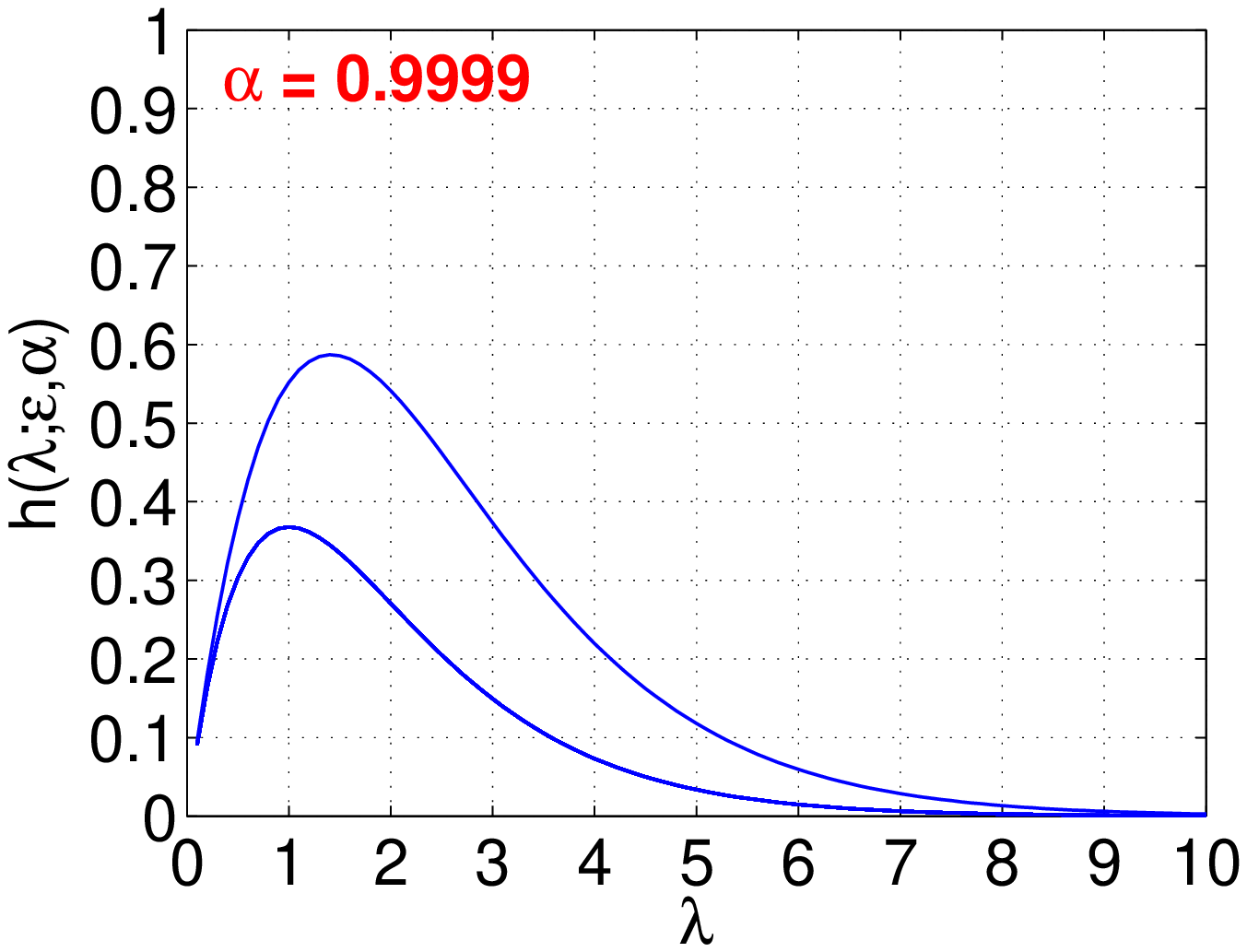}
}
\end{center}

\vspace{-0.25in}
\caption{$h(\lambda;\epsilon,\alpha)$ as defined in (\ref{eqn_h}) for $\alpha$ close to 1. As $\alpha\rightarrow1-$, the maximum of $h(\lambda;\epsilon,\alpha)$ approaches $e^{-1}$ attained at $\lambda=1$, for all $\epsilon<1$.  When $\epsilon=1$, the maximum approaches 0.5869, attained at $\lambda=\sqrt{2}$.}\label{fig_h1}
\end{figure}

This mean, the maximum of $h(\lambda;\epsilon<1,\alpha=1-)$ is $e^{-1}$ attained at $\lambda =1$, and the maximum of $h(\lambda;\epsilon=1,\alpha=1-)$ is $e^{-\sqrt{2}}(1+\sqrt{2})=0.5869$, attained at $\lambda=\sqrt{2}$, as confirmed by Figure~\ref{fig_h1}. In other words, it suffices to choose  the number of  measurements to be
\begin{align}
M= eK\log N/\delta\hspace{0.1in} \text{ if } \epsilon<1,\hspace{0.5in} M= 1.7038K\log N/\delta\hspace{0.1in} \text{ if } \epsilon=1
\end{align}

\section{Conclusion}

In this paper, we extend the prior work on {\em Compressed Counting meets Compressed Sensing}~\cite{Report:CCCS} and {\em very sparse stable random projections}~\cite{Proc:Li_Hastie_Church_KDD06,Proc:Li_KDD07} to the interesting problem of sparse recovery of nonnegative signals. The design matrix is highly sparse in that on average only $\gamma$-fraction of the entries are nonzero; and we sample the nonzero entries from an $\alpha$-stable maximally-skewed distribution where $\alpha\in(0,1)$. Our theoretical analysis demonstrates that the design matrix can be extremely sparse, e.g., $\gamma = \frac{1}{K}\sim \frac{2}{K}$. In fact, when $\alpha$ is away from 0, it is much more preferable to use a very sparse design.

\newpage\clearpage

\appendix

\section{Proof of Lemma~\ref{lem_Err_gamma}}~\label{app_lem_Err_gamma}

\begin{align}\notag
&\mathbf{Pr}\left(\hat{x}_{i,min,\gamma}> x_i + \epsilon\right)\\\notag
=&E\left(\mathbf{Pr}\left(\frac{y_j}{s_{ij}} > x_i + \epsilon,\ j\in T_i|T_i\right)\right)\\\notag
=&E\prod_{j\in T_i} \left[ \mathbf{Pr}\left(\frac{S_2}{S_1} >\frac{\epsilon}{\eta_{ij}^{1/\alpha}}\right)\right]\\\notag
=&E\prod_{j\in T_i}\left[1-F_\alpha\left(\left(\frac{\epsilon^\alpha}{\eta_{ij}}\right)^{1/(1-\alpha)}\right)\right]\\\notag
=&E\left\{\left[1-E\left\{F_\alpha\left(\left(\frac{\epsilon^\alpha}{\eta_{ij}}\right)^{1/(1-\alpha)}\right)\right\}\right]^{|T_i|}\right\}\\\notag
=&\left[1-\gamma + \gamma\left\{1-E\left\{F_\alpha\left(\left(\frac{\epsilon}{\eta_{ij}}\right)^{\alpha/(1-\alpha)}\right)\right\}\right\}\right]^M\\\notag
=&\left[1- \gamma E\left\{F_\alpha\left(\left(\frac{\epsilon^\alpha}{\eta_{ij}}\right)^{1/(1-\alpha)}\right)\right\}\right]^M
\end{align}

When $\alpha=0.5$, we have $F_\alpha(t) = \frac{2}{\pi}\tan^{-1}\sqrt{t}$ and hence
\begin{align}\notag
&\mathbf{Pr}\left(\hat{x}_{i,min,\gamma}> x_i + \epsilon\right)\\\notag
=&\left[1- \gamma E\left\{F_\alpha\left(\left(\frac{\epsilon^\alpha}{\eta_{ij}}\right)^{1/(1-\alpha)}\right)\right\}\right]^M\\\notag
=&\left[1- \gamma E\left\{\frac{2}{\pi}\tan^{-1}\left(\left(\frac{\sqrt{\epsilon}}{\eta_{ij}}\right)\right)\right\}\right]^M\\\notag
\leq&\left[1- \gamma \left\{\frac{2}{\pi}\tan^{-1}\left(\left(\frac{\sqrt{\epsilon}}{E\eta_{ij}}\right)\right)\right\}\right]^M \hspace{0.3in} (\text{Jensen's Inequality})\\\notag
\leq&\left[1- \gamma \left\{\frac{2}{\pi}\tan^{-1}\left\{\frac{1}{\gamma}\frac{\sqrt{\epsilon}}{\sum_{t\neq i} x_{t}^{1/2}}\right\}\right\}\right]^M
\end{align}

When $\alpha=0+$, we have $F_{0+}(t) =\frac{1}{1+1/t}$ and hence
\begin{align}\notag
&\mathbf{Pr}\left(\hat{x}_{i,min,\gamma}> x_i + \epsilon\right)\\\notag
=&\lim_{\alpha\rightarrow0+}\left[1-\gamma E\left\{ F_{0+}\left(\frac{1}{\eta_{ij}}\right)\right\}\right]^M\\\notag
=&\lim_{\alpha\rightarrow0+}\left[1-\gamma E\left\{ \left(\frac{1}{1+\eta_{ij}}\right)\right\}\right]^M\\\notag
\leq&\lim_{\alpha\rightarrow0+}\left[1-\gamma \left\{ \left(\frac{1}{1+E\eta_{ij}}\right)\right\}\right]^M\\\notag
\leq&\lim_{\alpha\rightarrow0+}\left[1-\gamma \frac{1}{1+\gamma K}\right]^M\\\notag
=&\left[1- \frac{1}{1/\gamma+ K}\right]^M
\end{align}

This completes the proof.

\section{Proof of Lemma~\ref{lem_Err_gamma0+}}\label{app_lem_Err_gamma0+}

\noindent\textbf{Proof:}\  \ When $\alpha=0+$, we have $F_{0+}(t) =\frac{1}{1+1/t}$ and hence
\begin{align}\notag
&\mathbf{Pr}\left(\hat{x}_{i,min,\gamma}> x_i + \epsilon\right)
=\lim_{\alpha\rightarrow0+}\left[1-\gamma E\left\{ \left(\frac{1}{1+\eta_{ij}}\right)\right\}\right]^M\\\notag
\end{align}

Suppose $x_i=0$, then as $\alpha\rightarrow0+$, $\eta_{ij}\sim Binomial(K,\gamma)$, and
\begin{align}\notag
&E \left(\frac{1}{1+\eta_{ij}}\right)\\\notag
=&\sum_{n=0}^K \frac{1}{1+n}\binom{K}{n}\gamma^n(1-\gamma)^{K-n}\\\notag
=&\sum_{n=0}^K \frac{1}{1+n}\frac{K!}{n!(K-n)!}\gamma^n(1-\gamma)^{K-n}\\\notag
=&\sum_{n=0}^K \frac{K!}{(n+1)!(K-n)!}\gamma^n(1-\gamma)^{K-n}\\\notag
=&\frac{1}{K+1}\frac{1}{\gamma}\sum_{n=0}^K \frac{(K+1)!}{(n+1)!((K+1)-(n+1))!}\gamma^{n+1}(1-\gamma)^{(K+1)-(n+1)}\\\notag
=&\frac{1}{K+1}\frac{1}{\gamma}\sum_{n=1}^{K+1} \frac{(K+1)!}{(n)!((K+1)-(n))!}\gamma^{n}(1-\gamma)^{(K+1)-(n)}\\\notag
=&\frac{1}{K+1}\frac{1}{\gamma}\left\{\sum_{n=0}^{K+1} \frac{(K+1)!}{(n)!((K+1)-(n))!}\gamma^{n}(1-\gamma)^{(K+1)-(n)}-(1-\gamma)^{K+1}\right\}\\\notag
=&\frac{1}{K+1}\frac{1}{\gamma}\left\{1-(1-\gamma)^{K+1}\right\}
\end{align}
Similarly, suppose $x_i> 0$, we have
\begin{align}\notag
E \left(\frac{1}{1+\eta_{ij}}\right)=\frac{1}{K}\frac{1}{\gamma}\left\{1-(1-\gamma)^{K}\right\}
\end{align}

Therefore, as $\alpha\rightarrow0+$, when $x_i=0$, we have
\begin{align}\notag
&\mathbf{Pr}\left(\hat{x}_{i,min,\gamma}> x_i + \epsilon\right)
=\left[1-\frac{1}{K+1}\left(1-(1-\gamma)^{K+1}\right)\right]^M
\end{align}
and when $x_i>0$, we have
\begin{align}\notag
&\mathbf{Pr}\left(\hat{x}_{i,min,\gamma}> x_i + \epsilon\right)
=\left[1-\frac{1}{K}\left(1-(1-\gamma)^{K}\right)\right]^M
\end{align}

To conclude the proof, we need to show
\begin{align}\notag
&\left[1-\frac{1}{K+1}\left(1-(1-\gamma)^{K+1}\right)\right]^M
\leq\left[1-\frac{1}{1/\gamma + K}\right]^M\\\notag
\Longleftrightarrow&\frac{1}{K+1}\left(1-(1-\gamma)^{K+1}\right)\geq \frac{1}{1/\gamma + K}\\\notag
\Longleftrightarrow& h(\gamma, K) = 1/\gamma-(1-\gamma)^{K+1}/\gamma - K(1-\gamma)^{K+1}- 1\geq 0
\end{align}
Note that $0\leq\gamma\leq1$, $h(0,K) = h(1,K)=h(\gamma,1) =0$. Furthermore
\begin{align}\notag
\frac{\partial h(\gamma, K)}{\partial K} =& -(1-\gamma)^{K+1}\log(1-\gamma)/\gamma - (1-\gamma)^{K+1} -  K(1-\gamma)^{K+1} \log(1-\gamma)\\\notag
=&-(1-\gamma)^{K+1}\left(\log(1-\gamma)/\gamma +1+K\log(1-\gamma)\right)\geq 0
\end{align}
as $\log(1-\gamma)/\gamma<-1$.  Thus, $h(\gamma,K)$ is a monotonically increasing function of $K$ and this completes the proof.

\section{Proof of Theorem~\ref{thm_worst_case}}\label{app_thm_worst_case}

\begin{align}\notag
\mathbf{Pr}\left(\hat{x}_{i,min,\gamma}> x_i + \epsilon\right) =& \left[1- \gamma E\left\{F_\alpha\left(\left(\frac{\epsilon^\alpha}{\eta_{ij}}\right)^{1/(1-\alpha)}\right)\right\}\right]^M\\\notag
\geq& \left[1-\gamma\mathbf{Pr}\left(\eta_{ij}=0\right)\right]^M\\\notag
=& \left[1-\gamma\left(1-\gamma\right)^{K-1+1_{x_i=0}}\right]^M\\\notag
\geq&\left[1-\gamma\left(1-\gamma\right)^{K}\right]^M
\end{align}
The minimum of $\gamma\left(1-\gamma\right)^{K-1+1_{x_i=0}}$ is attained at $\gamma = \frac{1}{K+1_{x_i=0}}$. If we choose $\gamma^* = \frac{1}{K+1}$, then
\begin{align}\notag
\mathbf{Pr}\left(\hat{x}_{i,min,\gamma}> x_i + \epsilon\right)
\geq&\left[1-\gamma^*\left(1-\gamma^*\right)^{K}\right]^M = \left[1-\frac{1}{K+1}\left(1-\frac{1}{K+1}\right)^K\right]^M
\end{align}
and it suffices to choose $M$ so that
\begin{align}\notag
M=\frac{1}{-\log \left[1-\frac{1}{K+1}\left(1-\frac{1}{K+1}\right)^K\right]}\log N/\delta
\end{align}
This completes the proof.

{

}

\end{document}